\documentclass{elsart}

\usepackage[square,comma]{natbib}
\usepackage{graphicx}
\usepackage{pxfonts}
\usepackage{lineno}

\usepackage{amssymb}
\usepackage{footnote}
\usepackage{multirow}
\usepackage{varwidth}
\usepackage{array}

\journal{}

\begin{document}

\thispagestyle{empty}
\begin{Large}
\textbf{DEUTSCHES ELEKTRONEN-SYNCHROTRON}

\textbf{\large{Ein Forschungszentrum der Helmholtz-Gemeinschaft}\\}
\end{Large}

DESY 13-040

March 2013

\begin{eqnarray}
\nonumber &&\cr \nonumber && \cr \nonumber &&\cr
\end{eqnarray}
\begin{eqnarray}
\nonumber
\end{eqnarray}
\begin{center}
\begin{Large}
\textbf{Grating monochromator for soft X-ray self-seeding the
European XFEL}
\end{Large}
\begin{eqnarray}
\nonumber &&\cr \nonumber && \cr
\end{eqnarray}

\begin{large}
Svitozar Serkez$^a$, Gianluca Geloni$^b$, Vitali Kocharyan$^a$ and
Evgeni Saldin$^a$
\end{large}

\textsl{\\$^a$Deutsches Elektronen-Synchrotron DESY, Hamburg}
\begin{large}

\end{large}
\textsl{\\$^b$European XFEL GmbH, Hamburg}
\begin{eqnarray}
\nonumber
\end{eqnarray}
\begin{eqnarray}
\nonumber
\end{eqnarray}
ISSN 0418-9833
\begin{eqnarray}
\nonumber
\end{eqnarray}
\begin{large}
\textbf{NOTKESTRASSE 85 - 22607 HAMBURG}
\end{large}
\end{center}
\clearpage
\newpage

\begin{frontmatter}



\title{Grating monochromator for soft X-ray self-seeding the European XFEL}


\author[DESY]{Svitozar Serkez \thanksref{corr},}
\thanks[corr]{Corresponding Author. E-mail address: svitozar.serkez@desy.de}
\author[XFEL]{Gianluca Geloni,}
\author[DESY]{Vitali Kocharyan}
\author[DESY]{and Evgeni Saldin}

\address[XFEL]{European XFEL GmbH, Hamburg, Germany}
\address[DESY]{Deutsches Elektronen-Synchrotron (DESY), Hamburg,
Germany}

\begin{abstract}
Self-seeding is a promising approach to significantly narrow the
SASE bandwidth of XFELs to produce nearly transform-limited pulses.
The implementation of this method in the soft X-ray wavelength range
necessarily involves gratings as dispersive elements. We study a
very compact self-seeding scheme with a grating monochromator
originally designed at SLAC, which can be straightforwardly
installed in the SASE3 type undulator beamline at the European XFEL.
The monochromator design is based on a toroidal VLS grating working
at a fixed incidence angle mounting without entrance slit. It covers
the spectral range from $300$ eV to $1000$ eV. The optical system
was studied using wave optics method (in comparison with ray
tracing) to evaluate the performance of the self-seeding scheme. Our
wave optics analysis takes into account the actual beam wavefront of
the radiation from the coherent FEL source, third order aberrations,
and errors from each optical element. Wave optics is the only method
available, in combination with FEL simulations, for the design of a
self-seeding monochromator without exit slit. We show that, without
exit slit, the self-seeding scheme is distinguished by the much
needed experimental simplicity, and can practically give the same
resolving power (about 7000) as with an exit slit.  Wave optics is
also naturally applicable to calculations of the self-seeding scheme
efficiency, which include the monochromator transmittance and the
effect of the mismatching between seed beam and electron beam.
Simulations show that the FEL power reaches $1$ TW and that the
spectral density for a TW pulse is about two orders of magnitude
higher than that for the SASE pulse at saturation.
\end{abstract}

%
%
%
\end{frontmatter}



\section{\label{sec:intro} Introduction}

Self-seeding is a promising approach to significantly narrow the
SASE bandwidth and to produce nearly transform-limited pulses
\cite{SELF}-\cite{OURY5b}. Considerable effort has been invested in
theoretical investigation and $R\&D$ at the LCLS leading to the
implementation of a hard X-ray self-seeding (HXRSS) setup that
relies on a diamond monochromator in transmission geometry.
Following the successful demonstration of the HXRSS setup at the
LCLS \cite{EMNAT}, there is a need for an extension of the method in
the soft X-ray range.

In general, a self-seeding setup consists of two undulators
separated by a photon monochromator and an electron bypass, normally
a four-dipole chicane. The two undulators are resonant at the same
radiation wavelength. The SASE radiation generated by the first
undulator passes through the narrow-band monochromator. A
transform-limited pulse is created, which is used as a coherent seed
in the second undulator. Chromatic dispersion effect in the bypass
chicane smears out the microbunching in the electron bunch produced
by the SASE lasing in the first undulator. The electrons and the
monochromatized photon beam are recombined at the entrance of the
second undulator, and radiation is amplified by the electron bunch
until saturation is reached. The required seed power at the
beginning of the second undulator must dominate over the shot noise
power within the gain bandpass, which is order of a kW in the soft
X-ray range.

For self-seeding in the soft x-ray range, proposed monochromators
usually consists of a grating \cite{SELF}, \cite{STTF}. Recently, a
very compact soft x-ray self-seeding (SXRSS) scheme has appeared,
based on grating monochromator \cite{FENG}-\cite{FENG3}.  The delay
of the photons in the last SXRSS version \cite{FENG3} is about $0.7$
ps only. The proposed monochromator is composed of only three
mirrors and a toroidal VLS grating. The design adopts a constant,
$1$ degree incidence-angle mode of operation, in order to suppress
the influence of movement of the source point in the first SASE
undulator on the monochromator performance.

In this article we study the performance of the soft X-ray
self-seeding scheme for the European XFEL upgrade. In order to
preserve the performance of the baseline undulator, we fit the
magnetic chicane within the space of a single $5$ m undulator
segment space at SASE3. In this way, the setup does not perturb the
undulator focusing system. The magnetic chicane accomplishes three
tasks by itself. It creates an offset for monochromator
installation, it removes the electron microbunching produced in the
upstream seed undulator, and it acts as an electron beam delay line
for compensating the optical delay introduced by the monochromator.
The monochromator design is compact enough to fit with this magnetic
chicane design. The monochromator design adopted in this paper is an
adaptation of the novel one by Y. Feng et al. \cite{FENG3}, and is
based on toroidal VLS grating, and has many advantages. It consists
of a few elements. In particular, it operates without entrance slit,
and is, therefore, very compact. Moreover, it can be simplified
further. Quite surprisingly, a monochromatic seed can be directly
selected by the electron beam at the entrance of the second
undulator. In other words, the electron beam plays, in this case,
the role of an exit slit. By using a wave optics approach and FEL
simulations we show that the monochromator design without exit slits
works in a satisfactory way.

With the radiation beam monochromatized down to the Fourier
transform limit, a variety of very different techniques leading to
further improvement of the X-ray FEL performance become feasible. In
particular, the most promising way to extract more FEL power than
that at saturation is by tapering the magnetic field of the
undulator  \cite{TAP1}-\cite{WANG}. A significant increase in power
is achievable by starting the FEL process from a monochromatic seed
rather than from shot noise \cite{FAWL}-\cite{LAST}. In this paper
we propose a study of the soft X-ray self-seeding scheme for the
European XFEL, based on start-to-end simulations for an electron
beam with 0.1 nC charge \cite{S2ER}. Simulations show that the FEL
power of the transform-limited soft X-ray pulses may be increased up
to 1 TW by properly tapering the baseline (SASE3) undulator. In
particular, it is possible to create a source capable of delivering
fully-coherent, 10 fs (FWHM) soft X-ray pulses with $10^{14}$
photons per pulse in the water window.

The availability of free undulator tunnels at the European XFEL
facility offers a unique opportunity to build a beamline optimized
for coherent diffraction imaging of complex molecules like proteins
and other biologically interesting structures. Full exploitation of
these techniques require 2 keV - 6 keV photon energy range and TW
peak power pulses. However, higher photon energies are needed to
reach anomalous edges of commonly used elements (such as Se) for
anomalous experimental phasing. Potential users of the bio-imaging
beamline also wish to investigate large biological structures in the
soft X-ray photon energy range down to the water window.  A
conceptual design for the undulator system of such a bio-imaging
beamline based on self-seeding schemes developed for the European
XFEL was suggested in \cite{BIO1}-\cite{BIO3}. The bio-imaging
beamline would be equipped with two different self-seeding setups,
one providing monochromatization in the hard x-ray wavelength range,
using diamond monochromators and one providing monochromatization in
the soft x-ray range using a grating monochromator. In relation to
this proposal, we note that the design for a soft x-ray self-seeding
scheme discussed here can be implemented not only at the SASE3
beamline but, as discussed in \cite{BIO1}-\cite{BIO3}, constitutes a
suitable solution for the bio-imaging beamline in the soft x-ray
range as well.

\section{\label{sec:descr} Self-seeding setup description}

\begin{figure}
\begin{center}
\includegraphics[clip, width=\textwidth]{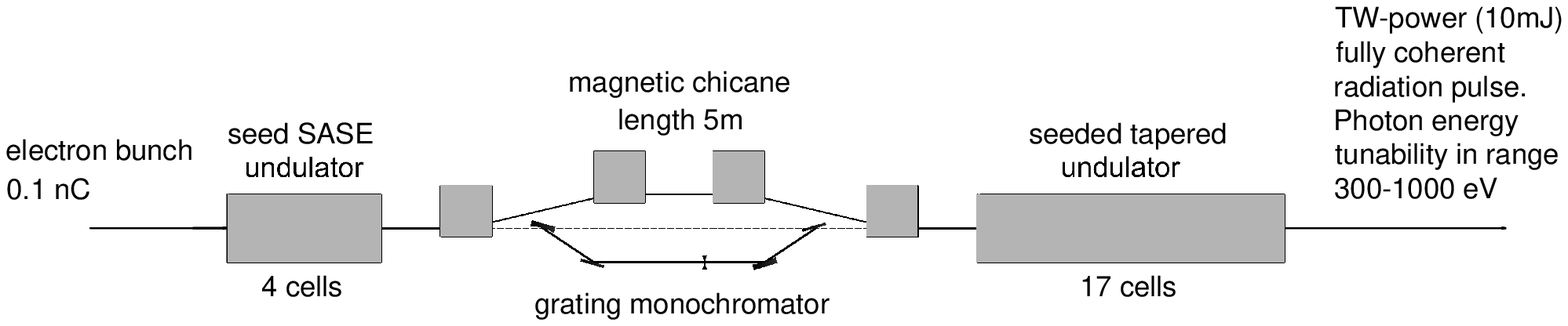}
\end{center}
\caption{Design of the SASE3 undulator system for TW mode of
operation in the soft X-ray range.} \label{mon_lay_ov}
\end{figure}

\begin{figure}
\begin{center}
\includegraphics[clip, width=\textwidth]{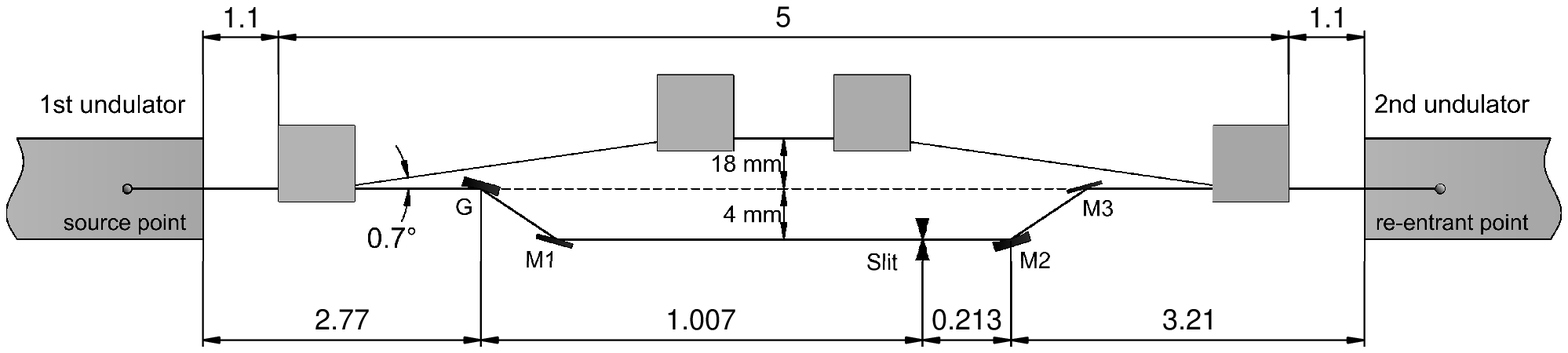}
\end{center}
\caption{Layout of the SASE3 self-seeding system, to be located in
the space freed after removing the undulator segment U5. The compact
grating monochromator design relies on a scheme originally proposed
at SLAC. G is a toroidal VLS grating. M1 is a rotating plane mirror,
M2 is a tangential cylindrical mirror, M3 is a plane mirror used to
steer the beam. The deflection of both electron and photon beams is
in the horizontal direction.} \label{mon_lay}
\end{figure}

\begin{figure}
\begin{center}
\includegraphics[clip,width=0.75\textwidth]{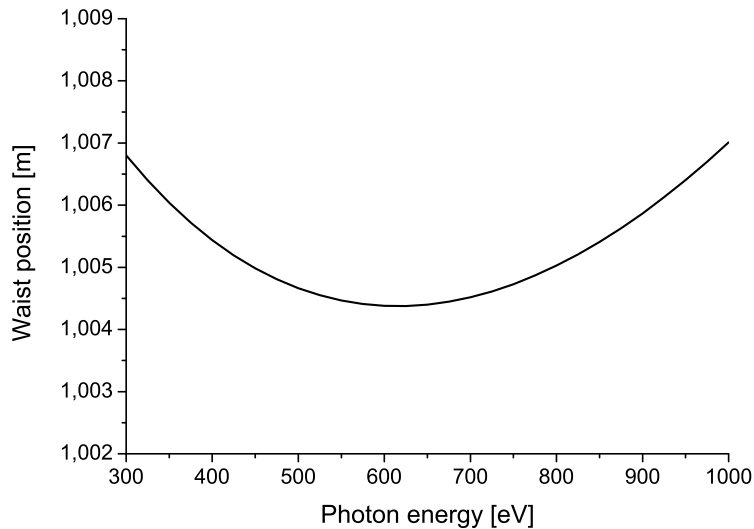}
\end{center}
\caption{Focusing at the slit. Distance between waist, characterized
by plane wavefront, and grating as a function of the photon energy.}
\label{waistpos}
\end{figure}

\begin{figure}
\begin{center}
\includegraphics[width=0.75\textwidth]{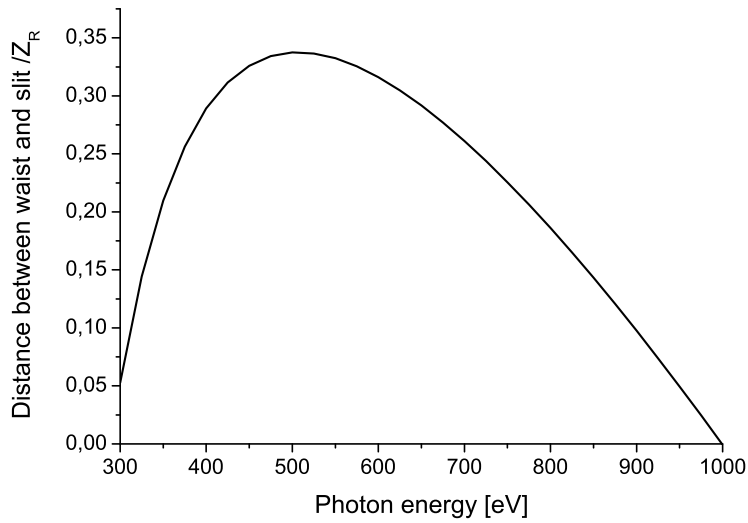}
\end{center}
\caption{Focusing at the slit. Variation of the distance between
waist and slit normalized on the Rayleigh range as a function of the
photon energy.} \label{waistpos_zr}
\end{figure}

\begin{figure}
\begin{center}
\includegraphics[clip,width=0.75\textwidth]{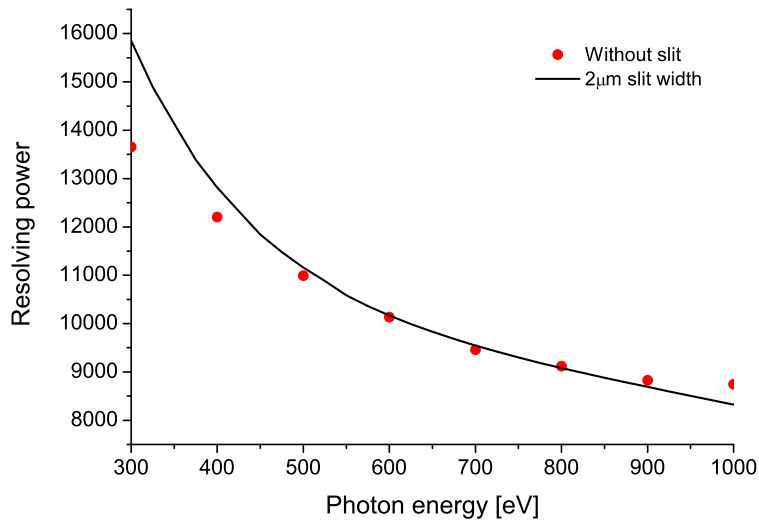}
\end{center}
\caption{Resolving power as a function of the photon energy for a
monochromator equipped with exit slit (bold curve) and without exit
slit (circles). The calculation with exit slit is for a slit width
of $2 \mu$m.} \label{resol_compare}
\end{figure}

\begin{figure}
\begin{center}
\includegraphics[clip,width=0.75\textwidth]{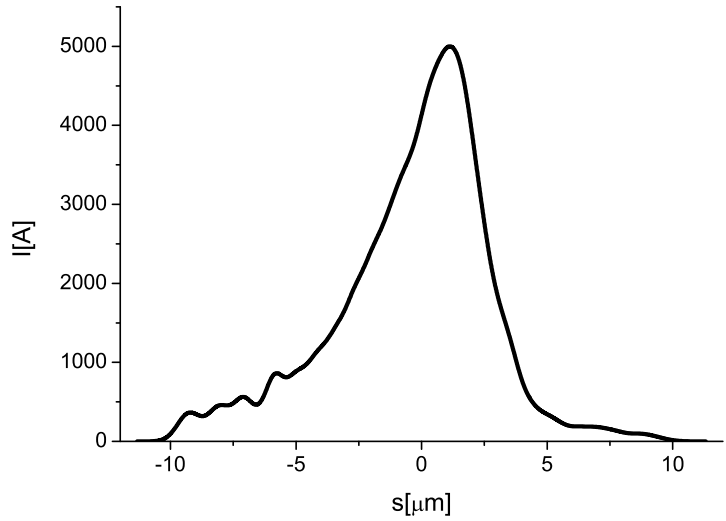}
\end{center}
\caption{Current profile for a $100$ pC electron bunch at the
entrance of the first undulator.} \label{curr}
\end{figure}

\begin{figure}
\begin{center}
\includegraphics[width=0.75\textwidth]{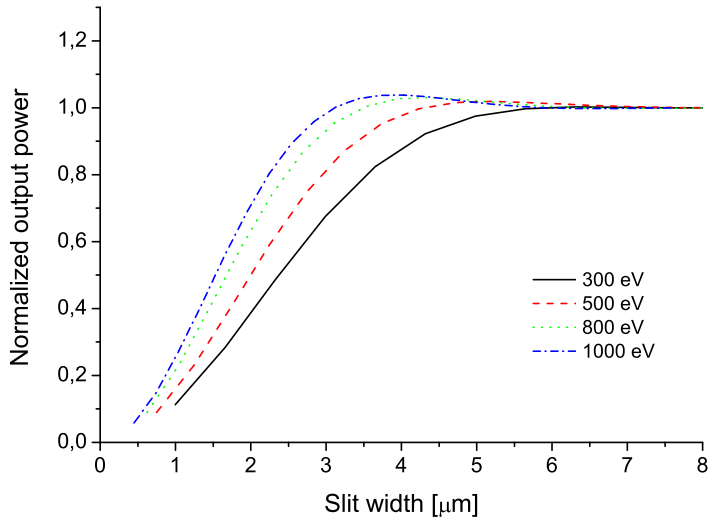}
\end{center}
\caption{Results of seeding efficiency simulations, showing the
normalized output power from the second FEL amplifier as a function
of the exit slit width for different photon energies. The FEL
amplifier operates in the linear regime. Results are obtained by
wave optics and FEL simulations.} \label{mismatch_slit_sec2}
\end{figure}

\begin{figure}
\begin{center}
\includegraphics[width=0.75\textwidth]{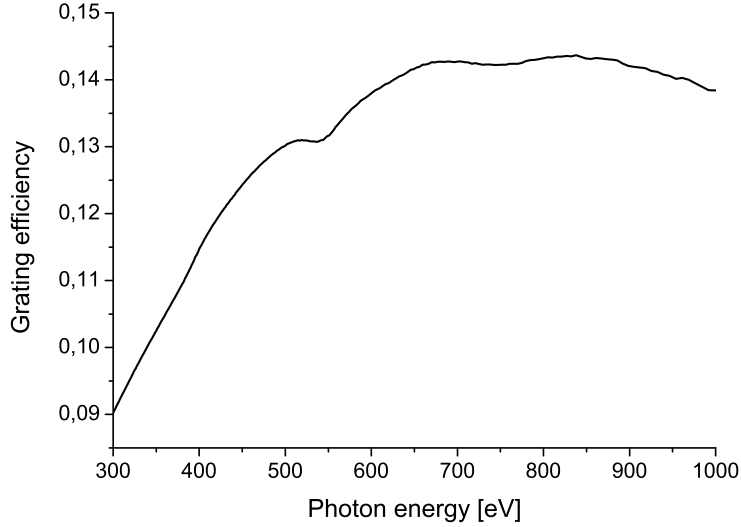}
\end{center}
\caption{First order efficiency of the blazed groove profile. Here
the groove density is $1100$ lines/mm, Pt coating is assumed, at an
incidence angle of $1^\circ$. The blaze angle is $1.2^\circ$; the
anti-blaze angle is $90^\circ$.} \label{gr_eff}
\end{figure}

\begin{figure}
\includegraphics[width=0.50\textwidth]{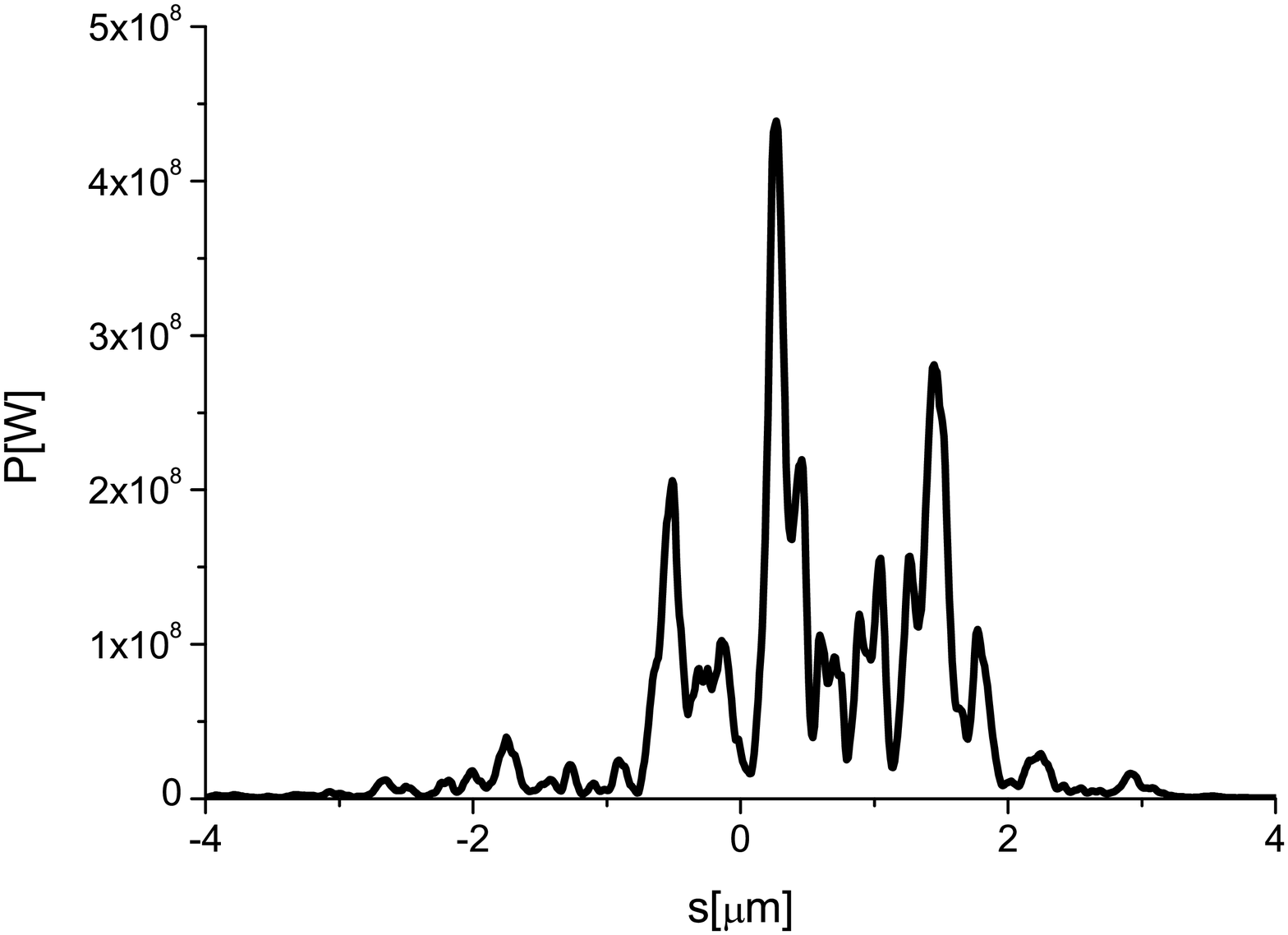}
\includegraphics[width=0.50\textwidth]{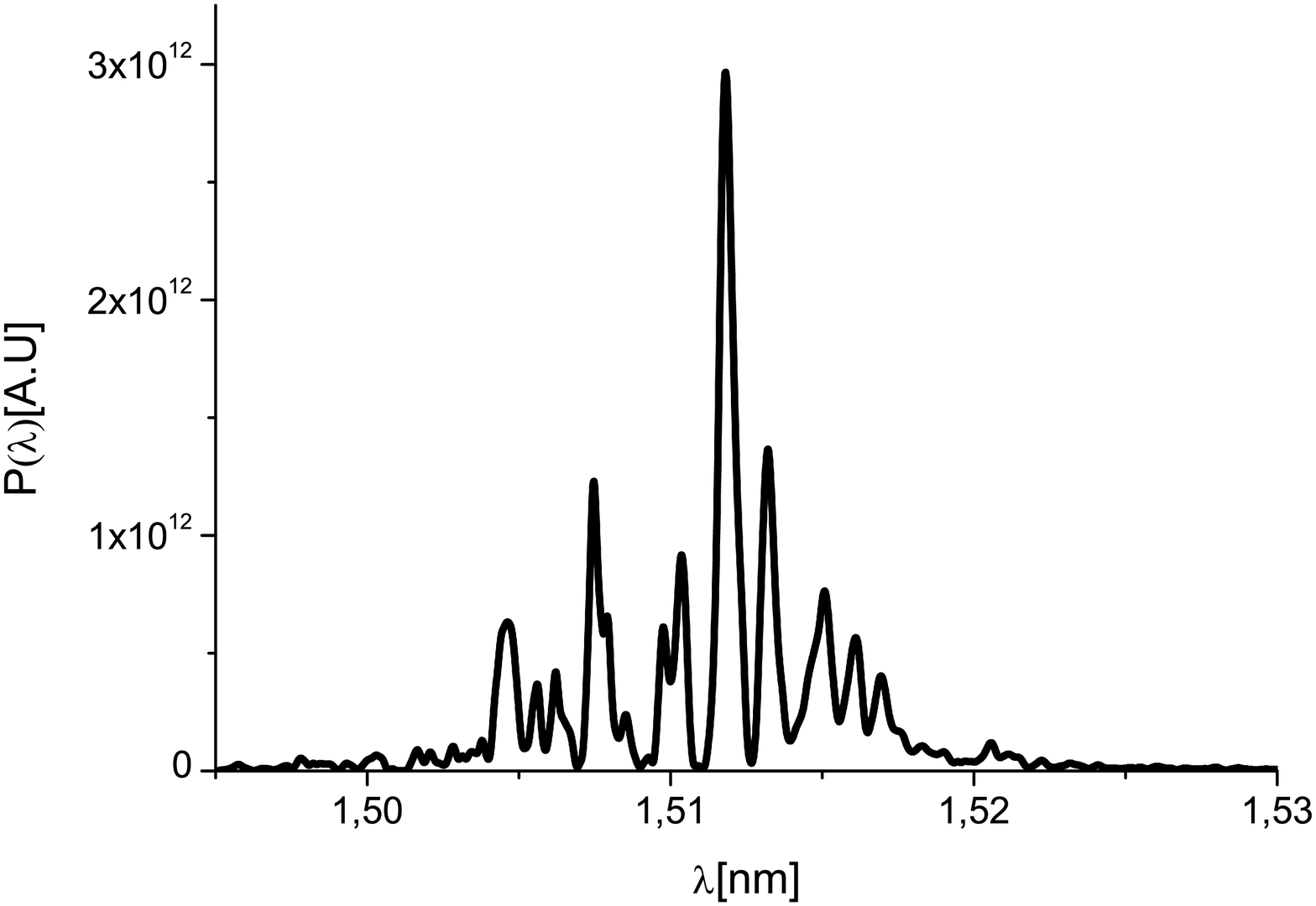}
\caption{Power distribution and spectrum of the SASE soft x-ray
radiation pulse at the exit of the first undulator.}
\label{inp_pow_spec}
\end{figure}

\begin{figure}
\begin{center}
\includegraphics[width=0.75\textwidth]{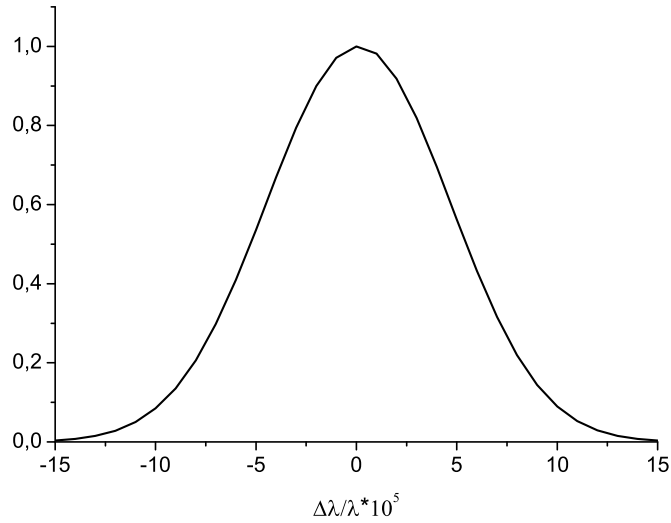}
\end{center}
\caption{Line profile of the self-seeding monochromator without exit
slit. The calculation is for a photon energy of $0.8$ keV. The
overall efficiency of the monochromator beamline is about $5\%$.}
\label{lineprof}
\end{figure}
\begin{figure}
\includegraphics[width=0.50\textwidth]{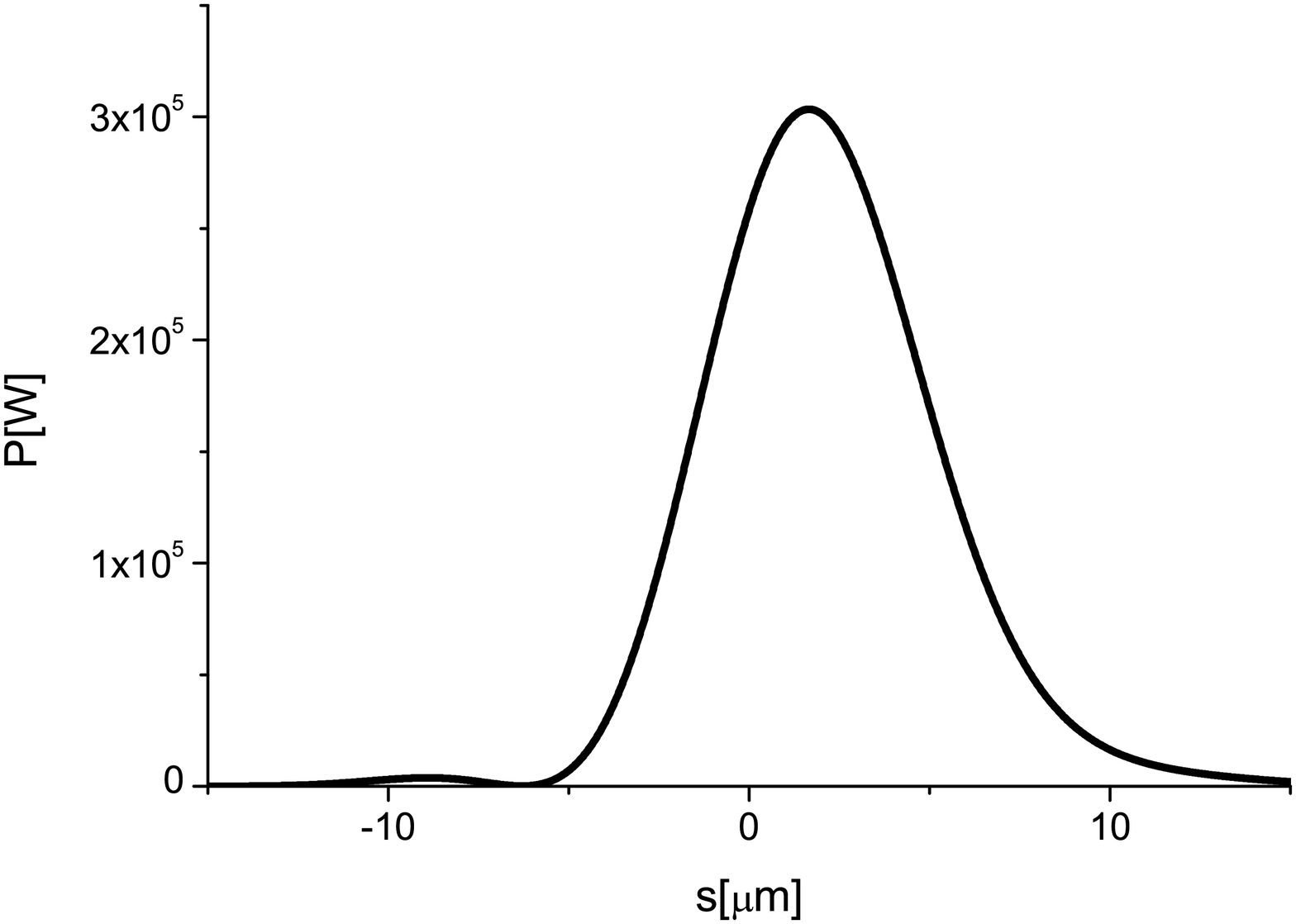}
\includegraphics[width=0.50\textwidth]{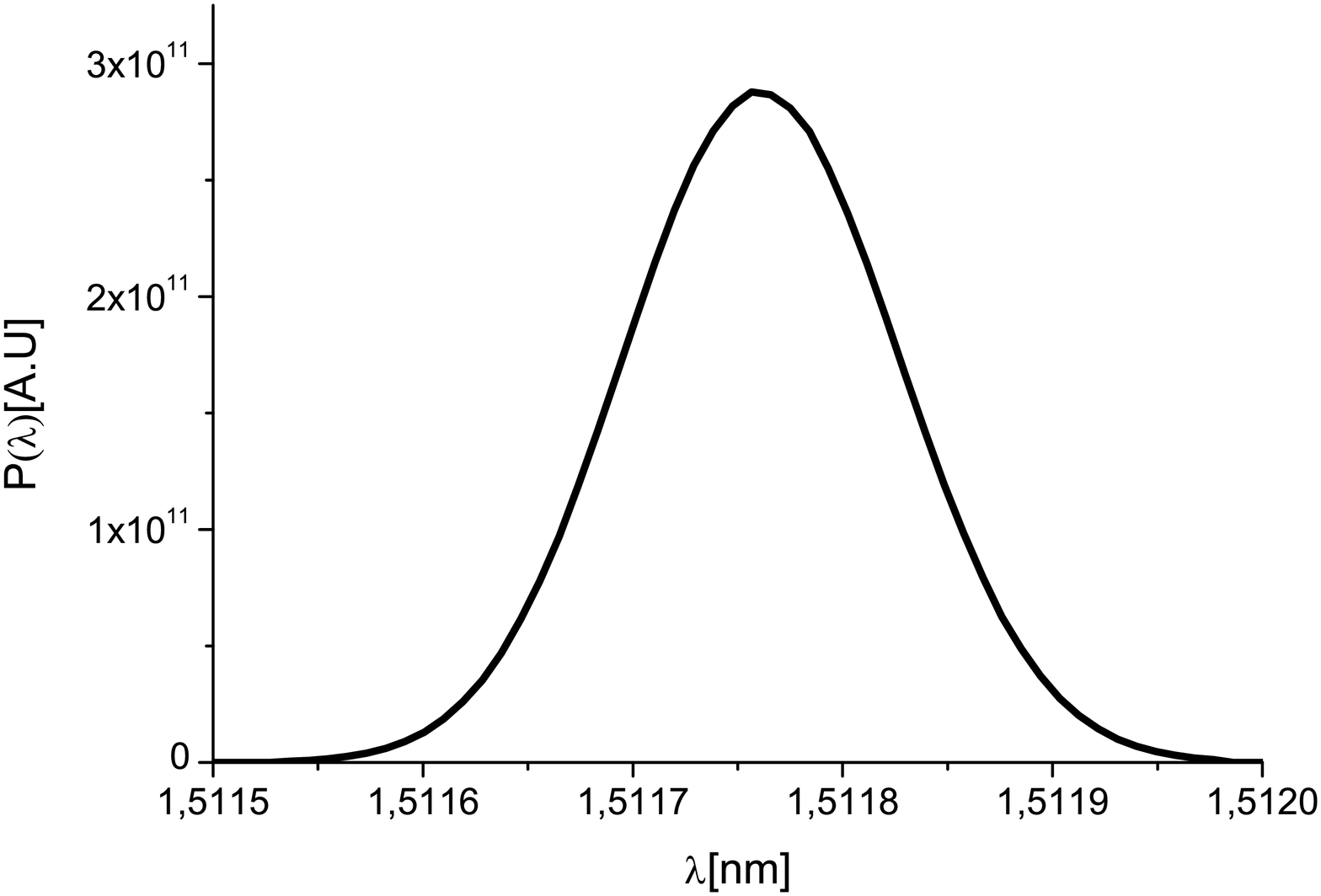}
\caption{Power distribution and spectrum of the SASE soft x-ray
radiation pulse after the monochromator. This pulse is used to seed
the electron bunch at the entrance of the second undulator.}
\label{mon_pow_spec}
\end{figure}

\begin{figure}
\includegraphics[width=0.50\textwidth]{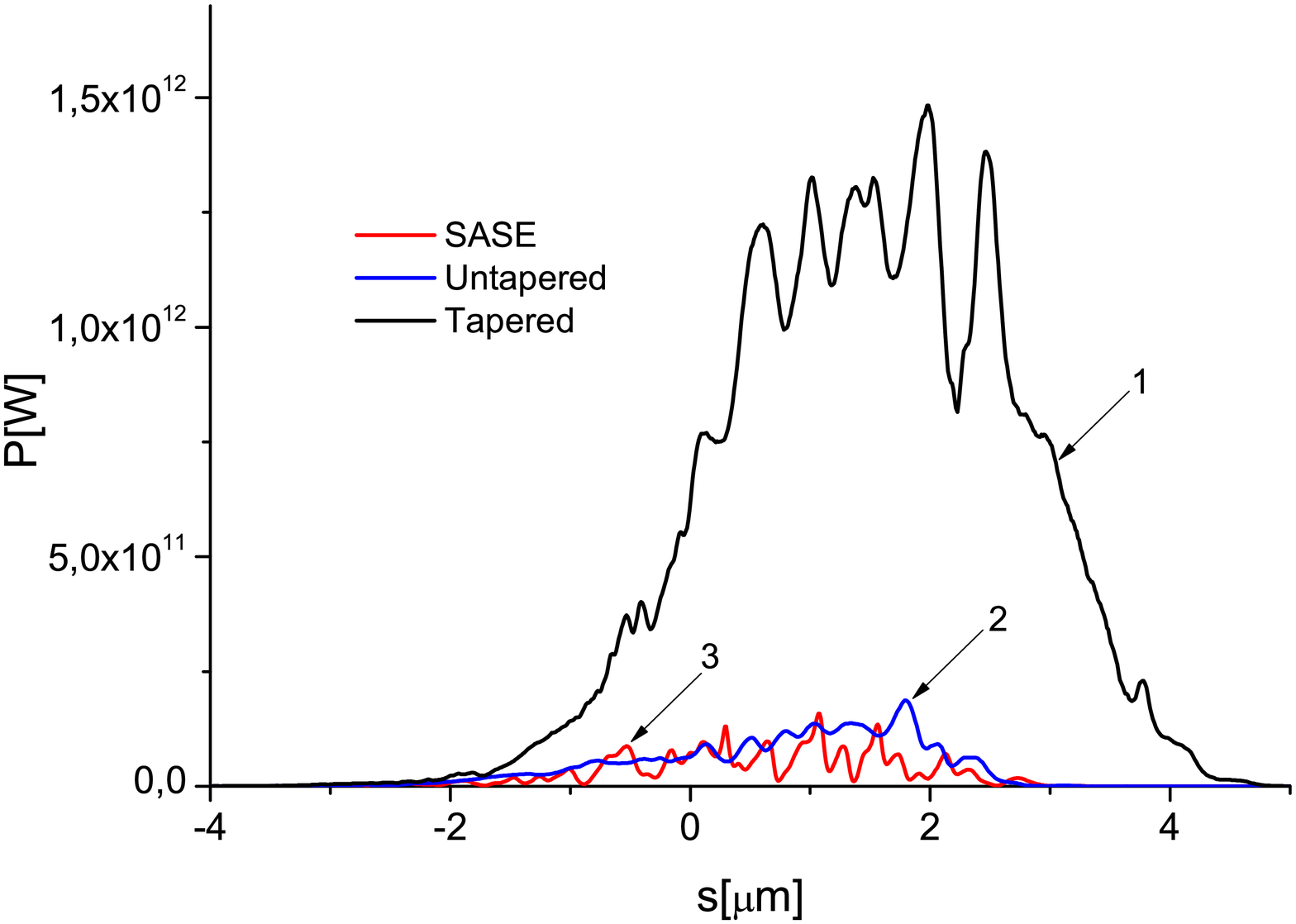}
\includegraphics[width=0.50\textwidth]{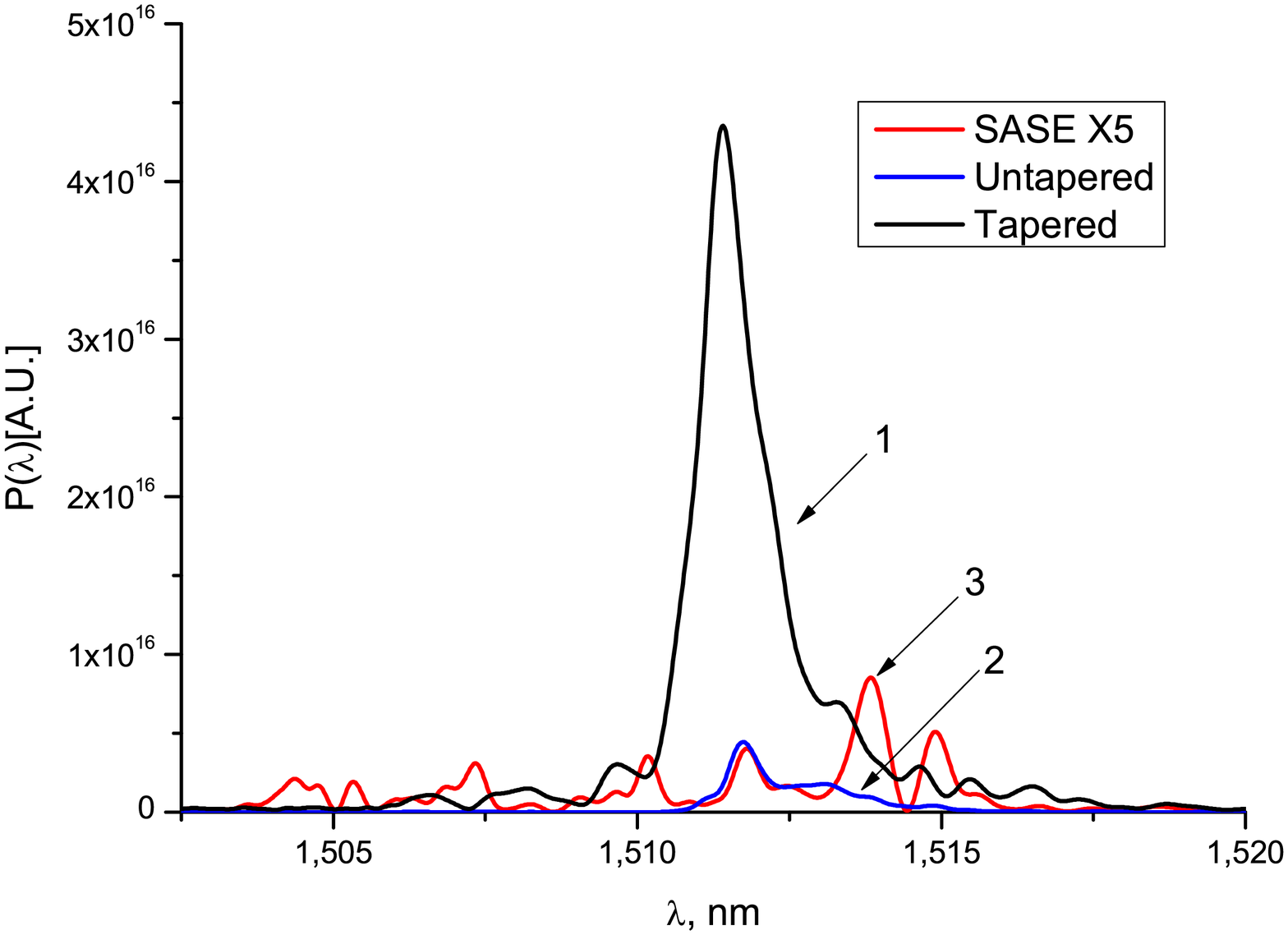}
\caption{Power distribution and spectrum of the output soft x-ray
radiation pulse. Curve 1 - seeded FEL output with tapering; curve 2
- seeded FEL output without tapering; curve 3 - SASE FEL output in
saturation. Here $\lambda=1.5$ nm, corresponding to $800$ eV.}
\label{outp_pow_spec}
\end{figure}

\begin{figure}
\begin{center}
\includegraphics[width=0.75\textwidth]{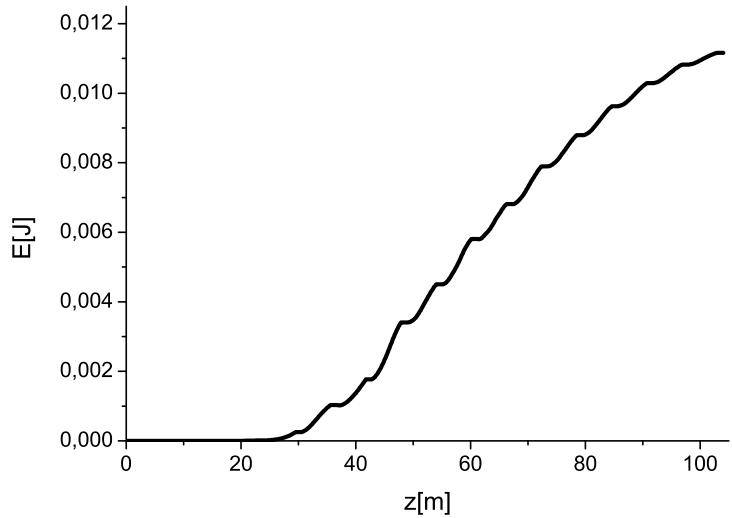}
\end{center}
\caption{Energy of the seeded FEL pulse as a function of the
distance inside the output undulator.} \label{nrg_gain}
\end{figure}
%


\begin{table}[htbp]
\caption{Parameters for the x-ray optical elements \newline $^1$
Distance to grating. \newline $^2$ Principal ray hit point.}

\begin{tabular}{|c |  c |  c | c | c | c | c |}
\hline \hline \multirow{2}{*}{\textbf{Element}} &
\multirow{2}{*}{\textbf{Parameter}} &
\multicolumn{3}{c|}{\textbf{Value at photon energy}} &
\multirow{2}{*}{\begin{minipage}{0.7in}\textbf{Required
precision}\end{minipage}} &
\multirow{2}{*}{\textbf{Unit}}\\
\cline{3-5}
&&\textbf{300 eV}&\textbf{600 eV}&\textbf{1000 eV}&&\\
\hline\hline
G   & Line density ($k$)                & \multicolumn{3}{c|}{1123}         &0.2\%  &l/mm\\
G   & Linear coeff ($n_1$)          & \multicolumn{3}{c|}{2.14}             &1\%        &l/mm$^2$\\
G   & Quad coeff ($n_2$)            & \multicolumn{3}{c|}{0.003}            &50\%   &l/mm$^3$\\
G   & Groove profile                & \multicolumn{3}{c|}{Blased 1.2$^\circ$}   &-  &-\\
G,M1    & Roughness (rms)               & \multicolumn{3}{c|}{-}                & 2 &nm\\
G   & Tangential radius                 & \multicolumn{3}{c|}{160}          & 1\%   &m\\
G   & Sagittal radius               & \multicolumn{3}{c|}{0.25}             & 10\%  &m\\
G   & Diffraction order                 & \multicolumn{3}{c|}{+1}               &   &-\\
G   & Incident angle                & \multicolumn{3}{c|}{1} & -
&deg\\ \cline{3-5}
G   & Exit angle                    & 5.615     & 4.028 & 3.816     &-  &deg\\
\hline \hline
    & Source distance\footnotemark[1]   & 3160  & 3470  & 3870      &-  &mm\\
    & Source size               & 30.3  & 27.7      & 24.2      &-  &$\mu$m\\
    & Image distance\footnotemark[1]    & 1007  & 1004  & 1007      &-  &mm\\
    & Image size                & 2.22  & 2.45  & 2.22      &-  &$\mu$m\\
\hline \hline
M1  &Location\footnotemark[1]\footnotemark[2]   & 33.2  & 43.8  & 52.6  &-  &mm\\
M1  &Incident angle             & 3.307 & 2.514 & 2.093     &-  &deg\\
\hline \hline
S   &Slit location\footnotemark[1]      & \multicolumn{3}{c|}{1007}         &0.5    &mm\\
S   &Slit width                 & \multicolumn{3}{c|}{2}                &5\%    &$\mu$m\\
\hline \hline
M2  &Location\footnotemark[1]       & \multicolumn{3}{c|}{1220}         &1  &mm\\
M2  &Incident angle             & \multicolumn{3}{c|}{0.859}            &-  &deg\\
M2  & Tangential radius                 & \multicolumn{3}{c|}{27.3}         & 1\%   &m\\
\hline \hline
M3  &Location\footnotemark[1]       & \multicolumn{3}{c|}{1348.3}           &-  &mm\\
M3  &Incident angle             & \multicolumn{3}{c|}{0.859}            &-  &deg\\
\hline \hline
    &Optical delay              & 935       & 757       &662            &-  &fs\\
\hline \hline
\end{tabular}

\label{table:components}

\end{table}

A design of the self-seeding setup based on the undulator system for
the European XFEL baseline is sketched in Fig. \ref{mon_lay_ov}. The
method for generating highly monochromatic, high power soft x-ray
pulses exploits a combination of a self-seeding scheme with grating
monochromator with an undulator tapering technique. The self-seeding
setup is composed by a compact grating monochromator originally
proposed at SLAC \cite{FENG3}, yielding about $0.7$ ps optical
delay, and a $5$ m-long magnetic chicane.

Usually, a grating monochromator consists of an entrance slit, a
grating, and an exit slit. The grating equation, which describes how
the monochromator works, relies on the principle of interference
applied to the light coming from the illuminated grooves. Such
principle though, can only be applied when phase and amplitude
variations in the electromagnetic field are well-defined across the
grating, that is when the field is perfectly transversely coherent.
The purpose of the entrance slit is to supply a transversely
coherent radiation spot at the grating, in order to allow the
monochromator to work with an incoherent source and with a given
resolution. However, an FEL source is highly transversely coherent
and no entrance slit is required in this case \cite{SVET,ROPE}.

Fig. \ref{mon_lay} shows the optical configuration of the
self-seeding monochromator. Table \ref{table:components} summarizes
the optical parameters of the setup. The design of the monochromator
was optimized with respect to the resolving power and the seeding
efficiency. The design energy range of the monochromator is in the
$0.3$ keV - $1$ keV interval with a resolution of about $7000$. It
is only equipped with an exit slit. A toroidal grating with variable
line spacing (VLS) is used for imaging the FEL source to the exit
slit of the monochromator. The grating has a groove density of
$1120$ lines/mm. The first coefficient $D_1$ of the VLS grating is
$D_1 = 2.1/ \mathrm{mm}^2$. The grating will operate in fixed
incident angle mode in the $+1$ order. The incident X-ray beam is
imaged at the exit slit and re-imaged at the entrance of the seed
undulator by a cylindrical mirror M2. In the sagittal plane, the
source is imaged at the entrance of the seed undulator directly by
the grating. The monochromator scanning is performed by rotating the
post-grating plane mirror. The scanning results on a
wavelength-dependent optical path. Therefore, a tunability of the
path length in the magnetic chicane in the range of $0.05$ mm is
required to compensate for the change in the optical path.

The choice was made to use a toroidal VLS grating similar to the
LCLS design \cite{FENG3}. As pointed out in that reference, the
source point in the SASE undulator moves upstream with the photon
energy. The proposed design has been chosen in order to minimize the
variation of the image distance. The object distance was based on
FEL simulations of the SASE3 undulator at the exit of the fourth
segment U4, Fig. \ref{mon_lay_ov}. The monochromator performance was
calculated using wave optics. The exact location of the waist,
characterized by a plane wavefront, Fig. \ref{waistpos} and Fig.
\ref{waistpos_zr}, was found to vary with the energy around the slit
within $2.7$ mm, which is small compared to the Rayleigh range, Fig.
\ref{waistpos_zr}. This defocusing effect was fully accounted for in
the wave optics treatment, and the impact of this effect on the
resolving power is negligible. The resolving power achievable with
the exit slit is shown in Fig. \ref{resol_compare}. It approaches
$8000$, and is sufficient to produce temporally transform-limited
seed pulses with FWHM duration between $25$ fs and $50$ fs over the
designed photon energy range. This duration is sufficiently longer
than the FWHM duration of the electron bunch, about $15$ fs in
standard mode of operation at $0.1$ nC charge, Fig. \ref{curr}. The
resolving power depends on the size of the FEL source inside the
SASE undulator, on the size of the exit slit (assumed fixed at $2
\mu$m) and on third order optical aberrations.

The operation of the self-seeding scheme involves simultaneous
presence of monochromatized radiation and electron beam in the seed
undulator. This suggests to consider a particularly interesting
approach to solve the task of creating a monochromatized seed. In
fact, the resolving power needed for seeding can be achieved without
exit slit by combining the presence of radiation and electron beam
in the seed undulator. The influence of the spatial dispersion in
the image plane at the entrance of the seed undulator on the
operation of the self-seeding setup without exit slit can be
quantified by studying the input coupling factor between the seed
beam and the ground mode of the FEL amplifier. A combination of wave
optics and FEL simulations is the only method available for
designing such self-seeding monochromator without exit slit. This
design has the advantage of a much needed experimental simplicity,
and could deliver a resolving power as that with the exit slit. The
comparison of resolving powers for these two designs is shown in
Fig. \ref{resol_compare}. The size of the beam waist near the slit
is about $2.2$-$2.4~\mu$m. The operation without exit slit would
give worse resolving power than the conventional mode of operation
only when the slit size is smaller than  $2~ \mu$m. Wave optics and
FEL simulations are naturally applicable also for calculating
suppression of the input coupling factor, due to the effect of a
finite size of the exit slit. The effect of the slit on the seeding
efficiency shown in Fig. \ref{mismatch_slit_sec2}. When the slit
size is smaller than $2 ~\mu$m, the effective seed power is reduced
by as much as a factor $2 - 3$. We conclude that the mode of
operation without exit slit is superior to the conventional mode of
operation, and a finite slit size would only lead to a reduction of
the monochromator performance.

The efficiency of the grating should be specified over the range of
photon energies where the grating will be used. The efficiency was
optimized by varying the groove shapes. Blazed grating was optimized
by adjusting the blaze angle; sinusoidal grating by adjusting the
groove depth, and rectangular grating by adjusting the groove depth,
and assuming a duty cycle of $50\%$. The blazed profile is
substantially superior to both sinusoidal and laminar alternatives.
For the specified operating photon energy range, the optimal blaze
angle is $1.2$ degree, and the expected grating efficiency with
platinum coating is shown in Fig. \ref{gr_eff}. This curve assumes a
constant incident angle of $1$ degree.

The electron beam chicane contains four identical dipole magnets,
each of them $0.5$ m-long. Given a magnetic field $B = 0.8 T$ and an
electron momentum $p = 10 $GeV/c, this length corresponds to a
dipole bending angle of $0.7$ degrees. The choice of the strength of
the magnetic chicane only depends on the delay that we want to
introduce. In our case, as already mentioned, it amounts to $0.23$
mm, or $0.7$ ps. Parameters discussed above fit with a short, $5$
m-long magnetic chicane to be installed in place of a single
undulator module. Such chicane, albeit very compact, is however
strong enough to create a sufficiently large transverse offset for
the installation of the optical elements of the monochromator.

Despite the unprecedented increase in peak power of the X-ray pulses
at SASE X-ray FELs some applications, including bio-imaging ,
require still higher photon flux \cite{HAJD}-\cite{BERG}. The most
promising way to extract more FEL power than that at saturation is
by tapering the magnetic field of the undulator. Tapering consists
in a slow reduction of the field strength of the undulator in order
to preserve the resonance wavelength, while the kinetic energy of
the electrons decreases due to FEL process. The undulator taper
could be simply implemented at discrete steps from one undulator
segment to the next. The magnetic field tapering is provided by
changing the undulator gap. Here we study a scheme for generating
TW-level soft X-ray pulses in a SASE3 tapered undulator, taking
advantage of the highly monochromatic pulses generated with the
self-seeding technique, which make the tapering very efficient. We
optimized our setup based on start-to-end simulations for an
electron beam with $100$ pC charge. In this way, the output power of
SASE3 could be increased from the baseline value of $100$ GW to
about a TW in the photon energy range between $0.3$ keV and $1$ keV.

Summing up, the overall self-seeding setup proposed here consists of
three parts: a SASE undulator, a self-seeding grating monochromator
and an output undulator in which the monochromtic seed signal is
amplified up to the TW power level. Calculations show that in order
not to spoil the electron beam quality and to simultaneously reach
signal dominance over shot noise, the number of cells in the first
(SASE) undulator should be equal to $4$. The output undulator
consists of two sections. The first section is composed by an
uniform undulator, the second section by a tapered undulator. The
transform-limited seed pulse is exponentially amplified passing
through the first uniform part of the output undulator. This section
is long enough, $6$ cells, in order to reach saturation, which
yields about $100$ GW power. Finally, in the second part of the
output undulator the monochromatic FEL output is enhanced up to the
TW power level taking advantage of a $3.5 \%$ taper of the undulator
magnetic field over the last $11$ cells after saturation.

Simulations were performed with the help of the Genesis code
\cite{GENE} running on a cluster in the following way: first we
calculated the 3D field distribution at the exit of the first
undulator, and downloaded the field file. Subsequently, we performed
a temporal Fourier transformation followed by filtering through the
monochromator, by using the filter amplitude transfer function.  The
electron microbunching is washed out by presence of non-zero chicane
momentum compaction factor $R_{56}$. Therefore, for the second
undulator we used a beam file with no initial microbunching, and
with an energy spread induced by the FEL amplification process in
the first SASE undulator. The amplification process in the second
undulator starts from the seed field-file. Shot noise initial
conditions were included, see section \ref{sec:FELsim} for details.
The output power and spectrum after the first (SASE) undulator tuned
at $1.5$ nm is shown in Fig. \ref{inp_pow_spec}. The instrumental
function is shown in Fig. \ref{lineprof}. The shape of this curve
was found as a response of the input coupling factor on the offset
of the seed monochromatic beam at the entrance of the seed undulator
due to spatial dispersion. The absolute value of the transmittance
accounts for the absorption of the monochromatic beam in the grating
and in the three mirrors, for a total of $5 \%$.  The influence of
the transverse mismatching of the seed beam at the entrance of the
seed undulator is accounted for by an additional suppression of the
input coupling factor. The resolution of the self-seeding
monochromator is good enough, and the spectral width of the filter
is a few times shorter than the coherent spectral interval (usually
referred to as "spike") in the SASE spectrum. Therefore, the seed
radiation pulse is temporally stretched in such way that the final
shape only depends on the characteristics of the monochromator. The
temporal shape and spectrum of the seed signal are shown in Fig.
\ref{mon_pow_spec}. The overall duration of the seed pulse is
inversely proportional to the bandwidth of the monochromator
transmittance spectrum. The particular temporal shape of the seed
pulse simply follows from a Fourier transformation of the
monochromator transfer function. The output FEL power and spectrum
of the entire setup, that is after the second part of the output
undulator is shown in Fig. \ref{outp_pow_spec}. The evolution of the
output energy in the photon pulse as a function of the distance
inside the output undulator is reported in Fig. \ref{nrg_gain}. The
photon spectral density for a TW pulse is about two orders of
magnitude higher than that for the SASE pulse at saturation (see
Fig. \ref{outp_pow_spec}). Given the fact that the TW-pulse
FWHM-duration is about $10$ fs, the relative bandwidth is 3 times
wider than the transform-limited bandwidth. There is a relatively
large energy chirp in the electron bunch due to wakefields effect.
Nonlinear energy chirp in the electron bunch induces nonlinear phase
chirp in the seed pulse during the amplification process in the
output undulator. Our simulations automatically include this effect.
This phase chirp increases the time-bandwidth product by broadening
the seeded FEL spectrum (see section \ref{sec:FELsim} for details).

\section{\label{sec:background} Theoretical background for designing a  grating
monochromator}

\subsection{\label{31} Wave optics approach}

In this section we derive the spatial frequency transfer function
for wave propagation and the Fresnel diffraction formula commonly
used in Fourier optics. We then analyze the propagation of a
Gaussian beam through ideal lenses and mirrors spaced apart from
each other.

\subsubsection{Spatial frequency transfer function and spatial impulse
response for wave propagation}

We start from the homogeneous wave equation for the electric field
in the space-time domain, $\vec{E}(t, \vec{r})$ expressed in
cartesian coordinates:

\begin{eqnarray}
\nabla^2 \vec{E} -\frac{1}{c^2} \frac{\partial^2 \vec{E}}{\partial
t^2} = 0~. \label{wavet}
\end{eqnarray}
Here $c$ indicates the speed of light in vacuum, $t$ is the time and
$\vec{r}$ is a 3D spatial vector identified by cartesian coordinates
$x,y,z$. As a consequence, the following equation for the field
$\vec{\bar{E}}(\omega, \vec{r})$ in the space-frequency domain
holds:

\begin{eqnarray}
\nabla^2 \vec{\bar{E}} + k_0^2 \vec{\bar{E}} = 0~, \label{wavef}
\end{eqnarray}
where $k_0 = \omega/c$. Eq. (\ref{wavef}) is the well-known
Helmholtz equation. Here $\vec{\bar{E}}(\omega, \vec{r})$  is
temporal Fourier transform of the electric field. We explicitly
write the definitions of the Fourier transform and inverse Fourier
transform for a function $f(t)$ in agreement with the notations used
in this paper as:

\begin{eqnarray}
&& \bar{f}(\omega )=\int_{-\infty }^{\infty }f(t) \exp[i \omega
t]dt~~, \cr && f(t)= \frac{1}{2\pi}\int_{-\infty }^{\infty
}\bar{f}(\omega ) \exp[-i \omega t]d\omega \label{ftift}
\end{eqnarray}
Similarly, the 2D spatial Fourier transform of $\vec{\bar{E}}(x,y,z,
\omega)$, with respect to the two transverse coordinates $x$ and $y$
will be written as

\begin{eqnarray}
\vec{\hat{E}}(\omega, k_x, k_y, z) =  \int_{-\infty}^{\infty} d x
\int_{-\infty}^{\infty} d y \vec{\bar{E}}(\omega, x, y) \exp[i k_x x
+ i k_y y] ~,\label{Ekkf}
\end{eqnarray}
so that

\begin{eqnarray}
\vec{\bar{E}}(\omega, x, y, z) = \frac{1}{4\pi^2}
\int_{-\infty}^{\infty} d k_x \int_{-\infty}^{\infty} d k_y
\vec{\hat{E}}(\omega, k_x, k_y) \exp[-i k_x x -i k_y y]
~.\label{Exxf}
\end{eqnarray}
With the help of this transformation the Helmholtz equation, which
is a partial differential equation in three dimensions, reduces to a
one-dimensional ordinary differential equation for the spectral
amplitude $\vec{\hat{E}}(\omega, k_x, k_y, z)$. In fact, by taking
the 2D Fourier transform of Eq. (\ref{wavef}),  we have

\begin{eqnarray}
\frac{d^2 \vec{\hat{E}}}{d z^2} + k_0^2 \left(1 - \frac{k_x^2}{k_0}
- \frac{k_y^2}{k_0^2}\right) \vec{\hat{E}} = 0~. \label{Eode}
\end{eqnarray}
We then obtain straightforwardly

\begin{eqnarray}
\vec{\hat{E}}(\omega,k_x, k_y, z) = \vec{\hat{E}}(\omega,k_x, k_y,
0) \exp\left[ i k_0 z \sqrt{1- \frac{k_x^2}{k_0^2} -
\frac{k_y^2}{k_0^2}}\right]~, \label{solEode}
\end{eqnarray}
where $\hat{E}(\omega,k_x, k_y, z)$ is the output field  and
$\hat{E}(\omega,k_x, k_y, 0)$ is the input field. Further on, when
the temporal frequency $\omega$ will be fixed, we will not always
include it into  the argument of the field amplitude and simply
write e.g. $\hat{E}(k_x, k_y, z)$. It is natural to define the
spatial frequency response of the system as

\begin{eqnarray}
H(k_x, k_y, z) = \frac{\vec{\hat{E}}(k_x, k_y,
z)}{\vec{\hat{E}}(k_x, k_y, 0)} = \exp\left[ i k_0 z \sqrt{1-
\frac{k_x^2}{k_0^2} - \frac{k_y^2}{k_0^2}}\right]~ . \label{Hresp}
\end{eqnarray}
Here the ratio between vectors has to be interpreted component by
component. $H$ is the spatial frequency transfer function related
with light propagation through a distance $z$ in free space. If we
assume that $k_x^2 + k_y^2 \ll k_0^2$, meaning that the bandwidth of
the angular spectrum of the beam is small we have

\begin{eqnarray}
H(k_x, k_y, z) \simeq \exp[ik_0 z] \exp\left[-\frac{i z}{2
k_0}(k_x^2 + k_y^2)\right] ~. \label{Hpar}
\end{eqnarray}
In other words, we enforce the paraxial approximation. In order to
obtain the output field distribution in the space-frequency domain
$\vec{\bar{E}}(x, y, z)$ at the distance z away from the input
position at $z=0$, we simply take the inverse Fourier transform of
Eq. (\ref{solEode}). If the paraxial approximation is now enforced
we obtain

\begin{eqnarray}
\vec{\bar{E}}(x, y, z) &&= \int_{-\infty}^{\infty}
\int_{-\infty}^{\infty} dx' ~ dy'~ \vec{\bar{E}}(x', y', 0) h(x-x',
y-y', z) \cr && = \vec{\bar{E}}(x, y, 0) \ast h(x, y, z)~,
\label{conv}
\end{eqnarray}

where

\begin{eqnarray}
h(x, y, z)&& =  \frac{1}{4 \pi^2} \exp [ik_0 z] \cr && \times
\int_{-\infty}^{\infty} \int_{-\infty}^{\infty} d k_x ~d k_y~
\exp\left[-\frac{i z}{2 k_0}(k_x^2 + k_y^2)\right] \exp[-ik_x x
-ik_y y] ~. \label{htr}
\end{eqnarray}
The result in Eq. (\ref{conv}) indicates that $h(x, y, z)$ is the
spatial impulse response describing the propagation of the system in
the formalism of Fourier optics. $h(x, y, z)$ is readily evaluated
as

\begin{eqnarray}
h(x, y, z) = - \frac{i k_0}{2 \pi z} \exp[ik_0 z]  \exp\left[
\frac{i k_0}{2 z} (x^2+y^2)\right]~ . \label{hxyz} \end{eqnarray}
Eq. (\ref{conv}) is  the Fresnel diffraction formula. In order to
obtain the output field distribution $\vec{\bar{E}}(x, y, z)$, we
need to convolve the input field distribution $\vec{\bar{E}}(x, y,
0)$ with the spatial impulse response $h(x, y, z)$.

\subsubsection{Gaussian beam optics}

\begin{figure}
\begin{center}
\includegraphics[ width=0.75\textwidth]{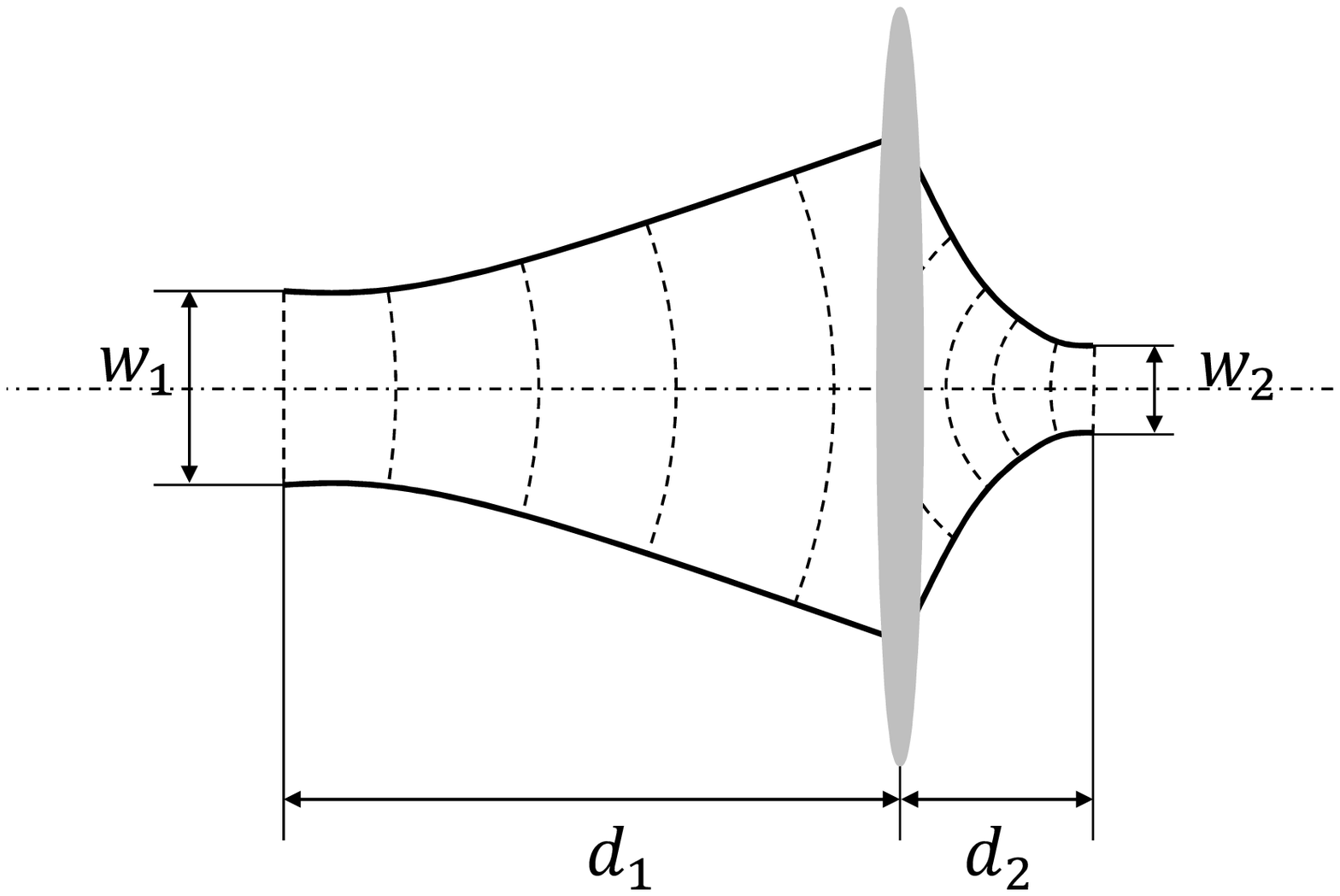}
\end{center}
\caption{Gaussian beam transformation by a lens. A Gaussian beam
with plane wavefront and waist $w_1$, located at a distance $d_1$
from the thin lens with focal length $f$ is transformed to a
Gaussian beam of plane wavefront and waist $w_2$, located at a
distance $d_2$, according to Eq. (\ref{qfinal}).} \label{gr_1}
\end{figure}

We now specialize our discussion considering a Gaussian beam with
initially (at $z=0$) plane wavefront in two transverse dimensions.
In order to simplify the notation, we will consider one component of
the field in the space-frequency domain only.

\begin{eqnarray}
\bar{E}(x, y, 0) = A \exp\left[-\frac{x^2+y^2}{w_0^2}\right] ~,
\label{waistbar}
\end{eqnarray}
where $w_0$ is the waist of the Gaussian beam. The spatial Fourier
transform of $\bar{E}$ is given by

\begin{eqnarray}
\hat{E}(k_x, k_y, 0) = A \pi w_0^2
\exp\left[-\frac{w_0^2}{4}\left(k_x^2+k_y^2\right)\right]~.
\label{FTwaist}
\end{eqnarray}
Using Eq. (\ref{Hpar}), after propagation over a distance $z$ one
obtains

\begin{eqnarray}
\hat{E}(k_x, k_y, z) &&= \hat{E}(k_x, k_y, 0) H(k_x,k_y,z) \cr &&= A
\pi w_0^2 \exp[ik_0 z] \exp\left[-\frac{i
z}{2k_0}\left(k_x^2+k_y^2\right)\right]
\exp\left[-\frac{w_0^2}{4}\left(k_x^2+k_y^2\right)\right] \cr &&  =
A \pi w_0^2 \exp[ik_0 z]\exp\left[-\frac{i q}{2
k_0}(k_x^2+k_y^2)\right] ~,\label{propG}
\end{eqnarray}
where $q$ is the so-called $q$-parameter of the Gaussian beam

\begin{eqnarray}
q = z - iz_R ~,\label{qpar}
\end{eqnarray}
where $z_R$ defines the Rayleigh range of the Gaussian beam

\begin{eqnarray}
z_R = k_0 w_0^2/2~. \label{ZR}
\end{eqnarray}
The spatial profile of the beam after propagation through a distance
$z$ can be found by taking the inverse Fourier transform of Eq.
(\ref{propG}):

\begin{eqnarray}
\bar{E}(x, y, z) = - \frac{i A k_0 w_0^2}{2 q} \exp[ik_0 z]
\exp\left[i\frac{k_0}{2 q}(x^2+y^2)\right] ~, \label{barE2}
\end{eqnarray}
which can also be written as

\begin{eqnarray}
\bar{E}(x, y, z)  && = A \frac{w_0}{w(z)} \exp[i \phi(z)+i k_0 z]
\cr && \times \exp\left[-\frac{(x^2+y^2)}{w^2(z)}\right] \exp\left[i
\frac{k_0 }{2 R(z)}(x^2+y^2)\right]~, \label{barE3}
\end{eqnarray}
where

\begin{eqnarray}
w^2(z) = w_0^2\left[1 + \left(\frac{z}{z_R}\right)^2\right]
~,\label{w2}
\end{eqnarray}
\begin{eqnarray}
R(z) = \frac{1}{z}\left(z^2 + z_R^2\right)~ , \label{Rz}
\end{eqnarray}
and

\begin{eqnarray}
\phi(z) = -\arctan\left[\frac{z}{z_R}\right] ~, \label{phiz}
\end{eqnarray}
with $z_R$ defined in Eq. (\ref{ZR}).  Note that the width $w(z)$ of
the Gaussian beam is a monotonically increasing function of the
propagation distance $z$, and reaches $\sqrt{2}$ times its original
width, $w_0$, at $z = z_R$.   The radius of curvature $R(z)$ of the
wavefront is initially infinite, corresponding to an initially plane
wavefront, but it reaches a minimum value of $2 z_R$ at $z = z_R$,
before starting to increase again. The slowly varying phase
$\phi(z)$, monotonically varies from $0$ at $z= 0$ to $-\pi/2$ as $z
\longrightarrow \infty$, assuming the value $\pi/4$ at $z = z_R$.

Note that the $q$-parameter contains all information about the
Gaussian, namely its curvature $R(z)$ and its waist $w(z)$. The
knowledge of the transformation of $q$ as a function of $z$ fully
characterizes the behavior of the Gaussian beam.

An optical system would usually comprise lenses or mirrors spaced
apart from each other. While Gaussian beam propagation in between
optical elements can be tracked using the translation law above, Eq.
(\ref{barE2}), we still need to discuss the law for the
transformation of $q$ by a lens. The transparency function for a
thin converging lens is of the form

\begin{eqnarray}
T_f(x, y) = \exp\left[-\frac{i k_0}{2 f}(x^2 + y^2)\right] ~.
\label{Tf}
\end{eqnarray}
The optical field immediately behind a thin lens at position $z$ is
related to that immediately before a lens by

\begin{eqnarray}
\bar{E}(x, y, z) && = T_f(x, y) \bar{E}_\mathrm{before}(x, y, z) \cr
&& = - \frac{i A k_0 w_0^2}{2 q} \exp[ik_0 z]
\exp\left[i\frac{k_0}{2 q}(x^2+y^2)\right] \exp\left[-\frac{i k_0}{2
f}(x^2 + y^2)\right] \cr && =  - \frac{i A k_0 w_0^2}{2 q} \exp[ik_0
z] \exp\left[i\frac{k_0}{2 q_l}(x^2+y^2)\right]  \label{lens}
\end{eqnarray}
where $\bar{E}_\mathrm{before}(x, y, z)$ is given by Eq.
(\ref{propG}) and $q_l$, the transformed of $q$, is defined by

\begin{eqnarray}
\frac{1}{q_l} = \frac{1}{q}  - \frac{1}{f}~ . \label{ql}
\end{eqnarray}
As example of application we analyze the focusing of a Gaussian beam
by a converging lens. We assume that a Gaussian beam with plane
wavefront and waist $w_1$, is located at distance $d_1$ from a thin
lens with focal length $f$. After propagation through a distance
$d_2$ behind the lens, it is transformed to a beam with plane
wavefront and waist $w_2$, Fig. \ref{gr_1}. Using Eq. (\ref{qpar})
and Eq. (\ref{ql}) we can find the transformed $q$-parameter at
distance $d_2$. From Eq. (\ref{qpar}), immediately in front of the
lens we have

\begin{eqnarray}
q(d_1) = q(0) + d_1 ~, \label{before}
\end{eqnarray}
Immediately behind the lens, the $q$-parameter is transformed to
$q_l$ according to Eq. (\ref{qpar}):

\begin{eqnarray}
\frac{1}{q_l} = \frac{1}{q(0) + d_1} - \frac{1}{f}~. \label{qlafter}
\end{eqnarray}
Finally, using again Eq. (\ref{qpar}), we find the $q$-parameter
after propagation through a distance $d_2$ behind the lens:

\begin{eqnarray}
q(d_2+d_1) = q_l + d_2  ~ . \label{qfinal}
\end{eqnarray}
The Gaussian beam is said to be focused at the point  $z = d_2 +d_1$
where  $q(d_2+d_1)$ becomes purely imaginary again, meaning that the
Gaussian beam has a planar wavefront. Thus, calculating explicitly
$q(d_2+d_1)$, setting $q(d_2+d_1) = ik_0 w_2^2/2$, and equating
imaginary parts we obtain

\begin{eqnarray}
w_2^2  =  \frac{w_1^2 f^2}{[(d_1 - f)^2 + (k_0 w_1^2/2)^2]} ~.
\label{w2}
\end{eqnarray}
Equating the real part of $q(d_2+d_1)$ to zero one obtains instead

\begin{eqnarray}
d_2 = f+  f^2 \frac{(d_1 - f)}{[(d_1 - f)^2 + (k_0 w_1^2/2)^2]} ~.
\label{d2}
\end{eqnarray}
Note that the Gaussian beam does not exactly focus at the
geometrical back focus of the lens. Instead, the focus is shifted
closer to the lens. In other words the "lensmaker" equation valid in
geometrical optics

\begin{eqnarray}
\frac{1}{d_1}+\frac{1}{d_2} = \frac{1}{f} \label{lensmaker}
\end{eqnarray}
is modified to

\begin{eqnarray}
\frac{1}{d_1 + z_R^2/(d_1-f)} + \frac{1}{d_2} = \frac{1}{f}~,
\label{lensmakermod}
\end{eqnarray}
which is just another way of writing Eq. (\ref{d2}) and is well
known from a long time (see e.g. \cite{SSLF}).

\subsection{Beam propagation in inhomogeneous media}

In section \ref{31}, we considered the problem of wave propagation
in a homogeneous medium, namely vacuum, characterized by constant
permittivity, $\epsilon = 1$. We specialized our investigations to
the case of a Gaussian beam and, additionally, we analyzed
propagation of a Gaussian beam through a thin lens using the wave
optics formalism. The description of wave propagation through a thin
lenses does not require the use of wave propagation theory in
inhomogeneous media. In fact, as we have seen, thin lenses
contribute to the wave propagation via a phase multiplication. In
other words, if we consider a wave field in front of and immediately
behind a lens, we find that the phase of the wave has changed, while
its amplitude has remained practically the same. A mirror may be
equivalently modeled by a similar phase transformation.

Of course, strictly speaking, the polarization of the light has an
influence on its reflection properties from the lenses. However, if
we are willing to disregard such reflection phenomena, we are
justified to use the scalar wave equation to describe the wave
optics of lenses, and to model a thin lens as described before. In
this section we will study, at variance, wave propagation in a
medium that is inhomogeneous. Therefore, we will be in position to
numerically analyze such effects as reflection of X-rays from
gratings or mirrors.

\subsubsection{Wave equation}

The fundamental theory of electromagnetic fields is based on Maxwell
equations. In differential form and in the space-time domain, these
can be written as

\begin{eqnarray}
&& \vec{\nabla}\cdot \vec{D} = 4 \pi \rho~, \cr && \vec{\nabla}\cdot
\vec{B} = 0~ , \cr && \vec{\nabla}\times \vec{E} =
-\frac{1}{c}\frac{\partial \vec{B}}{ \partial t}~,\cr &&
\vec{\nabla}\times \vec{H} = \frac{4\pi}{c}
\vec{j}+\frac{1}{c}\frac{\partial \vec{D}}{\partial t}~. \label{Max}
\end{eqnarray}
Here $\vec{j}$ is the current density and $\rho$ denotes the
electric charge density. $\vec{E}$ and $\vec{B}$ are the macroscopic
electric and magnetic field in the time domain, while $\vec{D}$ and
$\vec{H}$ are the corresponding derived fields, related to $\vec{E}$
and $\vec{B}$ by

\begin{eqnarray}
&& \vec{D} = \epsilon \vec{E} ~, \cr &&    \vec{B} = \mu \vec{H} ,
\cr && \vec{j} = \sigma \vec{E}~ , \label{constit}
\end{eqnarray}
where $\epsilon$ denotes the permittivity,  $\mu$ the permeability,
and $\sigma$ the conductivity of medium. In this article we do not
consider any magnetic or conductive media. Hence $\mu = 1$  and
$\sigma=0$. Moreover, $\rho=0$. The permittivity $\epsilon$ is,
instead, a function of the position, i.e., $\epsilon = \epsilon(x,
y, z)$, which allows us to consider inhomogeneous media such as a
mirror with rough surface. Maxwell equations can be manipulated
mathematically in many ways in order to yield derived equations more
suitable for certain applications.  For example, from Maxwell
equations we can obtain an equation which depends only on the
electric field vector $\vec{E}$:

\begin{eqnarray}
\vec{\nabla}\times (\vec{\nabla} \times \vec{E}) = -\frac{1}{c}
\frac{\partial(\vec{\nabla}\times \vec{B})}{\partial t} = -
\frac{\epsilon}{c^2}  \frac{\partial^2 \vec{E}}{\partial t^2} ~.
\label{nablaE}
\end{eqnarray}
It is worth noting that this equation holds even if $\epsilon$
varies in space.  However, the $\vec{\nabla}\times (\vec{\nabla}
\times (\cdot) )$ operator is not very easy to use, so that it is
advantageous to use the vector identity

\begin{eqnarray}
\vec{\nabla}\times (\vec{\nabla} \times \vec{E}) = \vec{\nabla}
(\vec{\nabla}\cdot \vec{E}) - \nabla^2 \vec{E}~ , \label{vecrel}
\end{eqnarray}
which holds if we use a cartesian coordinate system. Exploiting
$\vec{\nabla}\cdot \vec{D} = 0$ and $\vec{D} = \epsilon \vec{E}$ we
rewrite Eq. (\ref{nablaE}) as

\begin{eqnarray}
\nabla^2 \vec{E} + \vec{\nabla}\left[\vec{E}\cdot
\frac{\vec{\nabla}\epsilon}{\epsilon}\right] =\frac{\epsilon}{c^2}
\frac{\partial^2 \vec{E}}{\partial t^2} ~. \label{nablaE2}
\end{eqnarray}
The second term on the left-hand side of Eq. (\ref{nablaE2}) is in
general non-zero when there is a gradient in the permittivity of the
medium. However, if the spatial variation of  $\epsilon$ is small,
one can neglect the term in $\vec{\nabla} \epsilon/\epsilon$. If we
are content with this approximation, we can study the propagation of
light in inhomogeneous media using the wave equation

\begin{eqnarray}
\nabla^2 \vec{E}  =\frac{\epsilon}{c^2} \frac{\partial^2
\vec{E}}{\partial t^2} ~, \label{nablaE3}
\end{eqnarray}
where, once more, $\epsilon = \epsilon(x, y, z)$.  By taking the
\textit{temporal} Fourier transform of Eq. (\ref{nablaE3}) we
obtain, similarly as for Eq. (\ref{wavef})

\begin{eqnarray}
\nabla^2 \vec{\bar{E}} +k_0^2 \epsilon \vec{\bar{E}} = 0 ~,
\label{Efreq}
\end{eqnarray}
where, as before, $\vec{\bar{E}} = \vec{\bar{E}}(\omega, x, y, z)$
is the temporal Fourier transform of electric field, and $k_0 =
\omega/c$.

It is necessary to investigate under what conditions the wave
equation, Eq. (\ref{nablaE3}) is a good approximation of Eq.
(\ref{nablaE2}), since the latter equation is far more difficult to
handle and not very useful for actual calculations. The condition
for neglecting the term in $\vec{\nabla} \epsilon/\epsilon$ is
usually formulated as the requirement that the relative change of
$\epsilon$ over the distance of one wavelength be less than unity
\cite{MARC}, that is

\begin{eqnarray}
R =   |\epsilon_2 - \epsilon_1|/|\epsilon_1|  \ll 1 ~, \label{R}
\end{eqnarray}
where $\epsilon_2 - \epsilon_1$  is the difference in the dielectric
constants at two positions spaced by a wavelength. By examining the
arguments which lead to condition (\ref{R}), it can be usually found
that the gradient term is compared with the main term $\epsilon
\partial^2 E/\partial t^2$  in Eq. (\ref{nablaE2}).

However, a more careful look at Eq. (\ref{nablaE2}) reveals that
condition (\ref{R}) is not adequate. In order to see this, let us
present the main Eq. (\ref{nablaE2}) in another, equivalent form.
Consider the dielectric dipole moment density $\vec{P}$ related to
the electric field $\vec{E}$ according to $\vec{P} = \chi \vec{E}$,
where $\chi$ is the electric susceptibility. The field $\vec{D}$ is
basically the sum of $\vec{E}$ and $\vec{P}$ according to

\begin{eqnarray}
\vec{D} = \vec{E} + 4 \pi \vec{P} = (1 + 4 \pi \chi)\vec{E} =
\epsilon \vec{E}~ . \label{Dchi}
\end{eqnarray}

Using  Maxwell equations

\begin{eqnarray}
&& \vec{\nabla}\cdot \vec{D} = 0~, \vec{\nabla}\times \vec{H} =
\frac{1}{c} \frac{\partial \vec{E}}{\partial t}
+\frac{4\pi}{c}\frac{\partial \vec{P}}{\partial t}~, \label{Max2}
\end{eqnarray}
we can recast Eq. (\ref{nablaE2}) in the form

\begin{eqnarray}
c^2 \nabla^2 \vec{E} - \frac{\partial^2 \vec{E}}{\partial t^2} =
4\pi \frac{\partial^2 \vec{P}}{\partial t^2} -4\pi c^2
\vec{\nabla}\left(\vec{\nabla}\cdot \vec{P}\right) ~.
\label{nablaE2bis}
\end{eqnarray}
Eq. (\ref{nablaE2bis}) separates terms which are present in
free-space (on the left hand side) from terms related with the
propagation through the dielectric medium (on the right hand side).
When the gradient term in Eq. (\ref{nablaE2bis}) can be neglected,
one gets back Eq. (\ref{Efreq}). However, at variance with the
treatment in \cite{MARC}, in order for this approximation to be
applicable the gradient term must not introduce important changes to
the part of the equation relative to propagation through the
dielectric. In other words, the gradient term should be small
compared with $-4 \pi \omega^2 \vec{\bar{P}}$, and not with the
entire  term $-\omega^2 \vec{\bar{E}} -4 \pi \omega^2
\vec{\bar{P}}$. This hints to the fact that a correction of
condition (\ref{R}) to

\begin{eqnarray}
R =   |\epsilon_2 - \epsilon_1|/|\epsilon_1-1|  \ll 1 ~. \label{R2}
\end{eqnarray}
Note that for optical wavelengths and in general, in regimes where
$\epsilon$ is sensibly larger than unity, condition (\ref{R}) and
condition (\ref{R2}) will not lead to much different regions of
applicability. An important difference arises when one considers the
x-ray range, where $\epsilon$ is very near unity. In that case,
according to condition (\ref{R2}), the wave equation is not
applicable in such situations. However, in that case we can limit
ourselves to small angles of incidence. As we will see, condition
(\ref{R2}) will be modified under the additional small angle
approximation.

Instead of using directly the field equation in the form of Eq.
(\ref{nablaE2bis}), we can use the Green theorem to express the
Fourier-transformed of Eq. (\ref{nablaE2bis}) in integral form. We
first apply a temporal Fourier transformation to Eq.
(\ref{nablaE2bis}) to obtain the inhomogeneous Helmholtz equation

\begin{eqnarray}
c^2 \nabla^2 \vec{\bar{E}} +\omega^2 \vec{\bar{E}}  = -4\pi \omega^2
\vec{\bar{P}} - 4\pi c^2 \vec{\nabla}\left(\vec{\nabla}\cdot
\vec{\bar{P}}\right) ~. \label{nablaE2tris}
\end{eqnarray}
Note that here $\vec{\bar{P}}$ is the temporal Fourier transform of
$\vec{P}$. We now introduce a Green function for the Helmholtz wave
equation, $G(\vec{r}, \vec{r'})$, defined as

\begin{eqnarray}
\left(\nabla^2 + k_0^2\right) G(\vec{r}, \vec{r'}) = -  \delta
\left(\vec{r} - \vec{r'}\right) ~. \label{GreenH}
\end{eqnarray}
For unbounded space, a Green function describing outgoing waves is
given by

\begin{eqnarray}
G(\vec{r}, \vec{r'}) =  \frac{1}{4\pi} \frac{\exp\left[i k_0
|\vec{r} - \vec{r'}|\right]}{|\vec{r} - \vec{r'}|}  ~. \label{Grr1}
\end{eqnarray}
With the help of Eq. (\ref{Grr1}) we can write a formal solution for
the field equation Eq. (\ref{nablaE2tris}) as:

\begin{eqnarray}
\vec{\bar{E}}_d = \int d\vec{r'} ~G(\vec{r}, \vec{r'}) \left[
\omega^2 \vec{\bar{P}}(\vec{r'}) + c^2
\vec{\nabla}\left(\vec{\nabla}\cdot
\vec{\bar{P}}(\vec{r'})\right)\right]~, \label{grensol}
\end{eqnarray}
where we solve for the diffracted field only. Eq. (\ref{grensol}) is
the integral equivalent of the differential equation Eq.
(\ref{nablaE2tris}). This integral form is convenient to overcome
the difficulty of comparing the two terms on the right-hand side of
Eq. (\ref{nablaE2}). Integrating by parts the term in grad
$\vec{\nabla}\left(\vec{\nabla}\cdot \vec{\bar{P}}(\vec{r'})\right)$
twice we obtain

\begin{eqnarray}
\vec{\bar{E}}_d = \int d\vec{r'} ~G(\vec{r}, \vec{r'})\left[
\omega^2 \vec{\bar{P}}(\vec{r'}) - c^2 k_0^2 \vec{n} (\vec{n}\cdot
\vec{P}(\vec{r'})) \right] ~,\label{grensol2}
\end{eqnarray}
where $\vec{n} = (\vec{r} - \vec{r'})/|\vec{r} - \vec{r'}|$ is the
unit vector from the position of the "source" to the observer. We
assume that the condition $k_0 \gg 1/|\vec{r} - \vec{r'}|$ holds for
all values of $\vec{r}$ occurring in the integral in Eq.
(\ref{grensol2}). We thus account for the radiation field only. It
is then possible to neglect the derivative of $1/|\vec{r} -
\vec{r'}|$ compared to the derivative of $\exp[ik_0 |\vec{r} -
\vec{r'}|]$ when we integrate by parts. Moreover, the edge term in
the integration by parts vanishes since $\vec{P} = 0$ at infinity.
We note that the combination of the first and second term in the
integrand  obviously exhibits the property that the  diffracted
field $\vec{E}_d$ is directed transversely with respect to vector
$\vec{n}$, as it must be for the radiation field. Furthermore, one
can see that only the second term is responsible for the
polarization dependance.

Returning to X-ray optics, we can easily obtain that the second term
in the integrand of Eq. (\ref{grensol2}) includes, in this case, an
additional small factor proportional to the the diffraction angle
$\theta_d \sim (\vec{n}\cdot \vec{P})/|\vec{P}|\ll 1$, which can be
neglected under the grazing incidence approximation. Finally, we
conclude that for describing the reflection of a coherent X-ray beam
from the interface between two dielectrics, one can use the wave
equation Eq. (\ref{Efreq}) under the grazing incidence condition
with accuracy

\begin{eqnarray}
&& \theta_i^2 \ll 1~, \cr && \theta_d^2 \ll 1~ . \label{thid}
\end{eqnarray}
It is very important to realize that, in order for Eq. (\ref{Efreq})
to apply, it is not sufficient that the paraxial approximation for
X-ray propagation in vacuum or in a dielectric be satisfied.
Additionally, incident and diffracted angles relative to the
interface between dielectric and vacuum must be small compared to
unity, according to condition (\ref{thid}).

\subsubsection{The split-step beam propagation method}

Let us return to the model for inhomogeneous media given by the wave
equation, Eq. (\ref{Efreq}).  We can always write

\begin{eqnarray}
\vec{\bar{E}}(x, y, z) = \vec{A}(x, y, z) \exp[ik_0 z] ~.
\label{AExyz}
\end{eqnarray}
By substituting this expression into Eq. (\ref{Efreq}) we derive the
following equation for the complex field envelope:

\begin{eqnarray}
\nabla_\bot^2 \vec{A} + \frac{\partial^2\vec{A}}{\partial z^2} + 2 i
k_0 \frac{\partial \vec{A}}{\partial{z}} + k_0^2 \delta\epsilon
\vec{A}=0 ~,\label{Aequat}
\end{eqnarray}
where $\nabla_\bot^2$ denotes the transverse Laplacian, and $\delta
\epsilon \equiv \epsilon - 1$. If the electric field is
predominantly propagating along z-direction with an envelope
$\vec{A}$ which varies slowly with respect to the wavelength, Eq.
(\ref{AExyz}) separates slow from fast varying factors. We actually
assume that $\vec{A}$ is a slowly varying function of z in the sense
that

\begin{eqnarray}
\left|\frac{\partial\vec{A}}{\partial z}\right| \ll k_0 \vec{A} ~.
\label{Aslow}
\end{eqnarray}
This assumption physically means that, within a propagation distance
along $z$ of the order of the wavelength, the change in $\vec{A}$ is
much smaller than $\vec{A}$ itself. With this assumption, Eq.
(\ref{Aequat}) becomes the paraxial Helmholtz equation for $\vec{A}$
in inhomogeneous media, which reads

\begin{eqnarray}
\nabla_\bot^2 \vec{A}  + 2 i k_0 \frac{\partial
\vec{A}}{\partial{z}} + k_0^2 \delta\epsilon \vec{A}=0
~,\label{Aepar}
\end{eqnarray}
A large number of numerical methods can be used for analyzing beam
propagation in inhomogeneous media. The split-step beam propagation
method is an example of such methods. To understand the idea of this
method, we re-write Eq. (\ref{Aepar}) in the operator form
\cite{AGAR,TING}

\begin{eqnarray}
\frac{\partial \vec{A}}{\partial z} = (\mathcal{D} + \mathcal{S})
\vec{A}~, \label{Aop} \end{eqnarray}
where $\mathcal{D} = -(2ik_0)^{-1} \nabla_\bot^2$  is the linear
differential operator accounting for diffraction, also called the
diffraction operator, and $\mathcal{S} = (ik_0/2) \delta \epsilon$
is the space-dependent, or inhomogeneous operator. Both operators
act on $\vec{A}$ simultaneously, and a solution of Eq. (\ref{Aop})
in operator form is given by

\begin{eqnarray}
\vec{A}(x, y, z+ \delta z) = \exp\left[\left(\mathcal{D}+
\mathcal{S}\right)\delta z\right] \vec{A}(x, y, z) \label{solop}~.
\end{eqnarray}
Note that, in general, $\mathcal{D}$ and $\mathcal{S}$ do not
commute. In order to see this, it is sufficient to consider the
dependence of $\mathcal{S}$ on $z$. As a result,
$\exp\left[\left(\mathcal{D}+ \mathcal{S}\right)\delta z\right] \neq
\exp[\mathcal{D} \delta z]\exp[\mathcal{S} \delta z)]$. More
precisely, for two non-commuting operators $\mathcal{D}$ and
$\mathcal{S}$, we have

\begin{eqnarray}
\exp[\mathcal{D} \delta z]\exp[\mathcal{S} \delta z)] =
\exp[(\mathcal{D}  + \mathcal{S}) \delta z] +
[\mathcal{D},\mathcal{S}] \frac{\delta z^2}{2} + ...~, \label{twist}
\end{eqnarray}
where $[\mathcal{D},\mathcal{S}] = \mathcal{D}\mathcal{S} -
\mathcal{S}\mathcal{D}$ is the commutator of $\mathcal{D}$ and
$\mathcal{S}$. However, for an accuracy up to the first order in
$\delta z$, we can \textit{approximately} write:
\begin{eqnarray}
\exp[(\mathcal{D} + \mathcal{S}) \delta z] \simeq \exp[\mathcal{D}
\delta z]\exp[\mathcal{S} \delta z]~. \label{Itwist}
\end{eqnarray}
This means that, when the propagation step $\delta z$ is
sufficiently small, the diffraction and the inhomogeneous operators
can be treated independently of each other in Eq. (\ref{solop}), and
we obtain

\begin{eqnarray}
\vec{A}(x, y, z+\delta z) = \exp[\mathcal{S} \delta z]
\exp[\mathcal{D} \delta z] \vec{A}(x, y, z) ~. \label{Aopfin}
\end{eqnarray}
The role of the operator acting first, $\exp[\mathcal{D} \delta z]$,
is better understood in the spectral domain. This is the propagation
operator that takes into account the effect of diffraction between
the planes at position $z$ and $z + \delta z$. Propagation is
readily handled in the spatial-frequency domain using transfer
function for propagation given by

\begin{eqnarray}
H_A(k_x, k_y, \delta z) = \frac{\hat{A}(k_x, k_y, \delta
z)}{\hat{A}(k_x, k_y, 0)}= \exp[-i(k_x^2 + k_y^2) \delta z/(2k_0)]
~. \label{HA}
\end{eqnarray}
This is nothing but Eq. (\ref{Hpar}), specialized for the slowly
varying envelope of the field.

Hence, the action of the exponential operator $\exp[\mathcal{D}
\delta z]$ is carried out in the Fourier domain using the
prescription

\begin{eqnarray}
\exp[\mathcal{D} \delta z] \vec{A}(x,y,z) =
FT^{-1}\left\{\exp[-i(k_x^2+k_y^2) \delta z /(2k_0)] \hat{A}(k_x,
k_y, \delta z) \right\} ~, \label{Dzspace}
\end{eqnarray}
where "$FT^{-1}$" refers to the inverse spatial Fourier transform
defined as in Eq. (\ref{Exxf}). The second operator,
$\exp[\mathcal{S} \delta z]$, describes the effect of propagation in
the absence of diffraction and in the presence of medium
inhomogeneities, and is well-described in the spatial domain.

Summing up, a prescription for propagating $A(x,y,z)$ along a single
step in $\delta z$ can be written as

\begin{eqnarray}
A(x, y, z+ \delta z) =&& \exp[ ik_0 \delta \epsilon ~\delta z/2] \cr
&& \times  FT^{-1} \left\{\exp[-i(k_x^2+k_y^2) \delta z /(2k_0)]
\hat{A}(k_x, k_y, \delta z) \right\}~ . \label{2step}
\end{eqnarray}
The algorithm repeats the above process until the field has traveled
the desired distance. The usefulness of the Fourier transform lies
in the fact that one can reduce a partial differential operator to a
multiplication of the spectral amplitude $\hat{A}(k_x, k_y, z)$ with
a phase transformation function. Since $\mathcal{D}$ is just a
number in the spatial Fourier domain, the evaluation of Eq.
(\ref{Aopfin}) is straightforward.

\subsection{Grating Theory}
$
\\
\\
\\
\\
$

\begin{figure}[!h]
\begin{center}
\includegraphics[clip, width=0.75\textwidth]{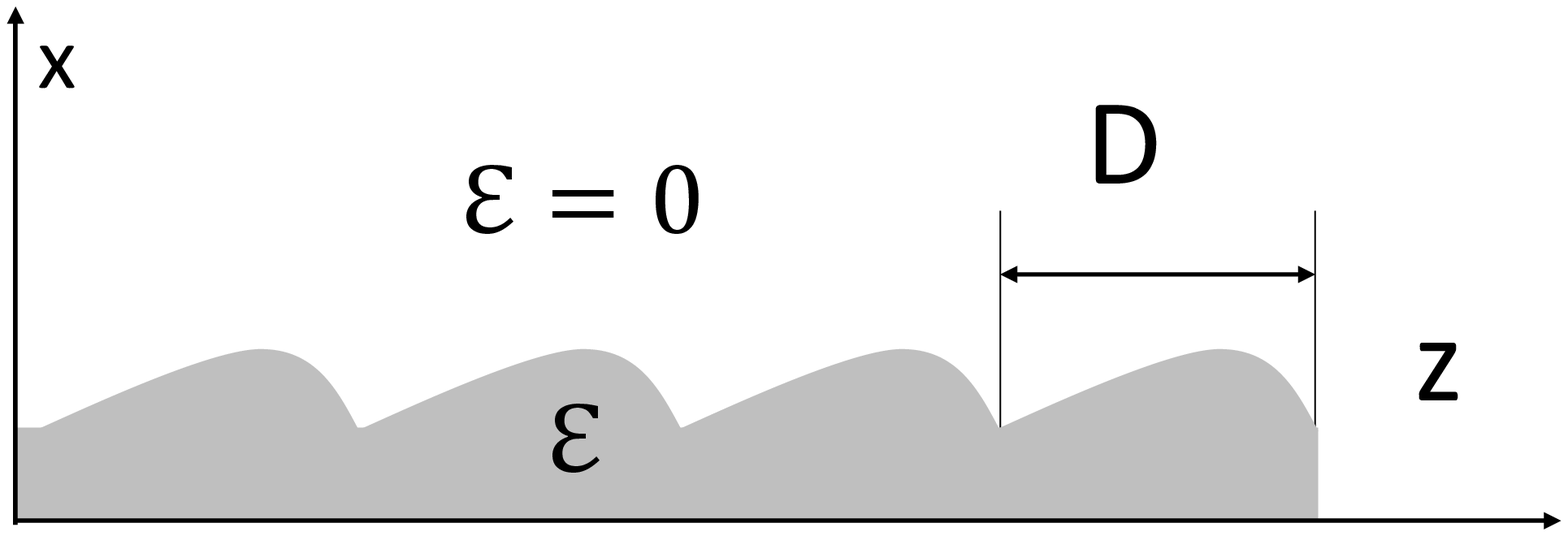}
\caption{Grating geometry and notation.} \label{gr_1}
\end{center}
\end{figure}
$
\\
\\
\\
\\
\\
\\
$

\begin{figure}[!h]
\begin{center}
\includegraphics[clip, width=0.75\textwidth]{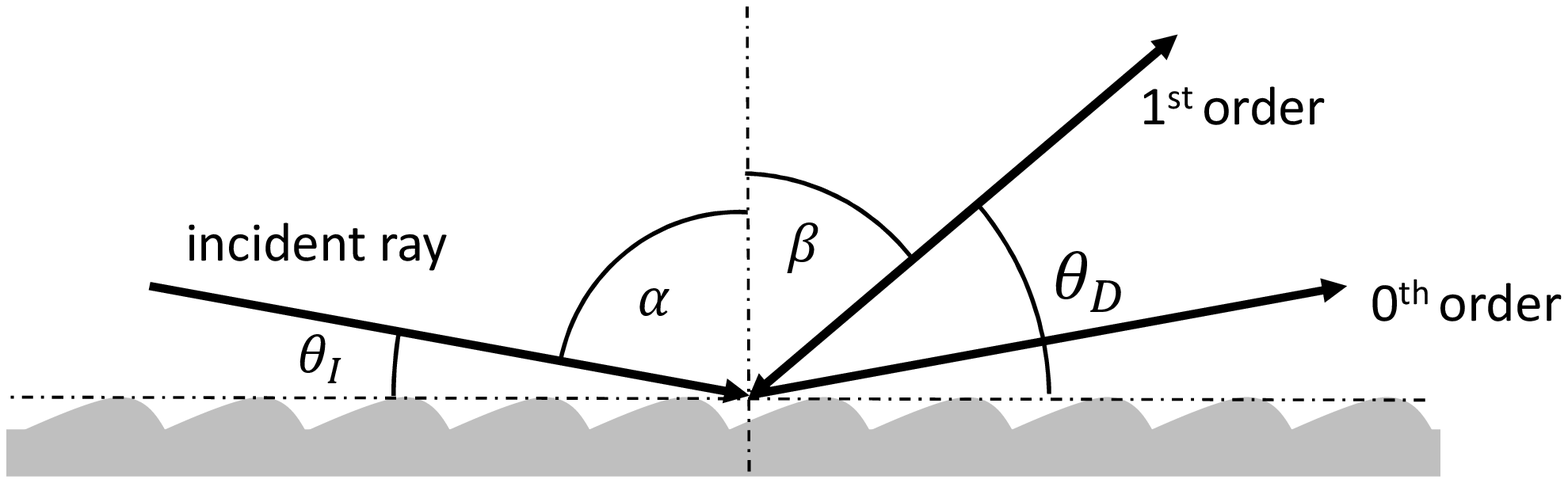}
\end{center}
\caption{Scattering geometry for a diffraction grating.}
\label{gr_2}
\end{figure}
$
\\
\\
\\
\\
\\
\\
\\
\\
\\
$
\begin{figure}[!h]
\begin{center}
\includegraphics[clip, width=0.75\textwidth]{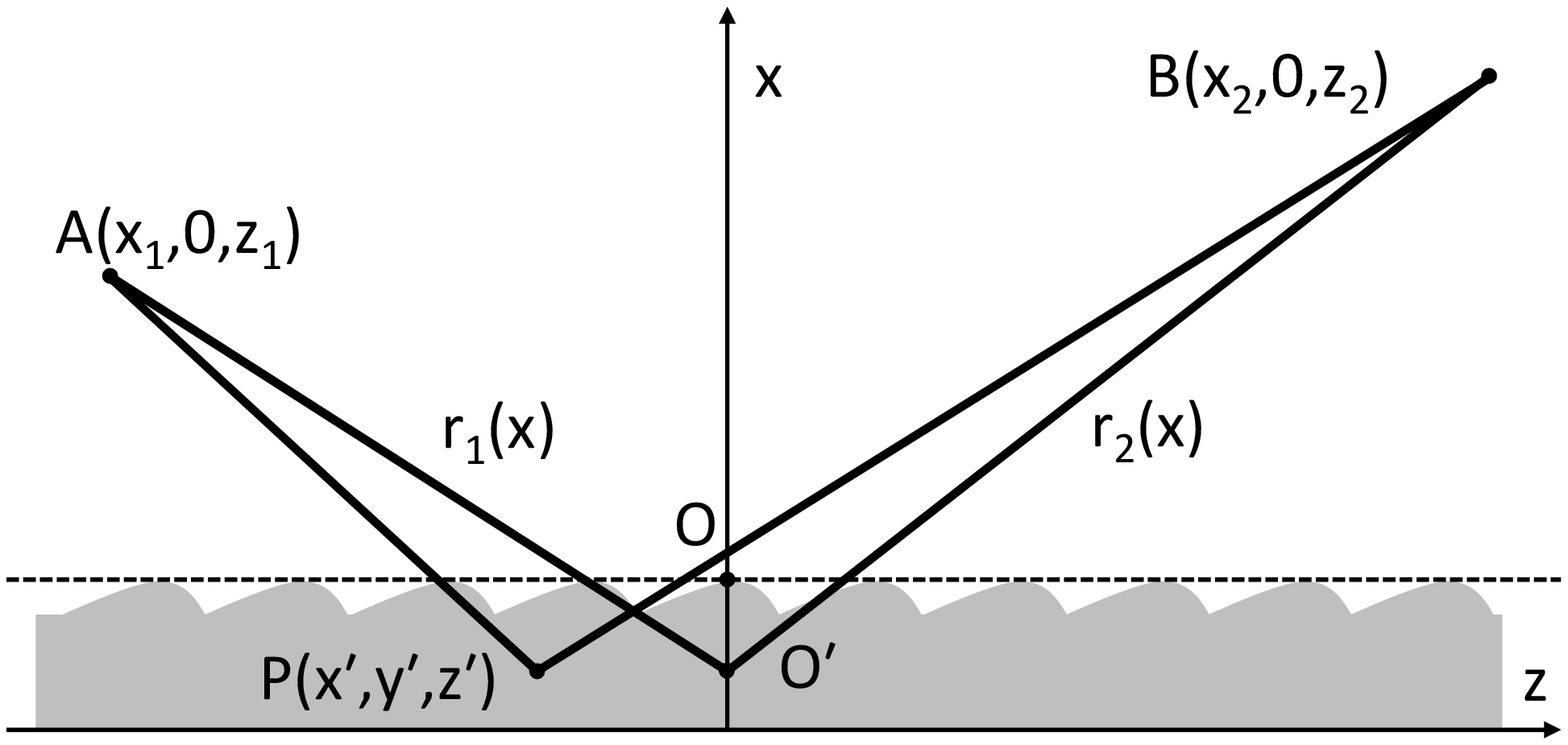}
\end{center}
\caption{Schematic diagram of diffraction from a plane grating. The
gray area represents the grating volume. A point source is located
at A. Point P is an arbitrary point inside the grating volume.
Grating can be divided into layers. Each layer is either homogeneous
or modulated with refractive index that changes periodically as a
function of $z$ at any given height $x$.} \label{gr_8}
\end{figure}

The derivation of the grating condition describing the geometry of
light diffraction by gratings presented in textbooks usually relies
on Huygens principle. At variance, our treatment of gratings theory
is based on first principles, namely Maxwell equations, still
retaining basic simplicity.

\begin{figure}[!h]
\begin{center}
\includegraphics[clip, width=0.75\textwidth]{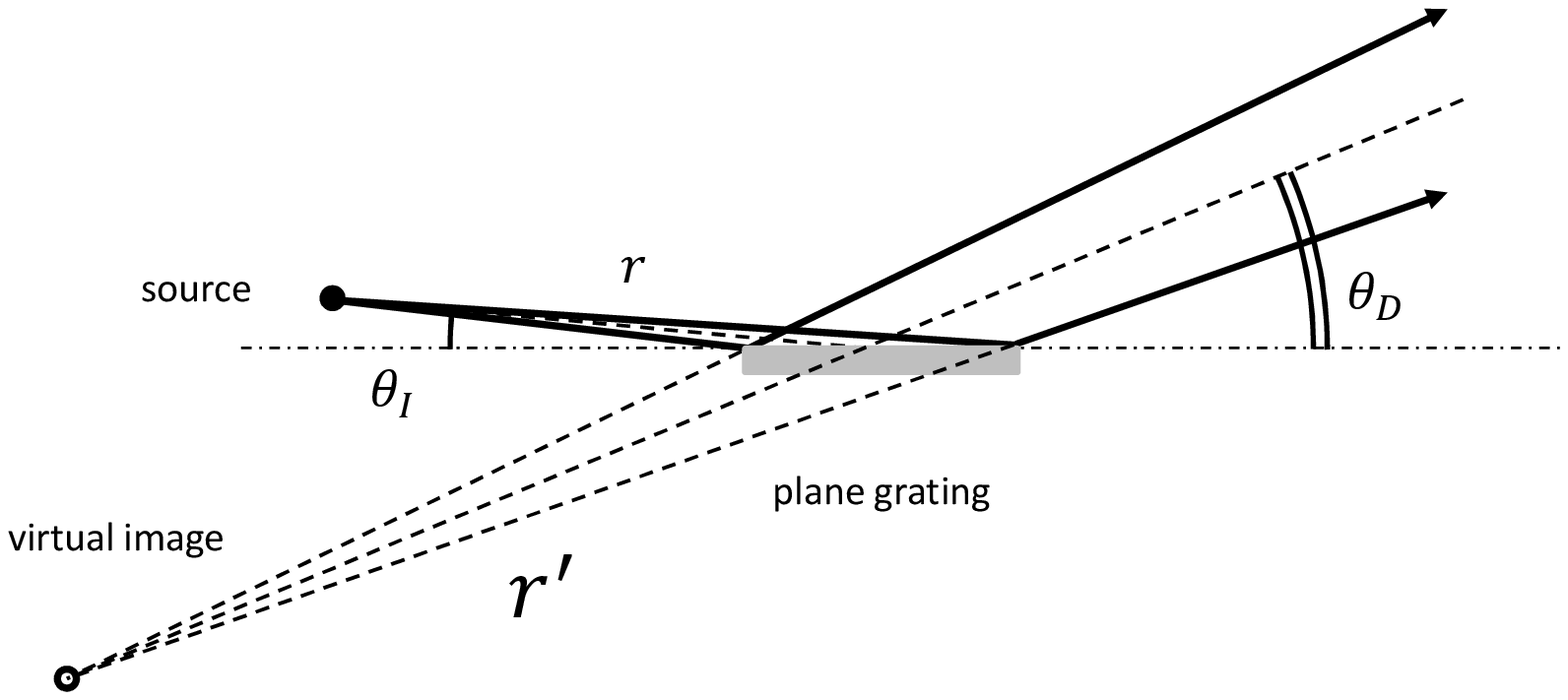}
\end{center}
\caption{Plane grating in the case of a monochromatic point source.
The virtual image of the real source is located at a distance
$r'=r\cdot\left(\sin(\theta_D)/\sin(\theta_I)\right)^2$ behind the
grating.} \label{gr_3}
\end{figure}
$
\\
\\
\\
\\
\\
\\
\\
\\
\\
\\
\\
\\
$

\begin{figure}[!h]
\begin{center}
\includegraphics[clip, width=0.75\textwidth]{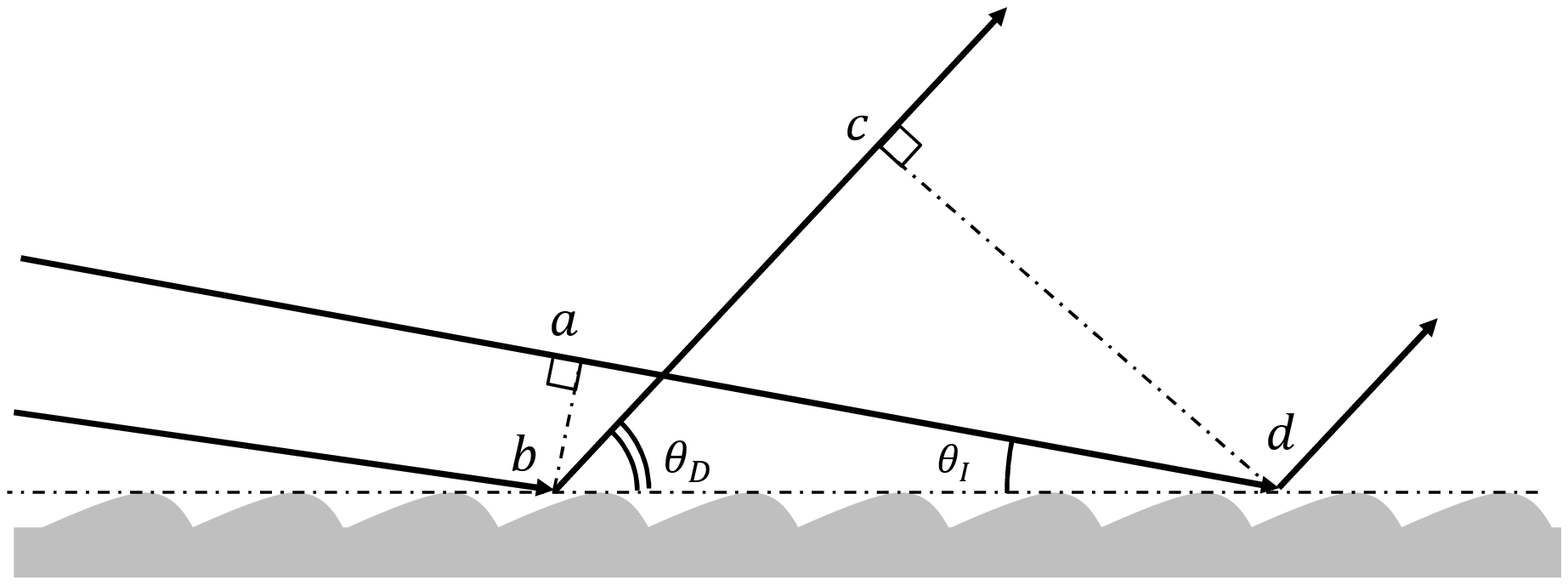}
\end{center}
\caption{Geometry of the reflection. The properties of the grating
are naturally described in terms of the asymmetry parameter
$b=\sin(\theta_D)/\sin(\theta_I)$} \label{gr_4}
\end{figure}
$
\\
\\
\\
\\
\\
\\
$

\begin{figure}[!h]
\begin{center}
\includegraphics[clip, width=0.75\textwidth]{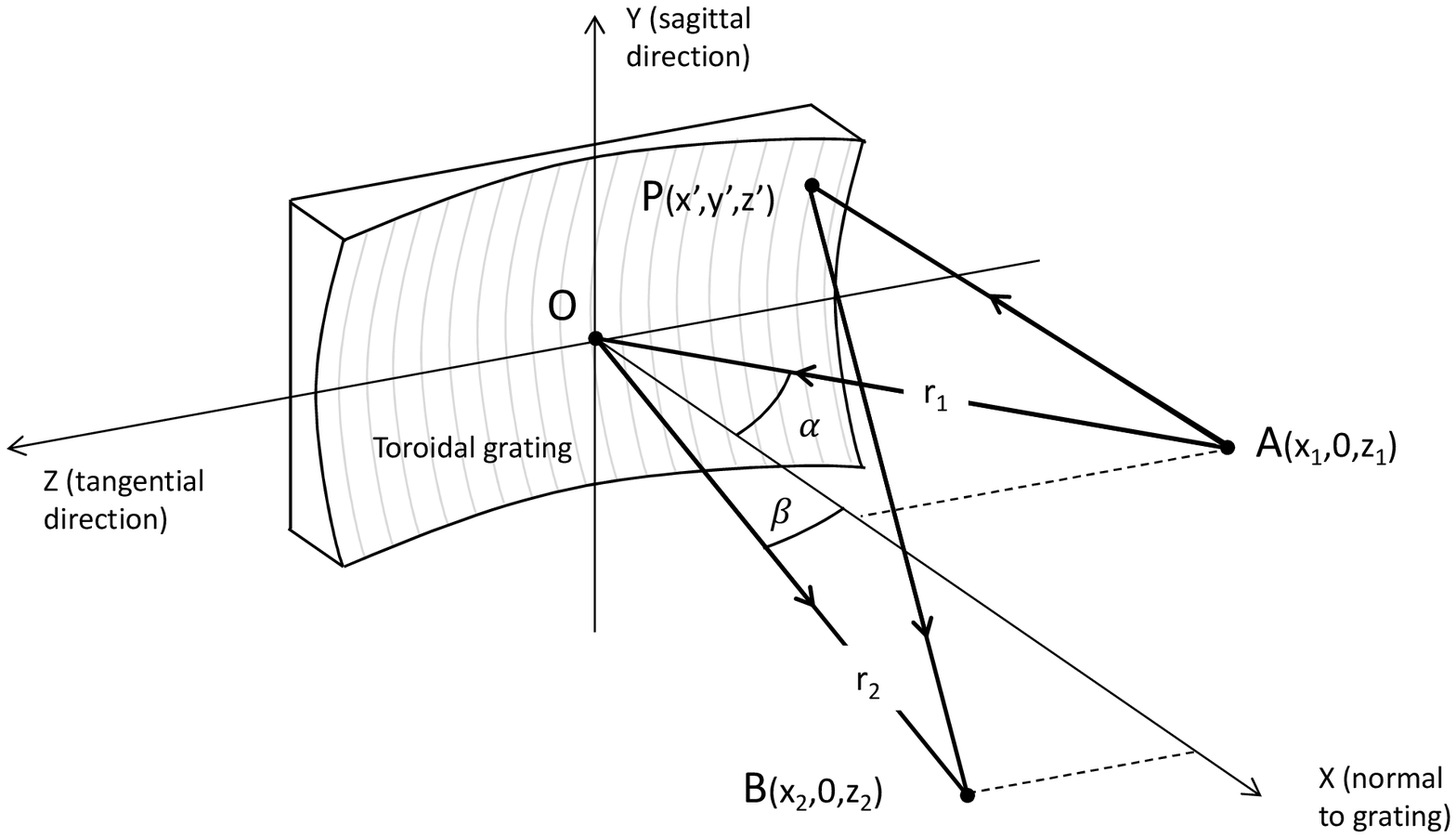}
\end{center}
\caption{Schematic diagram of a toroidal grating. A point source is
located at A. Point P is an arbitrary point of the grating.}
\label{gr_5}
\end{figure}
$
\\
\\
\\
\\
\\
\\
$

\begin{figure}[!h]
\begin{center}
\includegraphics[clip, width=0.75\textwidth]{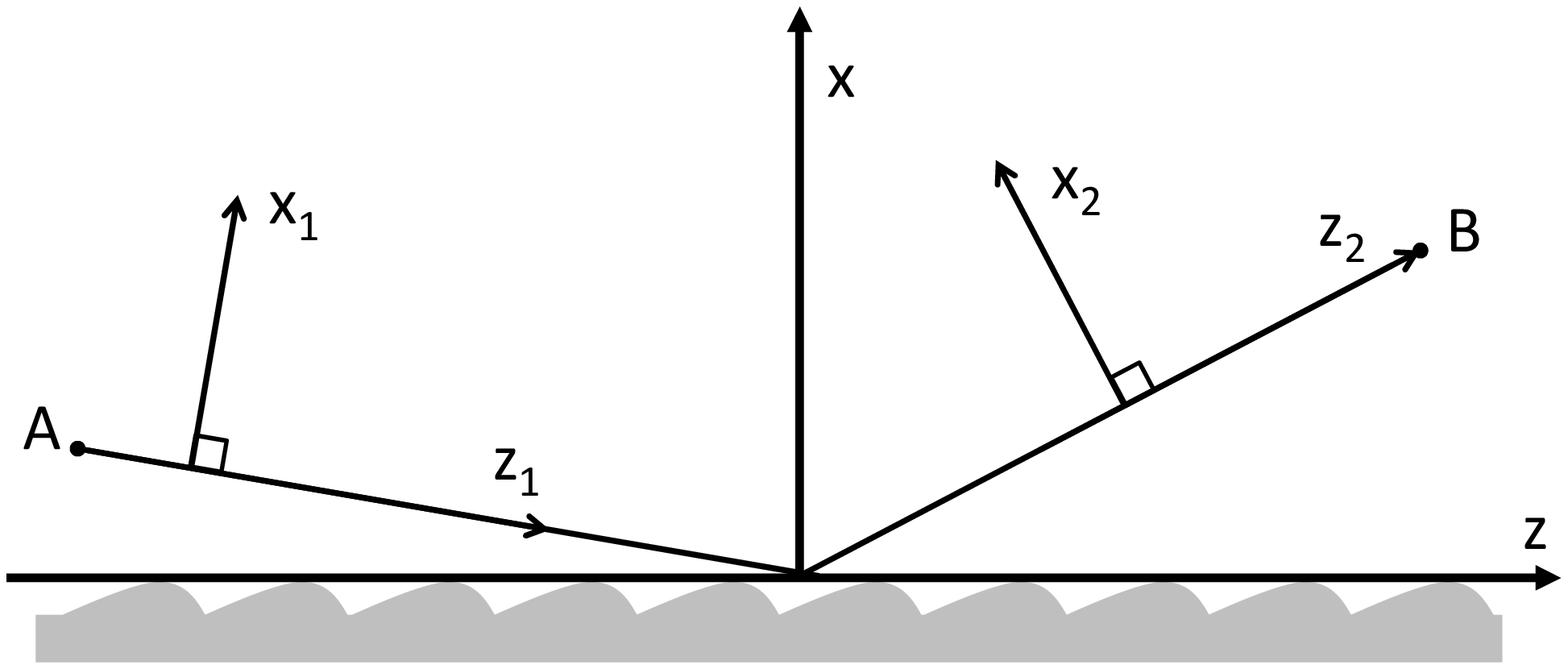}
\end{center}
\caption{Coordinate notation. The coordinate systems $(x,y,z)$,
$(x_1,y_1,z_1)$ and $(x_2,y_2,z_2)$ correspond to grating, incoming
beam and diffracted beam; the axes $z$, $z_1$ and $z_2$ are along
grating surface, incident and exit principal rays respectively.}
\label{gr_6}
\end{figure}
$
\\
\\
\\
\\
\\
\\
$

\begin{figure}[!h]
\begin{center}
\includegraphics[clip, width=0.75\textwidth]{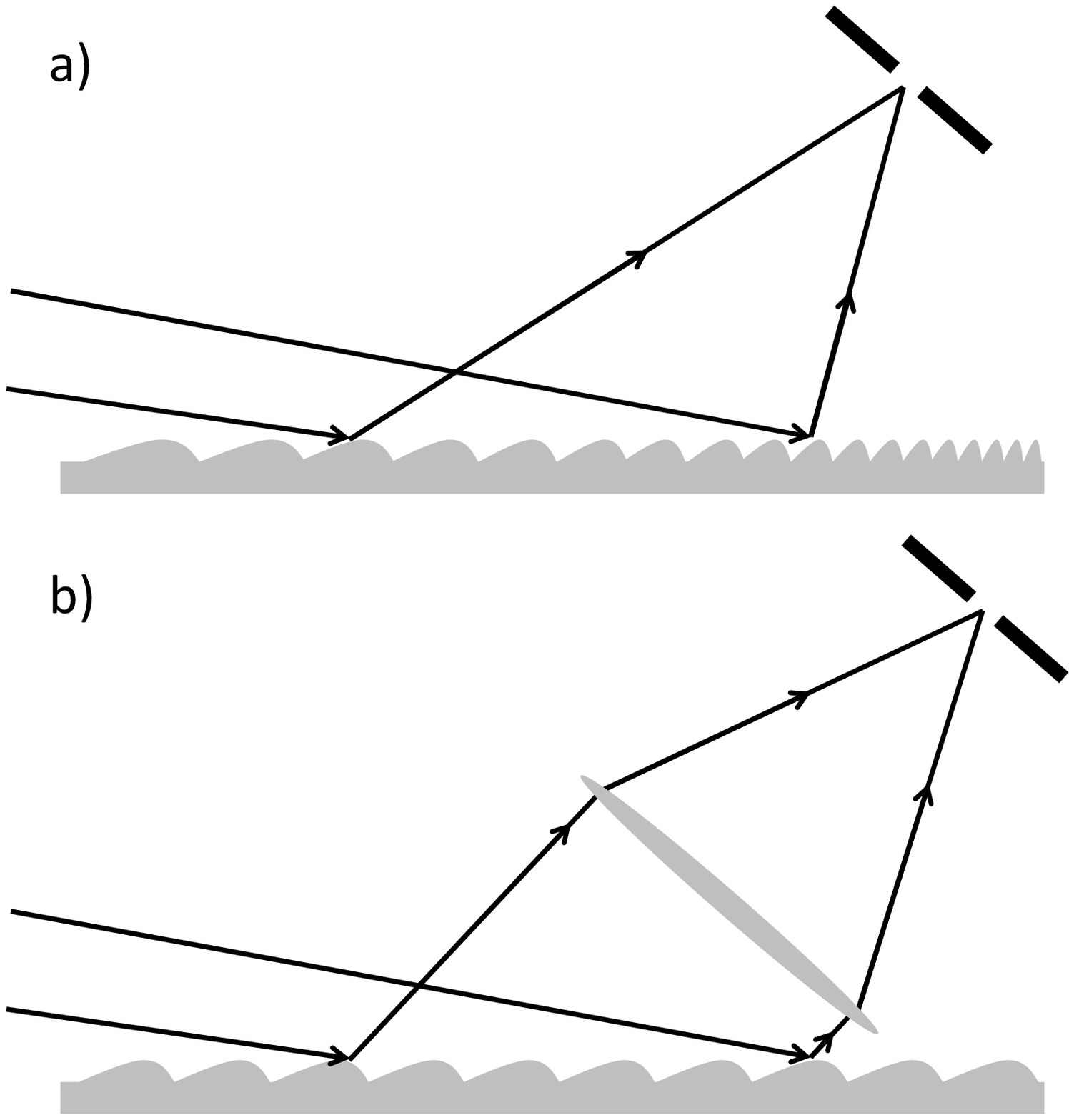}
\end{center}
\caption{Schematic diagram of a VLS grating element used in
theoretical analysis. The VLS grating is represented by a
contribution of a planar grating with fixed line spacing and a thin
lens.} \label{gr_7}
\end{figure}

\subsubsection{Plane grating}

Ruled gratings are essentially two-dimensional structures. As such,
their surface $S$ can be described by a function, e.g.  $x =
f(y,z)$, which expresses one of the three coordinates (in this case,
$x$) as a function of the other two, Fig. \ref{gr_1}. Let the beam
be incident from vacuum ($x > 0$) on the periodic cylindrical
interface illustrated in Fig. \ref{gr_2}. In this case, since S is
cylindrical, $f$ can be considered as the only function of $z$
independently on the value of $y$, and one has that $x = f(z)$ is a
periodic function of period $D$ (with spatial wave number $K =
2\pi/D$). Susceptibility is a periodic function of $z$ and can be
described by the Fourier series

\begin{eqnarray}
\delta \epsilon = 4\pi \chi = \sum_{m=-\infty}^{\infty} u_m(x)
\exp[i m K z]~ . \label{deps}
\end{eqnarray}
We want to obtain a diffracted wave, which we express in its most
general form as Eq. (\ref{grensol2}), from the knowledge of the
field incident on the grating. Using the relation between
$\vec{\bar{P}}$ and $\vec{\bar{E}}$, and the explicit expression for
$G$ in Eq. (\ref{Grr1}) we can write the following integral equation
for the electric field:

\begin{eqnarray}
\vec{\bar{E}}_d(\vec{r}) =&& k_0^2 \int d\vec{r'}  \frac{\exp[i k_0
|\vec{r}-\vec{r'}|]}{|\vec{r}-\vec{r'}|} \chi(x',z')\cr && \times
\left\{\left[\vec{\bar{E}}_d(\vec{r'}) + \vec{\bar{E}}_i
(\vec{r'})\right]- c^2 \vec{n}\left[\vec{n} \cdot
\left(\vec{\bar{E}}_d (\vec{r'}) + \vec{\bar{E}}_i
(\vec{r'})\right)\right]\right\} ~ . \label{scatt0}
\end{eqnarray}
It is customary to solve the scattering problem by a perturbation
theory, assuming that at all points in the dielectric medium the
diffracted field $\vec{E}_d$ is much smaller than the incident field
$\vec{E}_i$. This allows one to neglect the diffracted electric
field on the right hand side of Eq. (\ref{scatt0}) with the incident
field $\vec{E}_i$, yielding

\begin{eqnarray}
\vec{E}_d(\vec{r}) = k_0^2 \int d\vec{r'}  \frac{\exp[i k_0
|\vec{r}-\vec{r'}|]}{|\vec{r}-\vec{r'}|}
\chi(x',z')[\vec{E}_i(\vec{r'}) - c^2 \vec{n}(\vec{n} \cdot
\vec{E}_i(\vec{r'}))] ~ , \label{scatt}
\end{eqnarray}
where for simplicity we neglected the bar in the notation for the
field in the space-frequency domain.

In order to compute $\vec{E}_d$ in Eq. (\ref{scatt}) we need to
specify the incident field distribution $\vec{E}_i(\vec{r'})$ within
the dielectric medium. In fact, according to Eq. (\ref{scatt}) the
integration ranges over all coordinates $d\vec{r'}$, but $\chi$ is
different from zero inside the dielectric. Consider Fig. \ref{gr_8},
where we sketch the geometry for our problem. Monochromatic light
from a point source $A(x_1, y_1, z_1)$ is incident on a point
$P(x',y',z')$ located into the grating, i.e. into our dielectric
medium. Point $A$ is assumed, for simplicity, to lie in the $xz$
plane, i.e. $A=A(x_1,0,z_1)$. The plane $xz$ is called the
tangential plane (or the principal plane, or the dispersive plane).
The plane $yz$ is called the sagittal plane. As a first step we need
to express the incident field $\vec{E}_i$, appearing in Eq.
(\ref{scatt}), at the generic point $P$ inside the dielectric. In
order to do so, since we deal with a point source, we can take
advantage of the spatial impulse response of free-space. As we have
seen, this is nothing but the expression for a spherical wave
originating from $A$

\begin{eqnarray}
\vec{E}_i(x',y',z') = E_0 \frac{\exp[i k_0
|\vec{r'}-\vec{r}_1|]}{|\vec{r'}-\vec{r}_1|} ~.\label{Ei000}
\end{eqnarray}
After this, we consider that the beam is diffracted to the point
$B=B(x_2,0,z_2)$. Mathematically, diffraction is taken care of by
the Green's function in Eq. (\ref{scatt}), which represents a
secondary source from point $P$. Finally, an explicit expression for
$\chi$ is given in Eq. (\ref{deps}).

Even without explicit calculation of the integral in Eq.
(\ref{scatt}), a lot can be said analyzing the phase in the
integrand. In fact, since integration in Eq. (\ref{scatt}) involves
an oscillatory integrand, the integrand does not contribute
appreciably unless the arguments in the exponential functions
vanishes. We therefore calculate the total phase in the integrand of
Eq. (\ref{scatt}), and analyze it.

Calculations can be simplified by applying the paraxial
approximation. In fact, one can rely on it for writing expansions
for $\overline{AP} = |\vec{r'}-\vec{r}_1|$ and $\overline{PB} =
|\vec{r}_2-\vec{r'}|$ entering into the expression for the phase.
This can be done in terms of the distances $r_1(x) = \overline{AO'}$
and $r_2(x) = \overline{O'B}$, where $O'=(x',0,0)$, $x'$ being the
x-coordinate of point $P$. However, further simplifications apply by
noting that, in paraxial approximation, light actually traverses a
very small portion of material with susceptibility $\chi$. The range
of coordinates $x'$, $y'$, $z'$ inside the grating is much smaller
than the distances $r_1$ and $r_2$. In other words, the grating size
and its thickness are much smaller than $r_1$ and $r_2$.
Additionally, we assume that the grating thickness is much smaller
than the relevant transverse size. Thus, we can neglect the
dependence of distances $r_1$ and $r_2$ on $x'$ in the expansion for
the incident wave and in the Green function exponent, and use the
approximations $r_1 \simeq \overline{AO}$ and $r_2 \simeq
\overline{OB}$, where $O=(0,0,0)$ is a pole on the surface of
grating, Fig. \ref{gr_8}. Thus, the path $AOB$ defines the optical
axis of the beam, and the angle of incidence and of diffraction,
$\alpha$ and $\beta$ in Fig. \ref{gr_2}, are simply following that
optical axis. If points $A$ and $B$ lie on different sides of the
$xz$ plane, angles $\alpha$ and $\beta$ have opposite sign.

Starting from the expressions
\begin{eqnarray}
&&\overline{AP}^2 = [r_1 \sin \alpha + z']^2 + y'^2 + [r_1 \cos
\alpha ]^2 ~,\cr && \overline{PB}^2 = [r_2 \sin \beta + z']^2 + y'^2
+ [r_2 \cos \beta]^2~ . \label{APPB0}
\end{eqnarray}
and using a binominal expansion we can write the incident wave as

\begin{eqnarray}
&& \vec{E}_i (x', y', z') = \cr && E_0  \exp\left[ i k_0 \left(r_1 +
z' \sin \alpha + \frac{z'^2 \cos^2 \alpha }{2 (r_1 + z' \sin
\alpha)} + \frac{y'^2}{2(r_1+z' \sin \alpha)}\right) \right] =\cr &&
E_0 \exp\left[i k_0\left(r_1+ z' \sin \alpha + \frac{z'^2 \cos^2
\alpha}{2 r_1} + \frac{y'^2}{2r_1} - \frac{z'^3 \sin \alpha \cos^2
\alpha}{2r_1^2}  - \frac{z' y'^2 \sin \alpha}{2r_1^2}\right)\right]
,\cr && \label{binoinci}
\end{eqnarray}
The exponent of the Green function under the integral Eq.
(\ref{scatt}) as a function of the coordinates $x'$, $y'$ and $z'$
of the point $P$ on the grating. From Fig. \ref{gr_8}, one obtains

\begin{eqnarray}
&&\exp[ik_0|\vec{r}-\vec{r'}|] =\cr && \exp \left[ik_0\left(r_2 +
z'\sin \beta + \frac{z'^2 \cos^2 \beta }{2(r_2 + z'\sin \beta)} +
\frac{y'^2}{2(r_2 + z' \sin \beta)}\right)\right] = \cr &&
\exp\left[ik_0\left(r_2 + z'\sin \beta + \frac{z'^2 \cos^2 \beta}{2
r_2} + \frac{y'^2}{2r_2} - \frac{z'^3 \sin \beta \cos^2
\beta}{2r_2^2} - \frac{z'y'^2 \sin \beta}{2r_2^2}\right)\right]~
,\cr && \label{expesp}
\end{eqnarray}
We will now show that the periodic structure of the gratings
restricts the continuous angular distribution of the diffracted
waves to a discrete set of waves only, which satisfy the well-known
grating condition. In order to do so, we insert Eq. (\ref{deps}),
Eq. (\ref{binoinci}), and Eq. (\ref{expesp}) into Eq. (\ref{scatt}).
As noticed above,  the integrand does not contribute appreciably
unless the arguments in the exponential functions vanishes. From Eq.
(\ref{deps}), Eq. (\ref{binoinci}), and Eq. (\ref{expesp}) it
follows that the total phase in Eq. (\ref{scatt}) can be expressed
as a power series

\begin{eqnarray}
\phi = k_0 [ r_1 + r_2 + C_{10} z' + C_{20} z'^2 + C_{02} y'^2 +
C_{30} z'^3 + C_{12} z'y'^2  + ... ]  ~ . \label{Cexp}
\end{eqnarray}
Typically, third order aberration theory is applied to the analysis
of grating monochromators. In that case, the power series needs to
include third order terms. The explicit expressions for the
coefficients $C_{ij}$ are

\begin{eqnarray}
&&C_{10} = \frac{n K}{k_0} + \sin \alpha + \sin \beta ~,\cr &&C_{20}
=\frac{1}{2}\left[\frac{1}{r_1} \cos^2 \alpha + \frac{1}{r_2} \cos^2
\beta \right]~, \cr && C_{02} = \frac{1}{2} \left[ \frac{1}{r_1} +
\frac{1}{r_2}\right] ~, \cr && C_{30} = -\frac{1}{2 r_1^2} \sin
\alpha \cos^2 \alpha - \frac{1}{2 r_2^2} \sin \beta \cos^2 \beta~
,\cr && C_{12} = - \frac{1}{2 r_1^2} \sin \alpha  - \frac{1}{2
r_2^2} \sin \beta~.\label{Cij} \end{eqnarray}
$C_{20}$ and $C_{02}$ are the coefficients describing defocusing.
$C_{30}$ describes the coma, and $C_{12}$ the astigmatic coma
aberration\footnote{Differences in sign for $C_{10}$, $C_{30}$ and
$C_{12}$ with respect to literature are due to a different
definition of the direction of the $z$-axis, which points towards
$B$, and not towards $A$.}. In practice, the most important ones are
defocusing and coma. Ideal optics would require the phase $\phi$ to
be independent of $z$.

Note that the presence of the term $n K/k_0$ in the $C_{10}$
coefficient directly follows from the insertion of Eq. (\ref{deps})
into Eq. (\ref{scatt}). As said above, it is the periodic structure
of the gratings which restricts the continuous angular distribution
of the diffracted waves to a discrete set of waves.   In order to
find the direction of incident and diffracted beam, we impose the
condition $C_{10} = 0$, yielding:

\begin{eqnarray}
nK   +  k_0 (\sin \alpha + \sin \beta)   = 0 ~. \label{dir}
\end{eqnarray}
Eq. (\ref{dir}) is also valid for a plane mirror, if the grating
period is taken equal to infinity. This fact can be seen inspecting
Eq. (\ref{dir}), which yields $\alpha = -\beta$ for $D
\longrightarrow \infty$, which is nothing but the law of mirror
reflection.

Eq. (\ref{dir}) is known as the grating condition. This condition
shows how the direction of incident and diffracted wave are related.
Both signs of the diffraction order $n$ appearing into the equation
are allowed. Assuming for simplicity diffraction into first order,
i.e. $n = + 1$, one has

\begin{eqnarray}
\lambda = D (\cos \theta_i - \cos \theta_d) ~, \label{grat2}
\end{eqnarray}
where $\theta_i$ and  $\theta_d$  are the angles between the grating
surface and, respectively, the incident and the diffracted
directions. By differentiating this equation in the case of a
monochromatic beam one obtains

\begin{eqnarray}
b = \frac{d \theta_d}{d \theta_i} =  \frac{\sin \theta_i}{\sin
\theta_d}\label{b}
\end{eqnarray}
Note that $b = W_i/W_d$ is the ratio between the width of the
incident and of the diffracted beam. Fig. \ref{gr_4} shows the
geometry of this transformation. As has been pointed out elsewhere
this is just the consequence of Liouville's theorem.

The effect of the plane grating on the monochromatic beam is
twofold: first, the source size is scaled by the asymmetry factor
$b$ defined in Eq. (\ref{b}) and, second, the distance between
grating and virtual source behind the grating is scaled by the
square of the asymmetry factor $b$, Fig. \ref{gr_3}. In order to
illustrate this fact, we consider a 1D Gaussian beam with an
initially plane wavefront, described by the field amplitude (along a
given polarization component) $\psi(x, 0) =\ exp[-x^2/w_0^2]$.
Assuming that the plane grating is positioned at $z$,  the spatial
spectrum of the Gaussian beam immediately in front of the grating,
i.e. after propagation in free-space by a distance $z$ from the
waist point, is given by

\begin{eqnarray}
\psi(k_x,  z) =  \sqrt{\pi} w_0 \exp[ik_0 z] \exp\left[ -\frac{k_x^2
w_0^2}{4}\right] \exp\left[-\frac{i k_x^2 z}{2k_0}\right]~    .
\label{FTgauss}
\end{eqnarray}
However, according to Eq. ((\ref{b})), the transformation of the
angular spectrum performed by grating can be described with the help
of $k'_x = b k_x$, so that immediately after grating one obtains

\begin{eqnarray}
\psi(k_x, z) = \sqrt{\pi} w_0 \exp[ik_0z] \exp\left[- \frac{k_x^{'2}
w_0^2}{4 b^2}\right] \exp\left[-\frac{i k_x^{'2} z}{2 k_0
b^2}\right]~.\label{trasf}
\end{eqnarray}
We can interpret Eq. (\ref{trasf}) in the following way: the
Gaussian beam diffracted by the grating is characterized by a new
\textit{virtual} beam waist $w'_0 = w_0/b$ and by a new
\textit{virtual} propagation distance $z' = z/b^2 $. Introducing the
dimensionless distance through the relation $z/L_R$, where $L_R$ is
called the Rayleigh length, we can conclude that this dimensionless
distance is invariant under the transformation induced by the plane
grating.

The treatment of the diffraction grating given above yielded most of
the important results needed for further analysis. In particular, it
allowed us to derive the grating condition and it also allowed us to
study the theory of grating aberrations. Our theoretical approach
reaches into the foundation of electrodynamics, as is based on the
use of Maxwell equations. Note that the treatment considered so far
was carried out under the assumption of the validity of the first
order perturbation theory, i.e. we assumed that for all the points
in the dielectric medium, the diffracted field is negligible with
respect to the incident field. The properties of the field actually
exploited amount to the fact that in the $yz$ plane, the diffracted
field has the same phase as the incident field plus an extra-phase
contribution $n K z'$. If we go up to second and higher orders in
the perturbation theory we can see that this property remains valid,
and results derived above still hold independently of the
application of a perturbation theory. Note that inside the grating
the beam is attenuated with a characteristic length that is much
shorter compared to the range of the grating surface coordinates,
and can always be neglected in the phase expansion. We can
immediately extend the range of validity of our analysis to
arbitrary  values of the dielectric constant. The general proofs of
the grating condition and of the results of the theory of grating
aberration are derived from first principles as follows \cite{STRO}.

First let us note that two-dimensional problems are essentially
scalar in nature, and can be expressed in terms of only one single
independent electromagnetic field variable, either $E_y$  or $H_y$.
Here we will working considering the TE polarization, i.e. we will
be focusing on $E_y$. The action of the grating on the
electromagnetic field can be modeled, mathematically, as an operator
$\mathcal{G}$ that transforms an incident field into a diffracted
field, i.e. $E_d(z,y) = \mathcal{G}[E_i(z,y)]$. Since the grating is
periodic and extends to infinity, the action of the operator
$\mathcal{G}$ is invariant under translation by a grating period:
$E_d(z+D, y) = \mathcal{G}[E_i(z+D, y)]$. Since the incoming beam is
incident at an angle $\theta_i$, this translation adds an extra path
distance $D \cos \theta_i$ to the incident wave $E_i$, for a phase
change

\begin{eqnarray}
E_i(z+D, y) = \exp(ik_0 D \cos \theta_i) E_i(z, y)~  .
\label{phchange}
\end{eqnarray}
Also, since the set of Maxwell partial differential equations is
linear, any solution multiplied by a constant is still a solution
and one obtains

\begin{eqnarray}
\mathcal{G}[E_i(z+D, y)] = \mathcal{G}[\exp(i \delta \phi) E_i(z,
y)] = \exp(i \delta \phi) E_i (z, y) ~, \label{middleEi}
\end{eqnarray}
where $\delta \phi = k_0 D \cos \theta_i$. Now, since

\begin{eqnarray}
\mathcal{G}[E_i(z+D, y)] = E_d(z+D, y) \label{invaEd}
\end{eqnarray}
we must have

\begin{eqnarray}
\exp(i \delta \phi) E_d (z, y) = E_d (z+D, y)  ~. \label{phaseshift}
\end{eqnarray}
In other words, the diffracted field is a pseudo-periodic function.
Now, since  the product $E_d \exp[- i k_0 z \cos \theta_i]$  is a
periodic function, it can be represented as a Fourier series
expansion on the grating period $D$, and we can write the diffracted
field as

\begin{eqnarray}
E_d (z, y) = \sum_{m=-\infty}^{+\infty} E_m (y) \exp[ i m K z + ik_0
z \cos \theta_i] ~. \label{expsumEd}
\end{eqnarray}
This result is fully general, and all that is required to prove it
is that the grating is periodic. Eq. (\ref{expsumEd}) is sufficient
for describing the geometry of the beam diffraction by the grating.
We can use Eq. (\ref{expsumEd}) to derive once more the grating
condition.

In order to illustrate this fact, we see that the phase of the
integrand in the integral Eq. (\ref{scatt0}) consists of three
terms: the first term is the phase in the Green function, the second
is the phase in Eq. (\ref{deps}), and the third is the phase in
$\vec{E}_d$. The first and the second terms are known, and have
already been analyzed.  Eq. (\ref{expsumEd}) shows the structure of
the phase for $\vec{E}_d$ in the case for a plane wave impinging on
the grating with incident angle $\theta_i$. In principle, the
incident field $\vec{E}_i$ comes from a point source located in $A$,
and consists of a diverging spherical wave. Such spherical wave can
always been decomposed in plane waves and, due to the validity of
the paraxial approximation, only those plane wave components with
angle near to $\theta_i$ should be considered. Therefore, neglecting
small corrections in $\Delta \theta_i$, one can take the phase in
Eq. (\ref{expsumEd}) as a good approximation for the phase of the
diffracted field. Then, considering the expansion in Eq.
(\ref{expesp}) to the first order in $z'$ one obtains, without using
a perturbative approach, that the term in $z'$ in the integrand in
Eq. (\ref{scatt0}) is given by $(m+n) K   +  k_0 (\cos \theta_i
-\cos \theta_d)$. Imposing that this term be zero, and remembering
that $\alpha = \pi/2 - \theta_i$, one gets back Eq. (\ref{dir}).

This result, albeit very general, still says nothing about the
grating efficiency. We still do not know anything about the
amplitudes of the diffracted waves. In order to determine these
coefficients we need to model the grooves of the grating.  At this
point, we need to apply classical numerical integration techniques
\cite{PETI,BOOT}.

\subsubsection{VLS plane grating}

A diffractive plane grating can focus a diffracted beam when the
groove spacing properly varies with the groove position; such a
grating is called a variable-line-spacing (VLS) grating. A VLS plane
grating can be incorporated into the monochromator to act as both
dispersive and  spectrally focusing component. The working principle
of such kind of grating can be understood by expressing the groove
spacing $D(z)$ as a function of the coordinate $z$ along the
perpendicular to the grooves. So it can be expanded as a polynomial
series\footnote{Another choice of line-spacing parametrization found
in literature is the expansion of the line density $n(z) = 1/D(z) =
n_0 + n_1 z + n_2 z^2 + ...$. With these definitions, $n_1$ and
$n_2$ are the same as in Table \ref{table:components}.}:

\begin{eqnarray}
D(z) = D_0 + D_1 z + D_2 z^2 + ... ~, \label{VLSexp}
\end{eqnarray}
where the term $D_0$ is the spacing at the pole of the grating
(located, by definition, at $z = 0$), while $D_1$ and $D_2$ are the
parameters for the variation of the ruling with $z$. Now
susceptibility is not a periodic function of  $z$ anymore, and can
be described by the Fourier integral:

\begin{eqnarray}
\delta \epsilon = 4 \pi \chi =  \int_{-\infty}^{\infty} B(K, x)
\exp[iKz] d K ~. \label{deleps}
\end{eqnarray}
Let us assume, for simplicity, that the distance between grooves
varies according to the linear law: $D(z) = D_0 + D_1 z$. Now  we
also assume that $D_1 \ll 1$ and we apply the so-called adiabatic
approximation imposing that the width of the peaks in the spectrum
$B(K, x)$ is much narrower than the harmonic separation $K_0 =
2\pi/D_0$ between the peaks. In this case, Eq. (\ref{deleps}) can be
represented in the form

\begin{eqnarray}
\delta \epsilon = \sum_{-\infty}^{+\infty} B_n (x, z) \exp[ inK_0 z]
~, \label{deleps2}
\end{eqnarray}
where the complex amplitudes $B_n(x, z)$ are all slowly varying
function of the $z$ coordinate on the scale of the period $D_0$.
This means that the terms in sum over $n$ in Eq. (\ref{deleps2}) can
be analyzed separately for each value of $n$. For the case of a
linearly chirped grating considered here, the slowly varying
amplitude of the $n$th harmonic is given by

\begin{eqnarray}
B_n = A_n(x) \exp\left[i \frac{z^2}{2} \frac{d K}{dz} \right] ~,
\label{Bnslow}
\end{eqnarray}
where $d K/dz =  2\pi d(1/D)/dz = - (2\pi/D_0^2) D_1$ is the chirp
parameter.

We now substitute Eq. (\ref{deleps2}) into Eq.(\ref{scatt}) and, as
before, we express the phase in the integrand as a power series.
Only the $C_{20}$ term differs, with respect to the expression in
Eq. (\ref{Cij}). In fact, for a linearly chirped grating we obtain
\cite{ITOU}

\begin{eqnarray}
C_{20} = - \frac{\lambda}{2 D_0^2}  D_1 + \frac{1}{2r_1} \cos^2
\alpha+ \frac{1}{2 r_2} \cos^2 \beta ~. \label{C20vls}
\end{eqnarray}
The condition $C_{20} = 0$ has to be verified in order to guarantee
imaging in the tangential plane.

Here we used Maxwell equations for studying the imaging properties
of VLS grating. However, certain aspects of this theory can be
derived in a simple way using ray optics. For convenient use in the
following discussions, it is necessary to make clear the reference
coordinate systems and rays describing the optical system. Fig.
\ref{gr_6} shows the VLS plane grating optical system with an object
point $A$. The coordinate systems $(x,y,z)$ , $(x_1, y_1, z_1)$ and
$(x_2, y_2, z_2)$ correspond, respectively, to the grating, to the
incident beam, and to the diffracted beam; the axes $z$, $z_1$ and
$z_2$ are along the grating surface, the incident and the exit
principal rays, respectively. As shown in Fig. \ref{gr_2}, the input
beam is incident on grating at angle $\theta_i$. The diffracted
angle $\theta_d$ is a function of the groove distances according to
the grating equation

\begin{eqnarray}
\lambda = D (\cos \theta_i - \cos \theta_d )~ . \label{grateq2}
\end{eqnarray}
By differentiating over $z$ for the case of a monochromatic beam we
obtain

\begin{eqnarray}
\frac{d D}{dz} (\cos \theta_i - \cos \theta_d) =  - \frac{d
\theta_d}{ dz} D \sin \theta_d  ~, \label{diffgrte}
\end{eqnarray}
yielding

\begin{eqnarray}
[\lambda/D_0^2 (\sin^2 \theta_d)] D_1 = - \frac{d \theta_d}{dx_2} ~,
\label{tre}
\end{eqnarray}
where we used the relation  $z \sin \theta_d = x_2$.

Let us now define a thin lens as a device that deflects every light
beam incident parallel to the optical axis in such a way that it
crosses the optical axis at a fixed distance $f$ after passing
through the lens. In paraxial approximation, the thin lens equation
assumes the familiar form $d \theta_d = - d x_2/f$ . The physical
meaning of Eq. (\ref{tre}) is that the VLS plane grating can be
represented by a combination of a planar grating with fixed line
spacing and a lens after the grating, with a focal length $f$ equal
to the focal length of the VLS grating

\begin{eqnarray}
f = [\lambda D_1 /D_0^2  (\sin^2 \theta_d)]^{-1}~  , \label{VLSf}
\end{eqnarray}
as shown in Fig. \ref{gr_6}. It may seem surprising that the focal
length depends on $\theta_d$ only. However, it is reasonable to
expect an influence of the assumption that the lens placed after the
grating. One intuitively expects that full transfer matrix for the
VLS grating should not depend on the choice of the lens position. It
will be shown below that indeed, the transfer matrix satisfies this
invariance.

An ABCD matrix is intended to represent any arbitrary paraxial
element, or optical system located between an input plane and an
output plane. In the present case, the optical element is the VLS
plane grating with the input plane corresponding to the plane
perpendicular to the incident beam and with the output plane the
plane perpendicular to the diffracted beam.  The most usual
application for ray matrices is to forming the image of the object
The most usual application for ray matrices is the determination of
the image of the object located at the input plane. In this case,
some important properties of optical system are obtained when any of
the ABCD parameters vanish \cite{MORE}.

The total optical system from the object plane (to which point A
belongs) to the image plane (to which point B belongs), see Fig.
\ref{gr_6}, is represented by the matrix:

\begin{eqnarray}
\left(
\begin{array}{l}
A_\mathrm{tot} ~~~~ B_\mathrm{tot}
\\
C_\mathrm{tot} ~~~~ D_\mathrm{tot}
\end{array}
\right) = \left(
\begin{array}{l}
1 ~~~~ r_2
\\
0 ~~~~ 1
\end{array}
\right) \left(
\begin{array}{l}
1 ~~~~~~~~~~~ 0
\\
-1/f ~~~~ 1
\end{array}
\right)\left(
\begin{array}{l}
b ~~~~~~ 0
\\
0 ~~~~ 1/b
\end{array}
\right)\left(
\begin{array}{l}
1 ~~~~ r_1
\\
0 ~~~~ 1
\end{array}
\right)~,\label{ABCD1}
\end{eqnarray}
where $b$ is the asymmetric parameter  $\sin \theta_i/\sin
\theta_d$, see \cite{SIEG}. The explicit expression for the total
matrix elements are

\begin{eqnarray}
&&A_\mathrm{tot} = b - b r_2/f ~,\cr && B_\mathrm{tot} = b r_1 - b
r_1 r_2/f + r_2 / b~, \cr && C_\mathrm{tot} = - b/f  ~,\cr &&
D_\mathrm{tot} = - b r_1/f + 1/b ~. \label{ABCD2}
\end{eqnarray}
The condition $B_\mathrm{tot}$ = 0 has to be verified in order to
guarantee imaging of the object at the output plane. In fact, when
$B_\mathrm{tot} = 0$, any point source at the input plane focuses at
the corresponding point in the output plane, regardless of the input
angle. Therefore, the output plane is the image plane. Dividing the
equation $B_\mathrm{tot} = 0$  by $r_1 r_2$ on the left hand side we
find the imaging equation \cite{APRI}

\begin{eqnarray}
\frac{b}{r_1} + \frac{1}{b r_2} = \frac{D_1 \lambda}{D_0^2 \sin
\theta_i \sin \theta_d}~, \label{imrel}
\end{eqnarray}
which is identical to the imaging condition $C_{20} = 0$ which we
derived above from first principles, because $\sin \theta_i = \cos
\alpha$ and $\sin \theta_d = \cos \beta$. It thus follows that the
ABCD matrix for the VLS plane grating in the tangential plane has
the general form

\begin{eqnarray}
\left(
\begin{array}{l}
A_\mathrm{tot} ~~~~ B_\mathrm{tot}
\\
C_\mathrm{tot} ~~~~ D_\mathrm{tot}
\end{array}
\right) = \left(
\begin{array}{l}
A_\mathrm{tot}  ~~~~~~~~ 0
\\
-1/f_\mathrm{tot} ~~~~ D_\mathrm{tot}
\end{array}
\right) \label{finalabcd}
\end{eqnarray}
with the effective focal length given by \cite{APRI}

\begin{eqnarray}
\frac{1}{f_\mathrm{tot}} = \frac{\lambda D_1}{D_0^2 \sin \theta_i
\sin \theta_d}~ , \label{feff}
\end{eqnarray}
which is symmetric in $\theta_i$ and $\theta_d$ as it must be. The
ABCD matrix elements can be used to characterize  width and
wavefront curvature of the Gaussian beam after its propagation
through the VLS grating.

\subsubsection{Toroidal grating}

A logical extension of the  plane VLS grating concept described
above follows from the idea to rule the VLS grooves on a toroidal
surface, producing a toroidal VLS grating \cite{THOM}. Additional
design parameters, namely tangential and sagittal radius, are then
available to control imaging aberrations and to optimize the grating
monochromator performance \cite{HABE}. We consider a curved VLS
grating and we assume that the surface of the grating is toroidal
with tangential and sagittal radius of curvature $R$ and $\rho$
respectively, see Fig. \ref{gr_5}. Let us assume that the distance
between the grooves varies according to quadratic law:

\begin{eqnarray}
D(z) = D_0 + D_1 z + D_2 z^2 ~. \label{quadgrov}
\end{eqnarray}
As before, the susceptibility is not a periodic function with
respect to $z$, and in adiabatic approximation can be represented in
the form

\begin{eqnarray}
\delta \epsilon = \sum_{-\infty}^{+\infty} B_n (x, z) \exp[ inK_0 z]
~, \label{deleps2bis}
\end{eqnarray}
which is identical to Eq. (\ref{deleps2}), where complex amplitudes
$B_n$ are slowly varying functions of the $z$ coordinate on the
scale of the period $D_0$. In the case of quadratically chirped
grating, the $n$th amplitude is given by

\begin{eqnarray}
B_n = A_n(x) \exp [ i K' z^2/2 + i K''z^3/6 ] ~, \label{Bn2}
\end{eqnarray}
where

\begin{eqnarray}
&&K' = 2\pi (1/D)' = - 2\pi D_1/D_0^2 = 2\pi n_1 ~,\cr && K'' = 2
\pi (1/D)'' = - 4\pi [D_2/D_0^2 - D_1^2/D_0^3] = 4\pi n_2~,
\label{K1K2}
\end{eqnarray}
are the linear and quadratic chirp parameters.

From the geometry (see Fig. \ref{gr_5}), and similarly as done
before we can write

\begin{eqnarray}
&&\overline{AP}^2 = [r_1 \sin \alpha + z]^2 + y^2 + [r_1 \cos \alpha
- x]^2 ~,\cr && \overline{PB}^2 = [r_2 \sin \beta + z]^2 + y^2 +
[r_2 \cos \beta - x]^2~ , \label{APPB}
\end{eqnarray}
where the coordinate $x$ on the toroidal surface is related to $z$
and $y$ by the equation of the torus

\begin{eqnarray}
x = R - R\left[ 1 - \frac{z^2+y^2}{R^2} + 2 \frac{\rho}{R}\left(
\frac{\rho}{R} - 1\right)\left( 1 - \left(1-
\frac{y^2}{\rho^2}\right)^{1/2}\right)\right]^{1/2} ~. \label{torus}
\end{eqnarray}

The integrand in Eq. (\ref{scatt}) is oscillatory, and does not
contribute appreciably to the total integral unless the arguments of
the exponential function vanishes. Using Eq. (\ref{scatt}), Eq.
(\ref{binoinci}) and Eq. (\ref{expesp}), together with Eq.
(\ref{deleps2bis}), Eq. (\ref{Bn2}), Eq. (\ref{APPB}), and Eq.
(\ref{torus}), it is then possible to expand the phase as a power
series such as

\begin{eqnarray}
\phi = k_0 ( r_1 + r_2 + C_{10} z + C_{20} z^2 +  C_{02} y^2 +
C_{30} z^3 + C_{12} z y^2  + ...  ) ~  . \label{phitor}
\end{eqnarray}
The explicit expressions for coefficients $C_{ij}$ are \cite{HARA}

\begin{eqnarray}
&&C_{10} =  \frac{n \lambda}{D_0} + (\sin \alpha + \sin \beta) ~,\cr
&& C_{20} = \frac{n \lambda n_1}{2} + \frac{1}{2}\left[ \frac{\cos^2
\alpha}{r_1} + \frac{\cos^2 \beta}{r_2} - \frac{\cos \alpha}{R} -
\frac{\cos \beta}{R} \right]~ , \cr && C_{30} = \frac{n \lambda
n_2}{3} - \frac{1}{2}\left[\left( \frac{\cos^2 \alpha}{r_1} -
\frac{\cos \alpha}{R} \right) \frac{\sin \alpha}{r_1} + \left(
\frac{\cos^2 \beta}{r_2} - \frac{\cos \beta}{R}\right) \frac{\sin
\beta}{r_2} \right] ~, \cr && C_{02} = \frac{1}{2}\left[
\frac{1}{r_1} - \frac{\cos \alpha}{\rho}\right] + \frac{1}{2} \left[
\frac{1}{r_2} - \frac{\cos \beta}{\rho}\right] ~, \cr && C_{12} =
-\frac{1}{2} \left[ \left( \frac{1}{r_1} - \frac{\cos \alpha}{\rho}
\right) \frac{\sin \alpha}{r_1} + \left( \frac{1}{r_2} - \frac{\cos
\beta}{\rho}\right) \frac{\sin \beta}{r_2} \right]~ ,
\label{Cijtorus}
\end{eqnarray}
where the condition $C_{10} = 0$ yields back the grating condition,
$C_{20} = 0$ yields the position of the tangential focus position,
$C_{02} = 0$ that of the sagittal focus, and the relation $C_{30} =
0$ minimizes the coma aberration.

In section \ref{Methabe} we will demonstrate that toroidal grating
aberrations can be modeled very straightforwardly using a
geometrical approach. This derivation is very different from the
analytical method used in literature. We heavily relied on
geometrical considerations, and we hope that calculations performed
in section \ref{Methabe} are sufficiently straightforward to give an
intuitive understanding of Eq. (\ref{Cijtorus}).

\section{\label{sec:model} Modeling of self-seeding setup with grating
monochromator}

\subsection{Source properties}

\begin{figure}
\begin{center}
\includegraphics[width=0.75\textwidth]{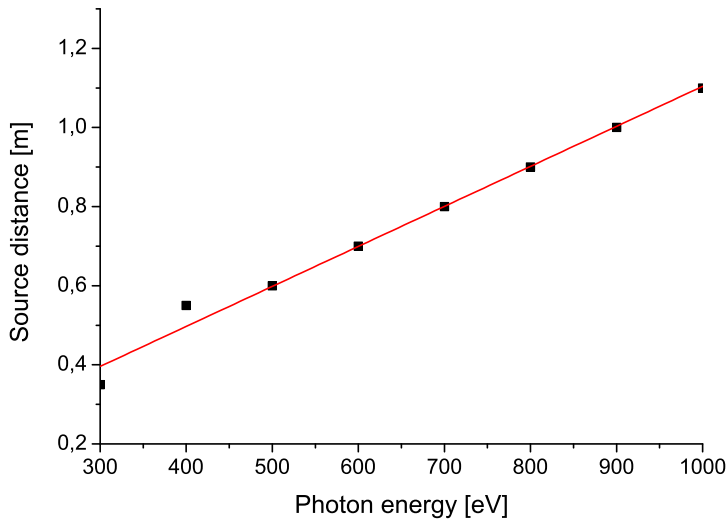}
\end{center}
\caption{Distance of the source from the SASE undulator exit as a
function of photon energy. Results are found by means of FEL
simulations.} \label{source_dist}
\end{figure}

\begin{figure}
\begin{center}
\includegraphics[width=0.75\textwidth]{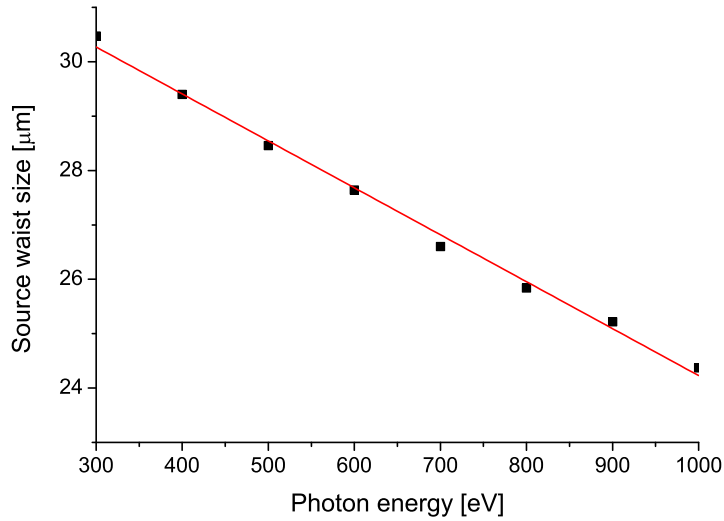}
\end{center}
\caption{Size of the source waist as a function of the photon
energy. Results found by means of FEL simulations.}
\label{source_size}
\end{figure}

In order to perform calculations of the grating beamline
performance, one needs the effective source size and position
through the operating photon energy range. The properties of the
effective source are found from steady-state simulations with the
help of the code Genesis 1.3 \cite{GENE}. The simulations include
electron beam parameters (emittance, energy spread, peak current)
found by start-to-end simulations for the $0.1$ nC electron bunch
mode of operation. Beam parameters for the steady-state simulations
have to be chosen to match the parameters of the bunch slice with
maximum peak current. The properties of the effective source can be
found from the simulated field at the SASE undulator exit. This is
accomplished by propagating the simulated field backwards from the
undulator exit in order to find the position of the waist. The field
must to be propagated in free-space. An in-house free-space
wavefront propagation code was used to this purpose. The code is
written in MATLAB and based on fast Fourier transform implementation
of the Fourier optics method discussed in section \ref{31}. Fig.
\ref{source_dist} shows the distance from the source to the SASE
undulator exit as a function of the photon energy. It is seen that
the source point moves upstream with increasing photon energy by as
much as one meter. The Gaussian fit gives the source waist size
$w_0$, as shown in Fig. \ref{source_size}.

\subsection{Focusing at the second undulator entrance}

\begin{figure}
\begin{center}
\includegraphics[width=0.75\textwidth]{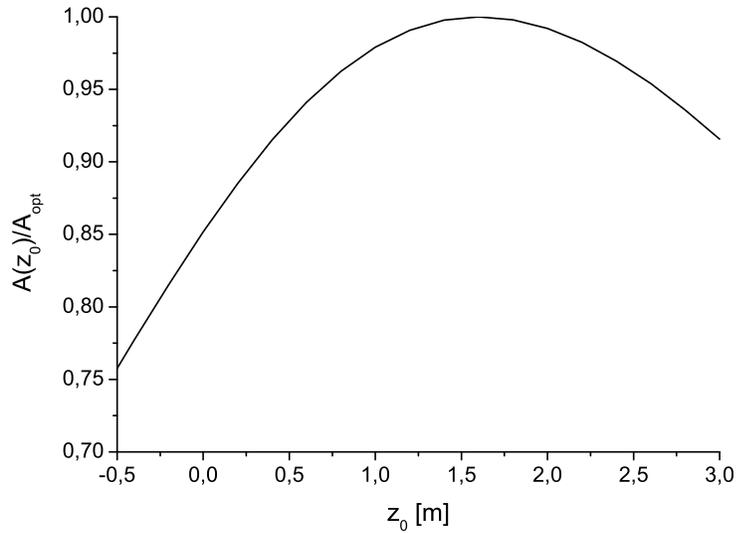}
\end{center}
\caption{Dependence of the input coupling factor A on the position
of the Gaussian beam waist. Here $\hbar \omega = 500$ eV.}
\label{nom_opt_compare}
\end{figure}
\begin{figure}
\begin{center}
\includegraphics[width=0.75\textwidth]{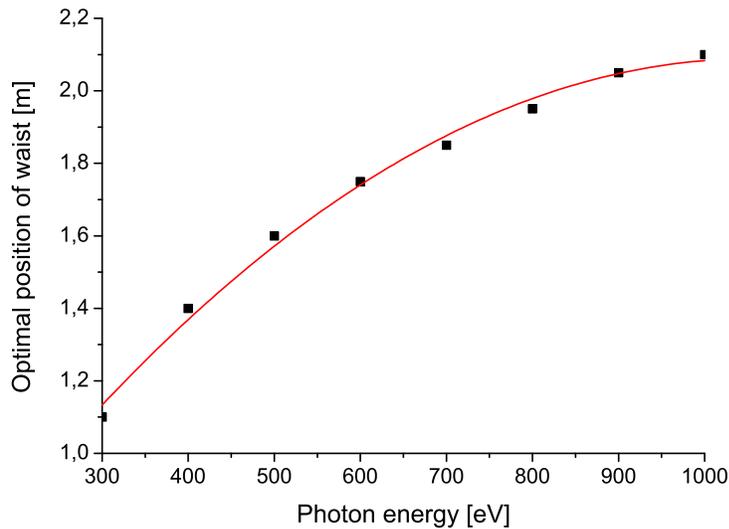}
\end{center}
\caption{Optimal position of the Gaussian beam waist (characterized
by plane wavefront) into the second undulator as a function of the
photon energy. The waist size of the seed beam is equal to the
source waist size in the first undulator. A usual figure of merit is
the optimal position of the waist for the maximal input coupling
factor. Results are obtained using wave optics and FEL simulations.}
\label{opt_pos}
\end{figure}
\begin{figure}
\begin{center}
\includegraphics[width=0.75\textwidth]{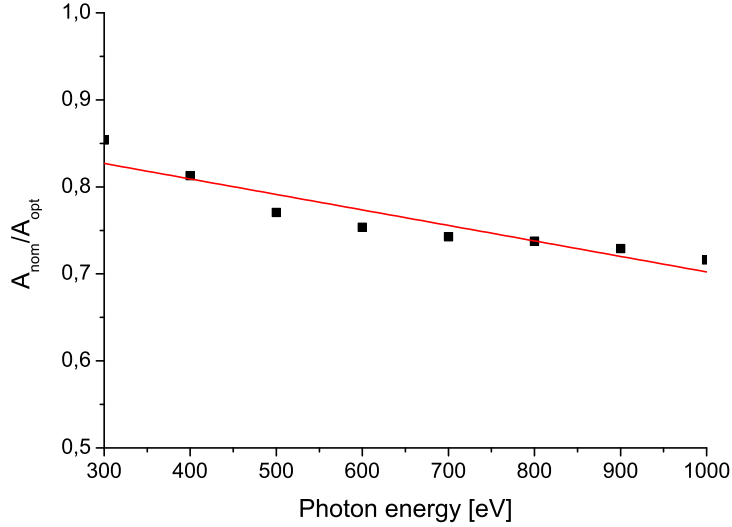}
\end{center}
\caption{Ratio of input coupling factor for nominal and optimal
seeding as a function of the photon energy.} \label{opt_coord}
\end{figure}

\begin{figure}
\begin{center}
\includegraphics[width=0.75\textwidth]{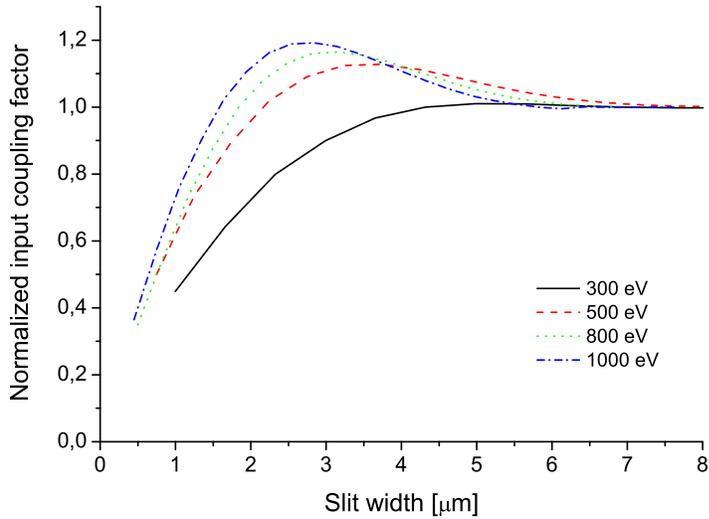}
\end{center}
\caption{Transverse mismatch between monochromatic seed and electron
beam as a function of exit slit width for different photon energies.
A useful figure of merit measuring the mismatch is the input
coupling factor normalized to the asymptotic case without exit slit.
Results obtained using wave optics and FEL simulations.}
\label{mismatch_slit_sec4}
\end{figure}

\begin{figure}
\begin{center}
\includegraphics[width=0.75\textwidth]{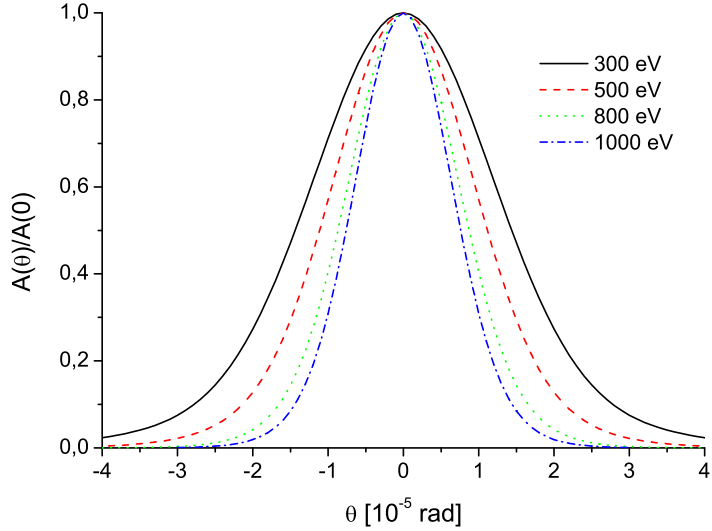}
\end{center}
\caption{Results of FEL amplifier simulations showing the influence
of the wavefront tilt in the seed beam. The normalized input
coupling factor is plot as a function of the tilt angle for two
photon energies.} \label{acceptance}
\end{figure}

Let us study the problem of optimal focusing of the seed radiation
on the electron beam at the undulator entrance. We consider the case
when the seed radiation has the form of a Gaussian beam, and when
the FEL operates at exact resonance. The optimal focusing conditions
can be found running steady-state simulations in Genesis 1.3
\cite{GENE}.

The waist of the Gaussian beam is located at position $z_0$, where
we have a plane phase front and a Gaussian distribution of
amplitude. When the undulator is sufficiently long, the output power
grows exponentially with undulator length, and the power gain, $G =
W_\mathrm{out}/W_\mathrm{seed}$, can be written as

\begin{eqnarray}
G = A \exp[z/L_g]~ , \label{gain}
\end{eqnarray}
where $z$ is the undulator length and $L_g$ is called the power gain
length. In the linear regime the power gain does not depend on the
input power $W_\mathrm{seed}$, so that the input coupling factor $A$
is a function of two parameters only: the coordinate of the waist
location, $z_0$, and the waist size, $w_0$. There are always optimal
values of Gaussian beam parameters, $w_0$ and $z_0$, when the input
coupling factor $A$ achieves its maximum. In order to simplify the
optimization problem, we will not study any change in $w_0$, but
rather set it equal to the waist size of the effective source in the
SASE undulator. Fig. \ref{nom_opt_compare} shows the dependence of
the input coupling factor $A$ on the focus coordinate $z_0$ at the
photon energy of $500$ eV. The optimal coordinate of the waist point
is a function of the photon energy. The plot of this function is
presented in Fig. \ref{opt_pos}. It is clearly seen that the optimal
position of the waist located $1-2$ m inside the seeding undulator.
The plots allow one to maximize the seeding efficiency at fixed
power of the seed beam.

From the above analysis follows that that a one-to-one imaging of
the radiation beam at the exit of the first undulator onto the
entrance plane of the second undulator (which is obviously optimal
in the case of negligible chicane influence) becomes non-optimal in
the case of our interest. This is a consequence of the fact that the
microbunching in the electron beam is washed out by the chicane and,
therefore, at the entrance of the second undulator the seed
radiation beam interacts with a "fresh" electron beam.  Numerical
simulations show that the reduction factor for the one-to-one
imaging case compared with the optimal case is about $30 \%$.

The main efforts in developing our design for a self-seeding
monochromator are focused on resolution and compactness. Therefore,
there is somewhat a residual mismatching between seed and electron
beam on the nominal mode of operation.  From Fig. \ref{opt_coord}
one can see that seed beam on the nominal mode of operation is
generated with a mismatching of only $10-20 \%$ .

Wave optics, together with FEL simulations are naturally applicable
also to the study the influence of finite slit size on the
amplification process into the second undulator. In particular, we
studied the influence of the exit slit size on the seeding
efficiency. Such effect is shown in Fig. \ref{mismatch_slit_sec2}.
One can see that decreasing the slit size drastically decreases the
efficiency. The reason for this is a reduction of the seed power and
the introduction of an additional mismatch between the seed beam and
the electron beam. It is instructive to study these two effects
separately. Fig. \ref{mismatch_slit_sec4} shows the ratio of the
input coupling factors for seeding with and without slit, as a
function of the slit size. When the slit size is smaller than $2
\mu$m, diffraction on the slit drastically decreases input coupling
factor. On the other hand, for a slit size of about $2 \mu$m,
perturbation of the Gaussian beam shape leads to about $10-15 \%$
increase  in the input coupling factor.

In order to calculate the tolerance on the wavefront tilt of the
seed beam, it is necessary to have knowledge of the angular
acceptance of the FEL amplifier.   Results of simulations performed
with the code Genesis \cite{GENE}are shown in Fig. \ref{acceptance}.
The minimum of the FWHM power amplification bandwidth ($0.2$ mrad)
is achieved at the photon energy of $1$ keV.

\subsection{Resolution}

\begin{figure}
\begin{center}
\includegraphics[width=0.75\textwidth]{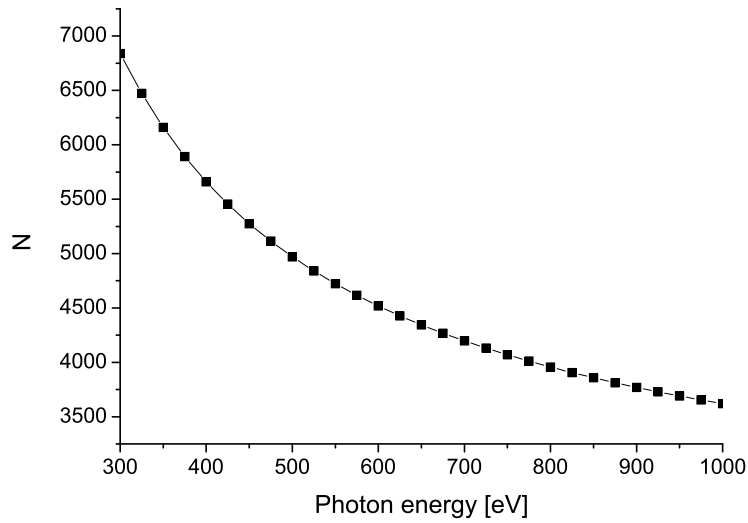}
\end{center}
\caption{The number of illuminated grooves (number of grooves per
waist of radiation beam illuminated) $N$ as a function of photon
energy.} \label{gr_illum}
\end{figure}

\begin{figure}
\begin{center}
\includegraphics[width=0.75\textwidth]{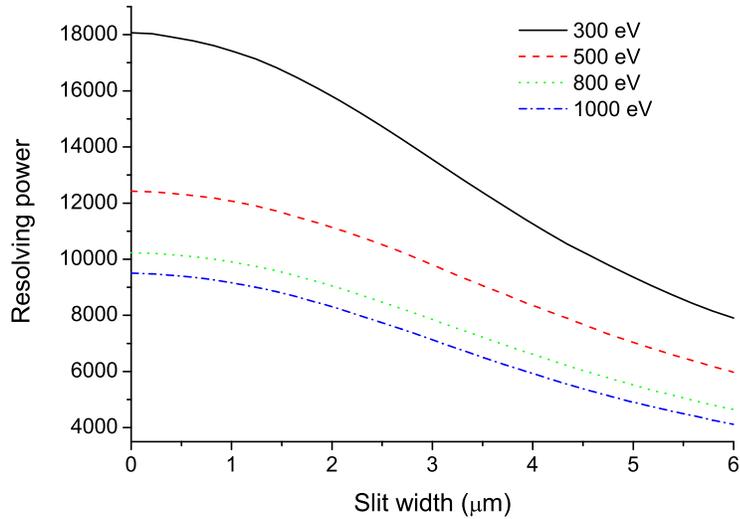}
\end{center}
\caption{Resolving power of the grating monochromator as a function
of the exit slit size for different photon energies.}
\label{Res_slit_vs_slitsize}
\end{figure}

\begin{figure}
\begin{center}
\includegraphics[width=0.75\textwidth]{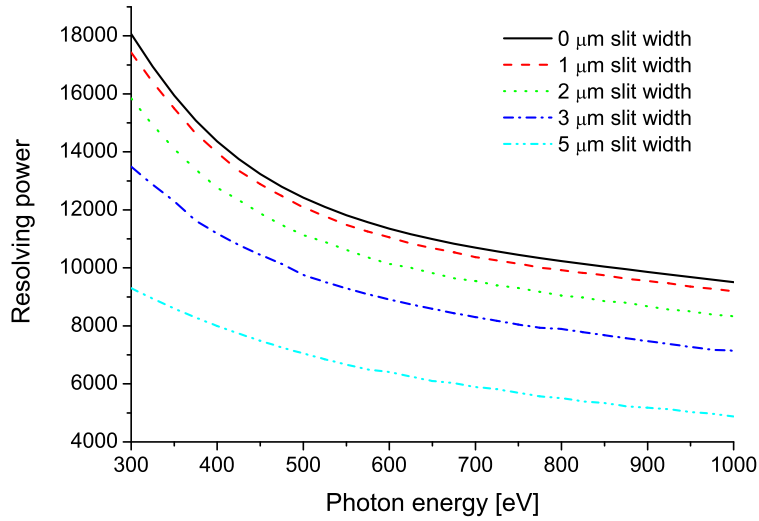}
\end{center}
\caption{Resolving power of the grating monochromator as a function
of the photon energy for different slit sizes.}
\label{Res_slit_vs_enrg}
\end{figure}

\begin{figure}
\begin{center}
\includegraphics[width=0.75\textwidth]{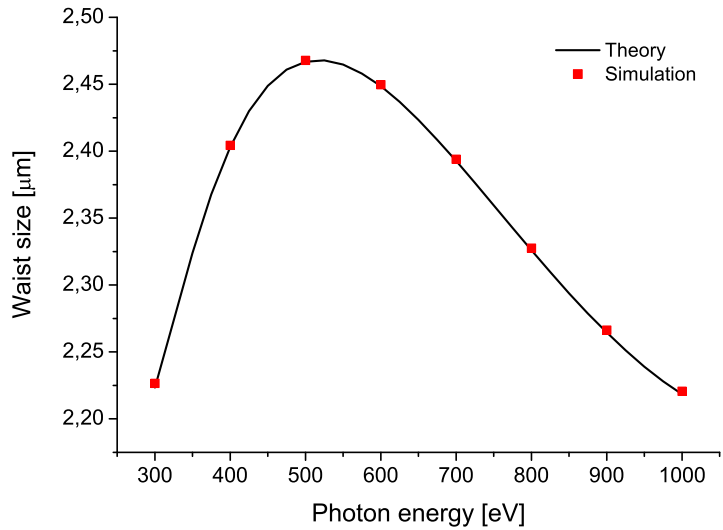}
\end{center}
\caption{Image waist size on the exit-slit plane as a function of
the photon energy. The curve are calculated with analytical
formulas. Squares are the result of numerical calculations with the
split-step beam propagation method.} \label{wsize_at_slit}
\end{figure}

\begin{figure}
\begin{center}
\includegraphics[width=0.75\textwidth]{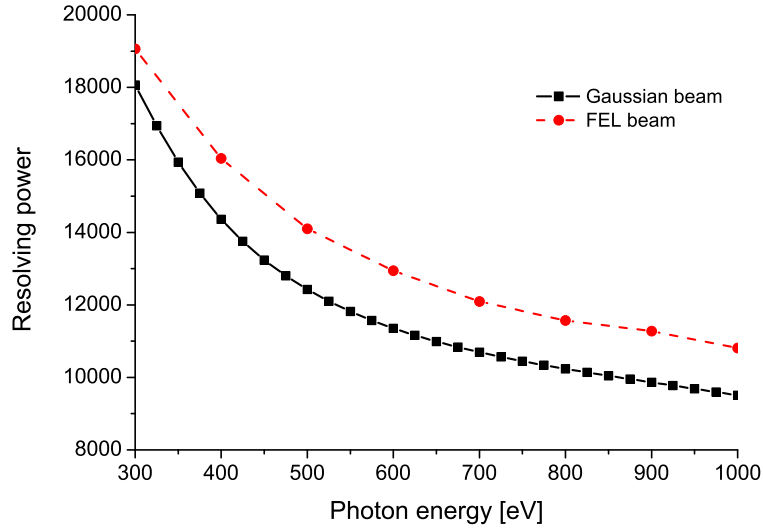}
\end{center}
\caption{Maximal resolving power, i.e. resolving power at closed
slits, as a function of photon energy. Results are obtained using
wave optics calculations. Squares are calculated using coherent
Gaussian beam, and circles are calculated using FEL beam.}
\label{real_resol}
\end{figure}

\begin{figure}
\begin{center}
\includegraphics[width=0.75\textwidth]{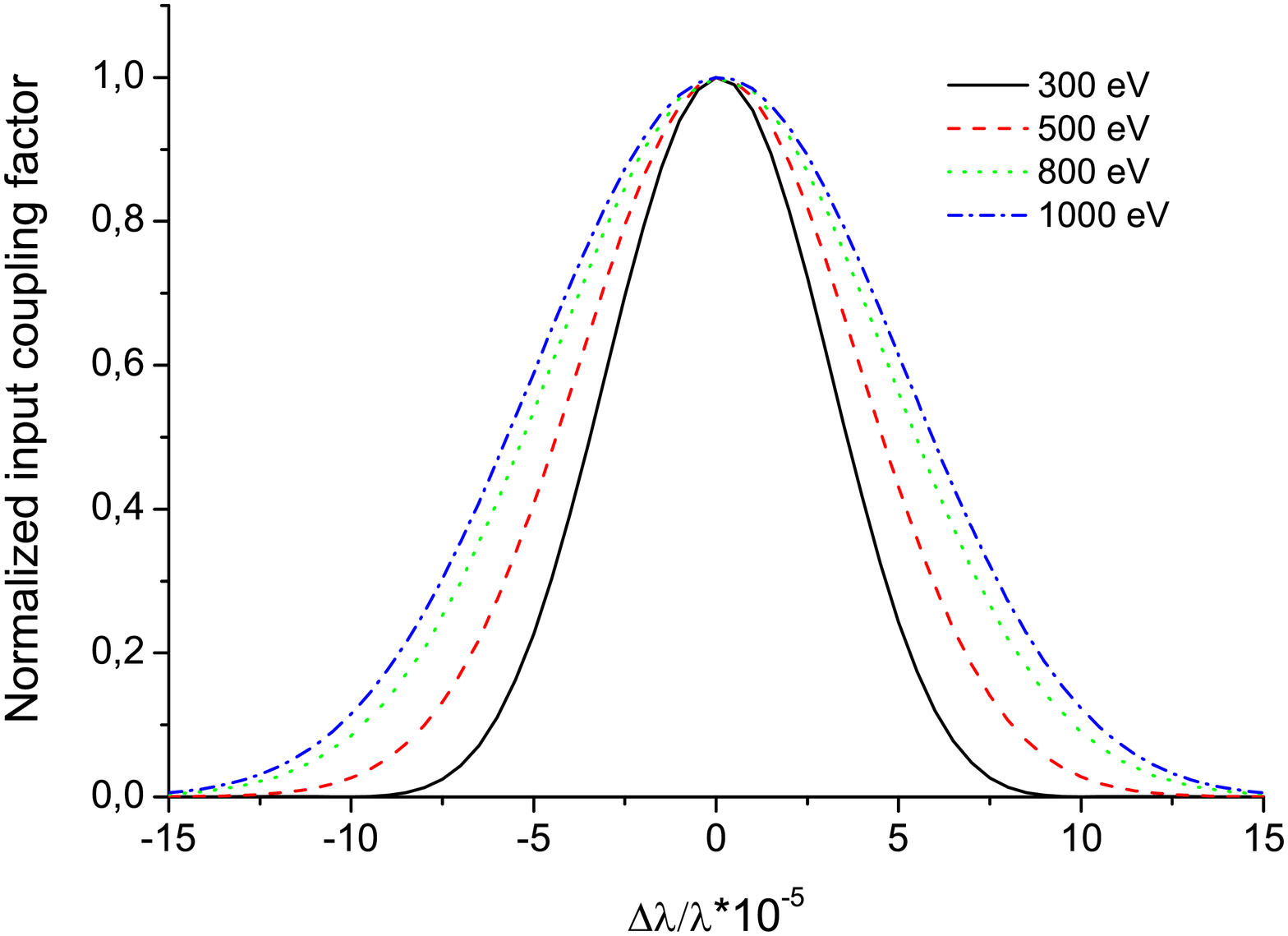}
\end{center}
\caption{Profile of the output spectral line from the grating
monochromator without exit slit at different photon energies.
Results are obtained by wave optics and FEL simulations. The FWHM of
the spectral line would indicates a resolving power of $7000$ in
photon energy range $0.3$ keV - $1$ keV.} \label{resolution_no_slit}
\end{figure}

A preliminary resolution study was first performed using Gaussian
optics calculations. Subsequently, in order to have a more realistic
wave optics simulation, after using Gaussian beam treatment, the
beam distribution was modeled using FEL simulations and accounting
for third order optical aberrations.Optimized specifications have
then been verified by ray-tracing simulations, accounting for all
geometrical aberrations, as reported in the end of this section. The
reason for first modeling the source as a Gaussian beam was to
obtain a completely analytical, albeit approximated description of
the self-seeding monochromator operation.

\subsubsection{Analytical  description}

Let us first assume that the incident FEL beam is characterized by a
Gaussian distribution. In this case, the ABCD matrix formalism is a
powerful tool to describe the propagation of the beam through an
arbitrary paraxial optical system.  The optical system for the
grating monochromator comprises grating, slit and mirrors spaced
apart from each other. All these optical elements (grating, mirrors
and free-space), with the exception of the slit, can be represented
with the help of ABCD matrices, which can be used to characterize
the width and the wavefront curvature of an optical Gaussian beam
after its propagation through a grating monochromator without exit
slit. Gaussian beam transformation due to mirrors, and translation
in between mirrors can be tracked using the law for the
transformation of $q$ in Eq. (\ref{qfinal}). It can be convenient to
describe the diffraction of a Gaussian beam from a toroidal VLS
grating using the ABCD matrix formalism too. The relevant geometry
is shown in Fig. \ref{gr_5}. The grating has a local groove spacing
$D(z) = D_0 + D_1 z$ at a position z on the grating surface, a
radius of curvature of the substrate $R$ in the tangential plane,
and $\rho$ in the sagittal plane. In the tangential plane, a
toroidal VLS grating can be represented by combination of a planar
grating with fixed line spacing and lens after the grating, Fig.
\ref{gr_7}, with a focal length equal to the focal length of the
toroidal VLS grating

\begin{eqnarray}
f_1 = \left[\frac{\lambda D_1}{D_0^2 \sin^2 \theta_d} + \frac{\sin
\theta_i}{2R \sin^2\theta_d} + \frac{1}{2R \sin\theta_d}\right]^{-1}
\label{f1tan}
\end{eqnarray}
In the sagittal plane the toroidal VLS grating can be represented by
a single lens with a focal length

\begin{eqnarray}
f_2 = \left[\frac{\sin \theta_i}{2\rho} + \frac{\sin
\theta_d}{2\rho} \right]^{-1} \label{f2sag}
\end{eqnarray}
In our analysis we calculate the propagation of the input signal to
different planes of interest within the monochromator. We start by
writing the input field in object plane, that is the source plane,
as

\begin{eqnarray}
\tilde{E}(x, y) = \exp\left[ - \frac{x^2}{w_0^2} -
\frac{y^2}{w_0^2}\right] ~. \label{Exyf}
\end{eqnarray}
As shown in Fig. \ref{gr_2}, the input beam is incident on the
grating at the angle $\theta_i$. The diffracted beam emerges at an
angle $\theta_d$, and is a function of the wavelength according to
grating equation. Assuming diffraction into $n=+1$ order, one has

\begin{eqnarray}
\lambda = D_0 (\cos \theta_i - \cos \theta_d) ~. \label{grat33}
\end{eqnarray}
By differentiating this equation one obtains

\begin{eqnarray}
\frac{d \theta_d}{d \lambda}  =  \frac{1}{D_0 \theta_d}~ ,
\label{diff33}
\end{eqnarray}
where we assume grazing incidence geometry, $\theta_i \ll 1$ and
$\theta_d \ll 1$. The physical meaning of this equation is that
different spectral components of the outcoming beam travel in
different directions. As said above, in the tangential plane the
toroidal VLS grating is represented as combination of plane grating
and convergent lens. We are interested  in determining the intensity
distribution in the image plane, i.e. at the slit position. The
grating introduces angular dispersion, which the lens transforms
into spatial dispersion in the slit plane. The spatial dispersion
parameter, which describes the proportionality between spatial
displacement and optical wavelength is given by

\begin{eqnarray}
\eta = \lambda \frac{d x}{d \lambda} =  \frac{d_2 \lambda}{D_0
\theta_d} ~, \label{spatdis}
\end{eqnarray}
where $d_2$ is the distance between grating and image plane. In our
case study, the relative difference between focal length and image
distance is about $1 \%$. As a result one may approximately write
$d_2 = f_1$. The spectral resolution of the monochromator equipped
with an exit slit depends on the spot size in the slit plane, is
related with the individual wavelengths composing the beam, and with
the rate of spatial dispersion with respect to the wavelength. For a
Gaussian input beam, the intensity distribution in the waist plane,
that is the slit plane, is given by $I = \exp( - 2x^2/w_s^2)$, where
$w_s$ is the waist size on the slit. A properly defined merit
function is indispensable for the design of a grating monochromator.
A merit function based on the spread of the radiation spots is a
suitable choice in our case of interest.  Let us consider the
limiting case of a slit with much narrower opening than the spot
size of the beam for a fixed individual wavelength centered at $x =
0$. In this case, the Gaussian instrumental function (i.e. the
spectral line profile of the beam after monochromatization) is given
by

\begin{eqnarray}
I = \exp\left[- 2 \left(\frac{f_1 \lambda}{w_s D_0
\theta_d}\right)^2 \left(\frac{\Delta
\lambda}{\lambda}\right)^2\right] ~. \label{I1}
\end{eqnarray}
The resolving power is often associated to the FWHM $\Delta \lambda$
of the instrumental function through the relation $R =
\lambda/(\Delta \lambda)$. In our case of interest the resolving
power is consequently given by

\begin{eqnarray}
R =   \frac{f_1 \lambda}{1.18 w_s D_0 \theta_d}~ . \label{R1}
\end{eqnarray}
The effect of a plane grating on the monochromatic beam is, as
previously discussed, twofold: first, the source size is scaled by
the asymmetry factor $b = \theta_i/\theta_d$ and, second, the
distance between grating and virtual source before the grating is
scaled by the square of the asymmetry factor. In our case, the waist
of the virtual source $w'_0$ and the distance $d'_1$ are thus given
by

\begin{eqnarray}
&&w'_0 = w_0 \theta_d/\theta_i  ~, \cr && d'_1 = d_1
(\theta_d/\theta_i)^2~ . \label{w'0}
\end{eqnarray}
After propagation through a distance $d_2$ behind the lens, the
Gaussian beam is said to be focused at the point where it has a
plane wavefront. Using Eq. (\ref{d2}), we obtain

\begin{eqnarray}
w_s = \frac{w'_0 f_1}{[(d'_1-f_1)^2 + z_R^{'2}]^{1/2}} ~, \label{ws}
\end{eqnarray}
where $z'_R  = z_R (\theta_d/\theta_i)^2$ is the Rayleigh range
associated with the virtual source. In our case of interest,
$f_1/d'_1 \sim 10^{-2}$, and the waist transforms as

\begin{eqnarray}
w_s = w_0 \frac{\theta_i}{\theta_d} \frac{f_1}{[d_1^2 +
z_R^2]^{1/2}}~ . \label{ws2}
\end{eqnarray}
Using this relation we can recast the expression for the resolving
power in the form

\begin{eqnarray}
R = \frac{\lambda z_R (1 + d_1^2/z_R^2)^{1/2}}{1.18 w_0 \theta_i
D_0} = \frac{\pi w_0 (1 + d_1^2/z_R^2)^{1/2}}{1.18 \theta_i D_0} ~,
\label{R333}
\end{eqnarray}
and with the help of Eq. (\ref{w2}), we finally obtain

\begin{eqnarray}
R = \frac{\pi w_g}{1.18 \theta_i D_0}~  , \label{f33fin}
\end{eqnarray}
where $w_g$ is the actual waist size of the Gaussian beam after
propagation through a distance $d_1$, i.e in the plane immediately
in front of the grating. In that plane beam has finite radius of
curvature and its intensity is given by

\begin{eqnarray}
I = I_0 \exp\left[-2 \frac{x^2}{w_g^2}\right] . \label{II0}
\end{eqnarray}
We now introduce the new parameter $N = w_g/(D_0\theta_i)$, which
may be identified as the number of illuminated grooves within the
projected beam-waist size $w_g/\theta_i$, and is related to the
resolving power by

\begin{eqnarray}
R = \frac{\pi N}{1.18}~ . \label{RN}
\end{eqnarray}
The number of illuminated grooves is plotted  against the photon
energy in Fig. \ref{gr_illum}. Influencing factors include the
variation of the source size, and the actual distance between source
and grating.

We now turn to consider the case with an arbitrary slit width.
Generally, the presence of the slit modifies the output spectrum,
and the instrumental function is essentially a convolution of the
diffraction-limited (Gaussian) instrumental function with the slit
transmission function. As in the case for a diffraction-limited
asymptotic, the resolving power is associated to the FWHM $\Delta
\lambda$ of the instrumental function through the relation $R =
\lambda/(\Delta \lambda)$. Fig. \ref{Res_slit_vs_slitsize} and Fig.
\ref{Res_slit_vs_enrg} illustrate the dependence of the resolving
power on slit width and photon energy. Note that in our particular
case study of self-seeding, the word "resolving power" presented on
Fig. \ref{Res_slit_vs_slitsize} and Fig. \ref{Res_slit_vs_enrg} is
to be understood in a narrow sense. Namely, as we will discuss
below, the electron beam, which interacts with the seed beam into
the second undulator, plays the role of the exit slit with some
effective width, and this additional spectral filtering is always
present. Here we are not to discuss about the overall modification
of the output spectrum, but only about how the presence of the slit
modifies the spectrum of the transmitted beam.

\subsubsection{Simulations using beam propagation method}

Above we analyzed the resolution of the grating monochromator using
an analytical method. Here we show simulation results using the beam
propagation method (BPM). We used a in-house developed MATLAB code
that calculates the propagation of the monoenergetic beam through
the monochromator. The accuracy of the beam propagation method could
be tested with analytical results for the Gaussian beam
approximation. We simulated the focusing of the Gaussian beam by a
toroidal VLS grating on the exit slit. Fig. \ref{wsize_at_slit}
shows the dependence of waist size as a function of photon energy.
From Fig. \ref{wsize_at_slit} it can be seen that there is a good
agreement between numerical and analytical results.

Most of the results presented in this article were obtained in the
framework of a Gaussian beam model. This is a very fruitful
approach, allowing one to study many features of the self-seeding
monochromator by means of relatively simple tools. However, it is
relevant to make some remarks on the applicability of the Gaussian
beam model. In practical situations the FEL beam has no Gaussian
distribution, and the question arises whether a Gaussian
approximation yields a correct design for a self-seeding
monochromator. We therefore performed the same analysis using BPM
simulations. With the help of the plots presented in Fig.
\ref{real_resol} one can give a quantitative answer to the question
of the accuracy of the Gaussian beam model. Numerical simulations
for the monochromator have been performed in the steady-state FEL
beam approximation using geometry parameters  (in particular, the
position of the slit) obtained from the Gaussian beam approximation.
One can see that the the characteristics of the monochromator
designed using a Gaussian beam approach do not differ significantly
from those based on a model exploiting steady-state FEL beam
distribution.

\subsubsection{Modeling the monochromator without slit}

When describing the operation of the self-seeding setup, we always
considered the exit slit as spectral filter. However, to some extent
this is a simplification since in reality, for sufficiently large
slit sizes, the filtering is automatically produced by the second
FEL amplifier. In fact, the angular dispersion of the grating causes
a separation of different optical frequencies at the entrance of the
second undulator. The spectral resolution without slit depends on
the radiation spot-size at the entrance of the second undulator
related with individual frequencies, and on the rate of the spatial
dispersion with respect to frequencies. The center frequency of the
passband filter is determined by the transverse position of the
electron beam. The resolving power is limited by the electron beam
transverse size, and can be high in the whole photon energy range
covered by the monochromator. This mode of operation has the an
advantage. In fact, it is important to maximize the transmission
through the monochromator in order to preserve both the beam power
and the transverse beam shape. It can easily be demonstrated that
such beam power loss and mismatching are minimized when the
monochromator operates without a slit.

\begin{figure}
\begin{center}
\includegraphics[width=0.75\textwidth]{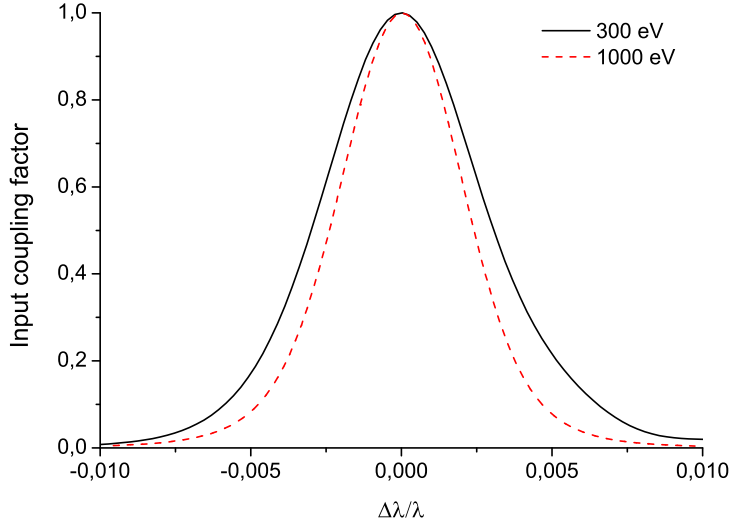}
\end{center}
\caption{Input coupling factor as a function of wavelength
detuning.} \label{icf_wave_detun}
\end{figure}
It is important to quantitatively analyze this filtering process.
The influence of the spatial dispersion at the entrance of the
second undulator on the operation of the self-seeding setup can be
quantified by studying the input coupling factor between seed beam
and FEL amplifier. In the linear regime, the input coupling factor
$A$ can be found independently for each  individual frequency, and
allows for a convenient measure of the influence of the seed-beam
displacement. In practice, it is sufficient to consider the limiting
case of an instrumental function bandwidth ($\sim 0.02 \%$) much
narrower than the FEL amplification bandwidth ($\sim 0.5 \%$). In
this case the resolution is defined by the response of the FEL
amplifier power on the seed displacement in the case of a
monochromatic beam transmitted through the monochromator without
slit. A spatial dispersion parameter, which describes the
proportionality between spatial displacement and frequency at the
entrance of the second undulator, can be found by monochromator
simulations using our BPM code. The instrumental functions of the
self-seeding setup without slit for different photon energies are
presented in Fig. \ref{resolution_no_slit}. In order to calculate
the tolerance on the frequency detuning of the seed beam, it is
necessary to have knowledge of the frequency response of the FEL
amplifier. Results of simulations are shown in Fig.
\ref{icf_wave_detun}.

\subsubsection{\label{Methabe} Method for computing third order aberrations for a toroidal grating}

\begin{figure}
\begin{center}
\includegraphics[clip,width=0.75\textwidth]{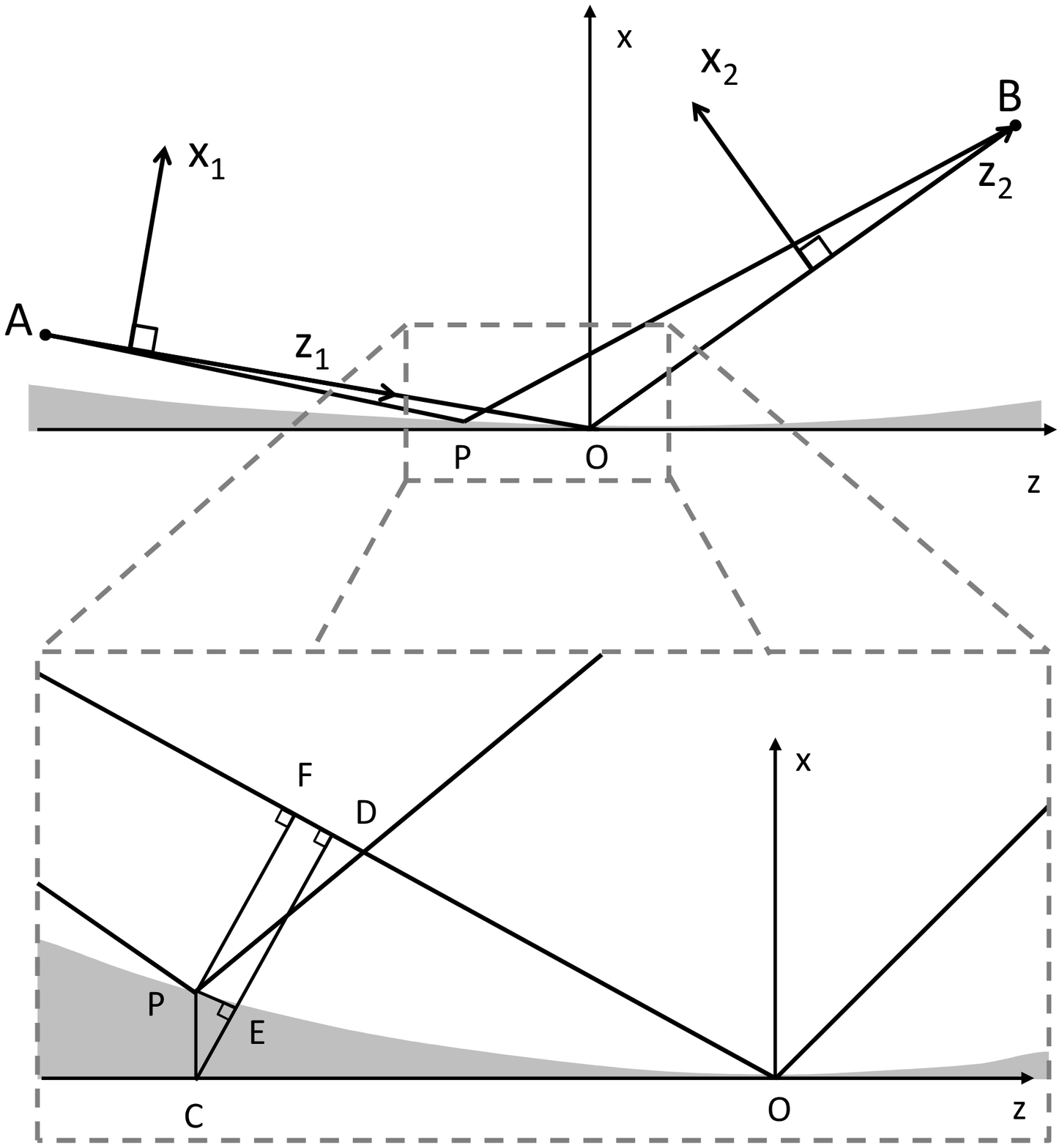} 
\end{center}
\caption{Optical scheme and coordinate systems for a toroidal
grating system. The lower sketch is an enlarged fraction of the
upper one.} \label{gr_9}
\end{figure}

\begin{figure}
\begin{center}
\includegraphics[width=0.75\textwidth]{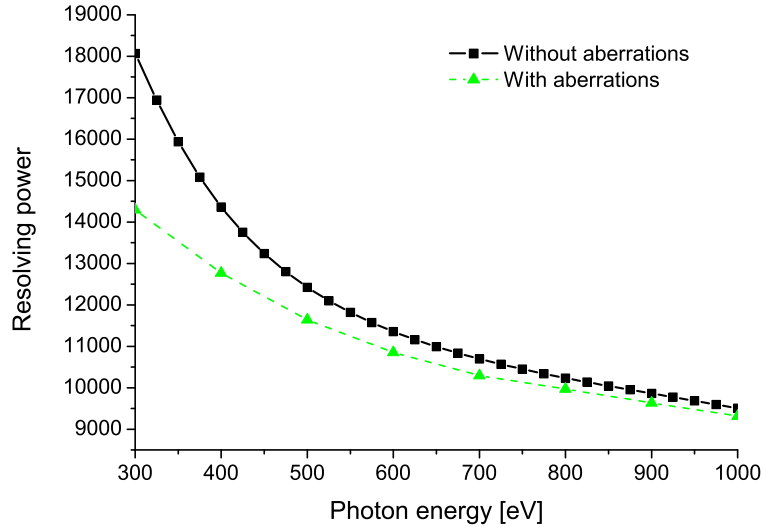} 
\end{center}
\caption{Resolving power of the grating monochromator at closed slit
as a function of the photon energy. Results are obtained using wave
optics calculations. Squares are calculated for an optical system
without aberrations, and triangles are results for an aberrated
optical system.} \label{abberation_resolution_slit}
\end{figure}
\begin{figure}
\begin{center}
\includegraphics[width=0.75\textwidth]{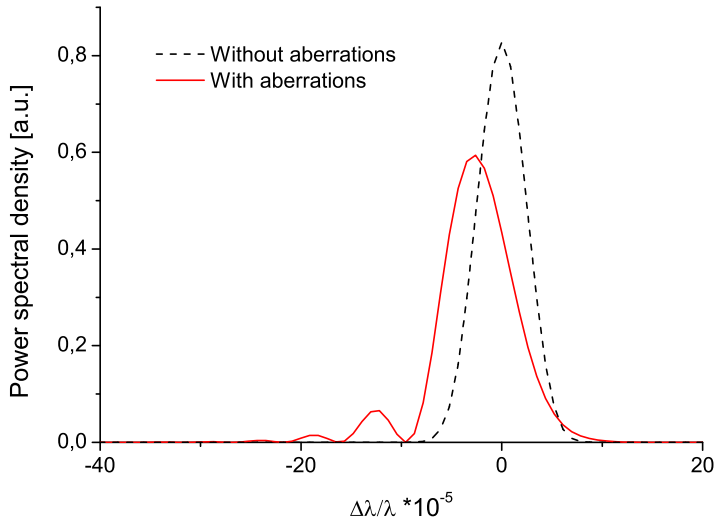} 
\end{center}
\caption{Effect of aberrations on the monochromator performance.
Results are obtained using wave optics calculations. Simulations of
the instrumental function at closed exit slit for aberrated and
non-aberrated optical system are compared. Here $\hbar\omega=300$
eV.} \label{abberation_on_slit}
\end{figure}

In this paragraph we study the theory of third-order aberrations
theory for a toroidal grating. It is first necessary to clearly
define the reference coordinate systems used to describe the optical
system. Fig. \ref{gr_9} shows the toroidal grating and the object
point $A$. The three coordinate systems $(x, y, z)$, $(x_1, y_1,
z_1)$, and $(x_2, y_2, z_2)$ are used to describe the position of
the wave on the optical surface, the incoming wavefront and the
diffracted wavefront, respectively. The ray AOB is referred to as
the principal ray. In the following, the wavefront aberrations, the
positions of the object and image plane are specified with respect
to this ray. The wavefront aberration $W$ for a spherical wave
passing through a point $P$ in the system is defined as the path
difference between the principal and auxiliary ray: $W =
\overline{APB} - \overline{AOB}$. Here we take advantage of the
paraxial approximation obtained by ignoring all terms but the first
quadratic terms in  $x_1$ and $y_1$. Let us assume that point $A$ is
in the tangential plane and $P(x_1, 0, z_1)$ is any point on the
grating surface satisfying the constraint $y_1 = 0$. The equation
for the path $\overline{AP}$ is

\begin{eqnarray}
\overline{AP}=  z_1 + \frac{x_1^2}{2 z_1}  =  \overline{AF} +
\frac{\overline{PF}^2}{2 \overline{AF}}~ , \label{APpath}
\end{eqnarray}
where  $\overline{AF} = \overline{AO} - \overline{OD} -
\overline{FD}$,  $\overline{OD} = |z| \cos \alpha$, $\overline{FD} =
x \cos \alpha$, $\overline{PF} = \overline{CD} - \overline{CE} = |z|
\cos \alpha - x \sin \alpha$. Here $(x, y, z)$ are the coordinates
of the point $P$ in the grating coordinate system and the coordinate
system is chosen in such a way that $x>0$. Neglecting all terms of
order higher than the second in $x$ and $y$, the form of a toroidal
surface can be expressed by the equation

\begin{eqnarray}
x = \frac{z^2}{2 R} + \frac{y^2}{2 \rho}~ , \label{torusb}
\end{eqnarray}
where $R$ and $\rho$ are tangential and sagittal radius of
curvature. Thus, the distance $PC$ is given by $x = z^2/(2 R)$.
Finally, we have

\begin{eqnarray}
\overline{AP} - \overline{AO} = z \sin \alpha - \frac{z^2}{2R} \cos
\alpha + \frac{\left[z \cos \alpha + z^2 \sin \alpha/(2
R)\right]^2}{2 \left[r_1 + z \sin \alpha - z^2 \cos \alpha/(2
R)\right]} ~, \label{APAO}
\end{eqnarray}
where we used the notation $\overline{AO} = r_1$. Expanding this
last difference as a power series in $z$ including the third order
yields

\begin{eqnarray}
&&\overline{AP} - r_1= \cr && z \sin \alpha + \frac{z^2}{2 r_1}
\cos^2 \alpha - \frac{z^2}{2R} \cos \alpha           - \frac{z^3}{2
r_1^2} \sin\alpha \cos^2 \alpha + \frac{z^3}{2 R r_1} \cos \alpha
\sin \alpha~ . \cr && \label{r1AP}
\end{eqnarray}
Extension to the case of nonzero sagittal coordinate can be derived
in the same way as above and results into an additional term $\Delta
T$

\begin{eqnarray}
\Delta T = \frac{y_1^2}{2 z_1} = \frac{y^2}{2 r_1}  +
\frac{y^2}{2\rho} \cos \alpha- \frac{y^2 z}{2 r_1^2} \sin \alpha~.
\label{DELT}
\end{eqnarray}
Note that the difference $\overline{BP} - \overline{BO}$ can be
obtained following the same procedure described above,  simply
replacing the incidence angle $\alpha$ with the diffraction angle
$\beta$. From Eq. (\ref{r1AP}), Eq. (\ref{DELT}), Eq.
(\ref{phitor}), and Eq. (\ref{Cijtorus}) one obtains that the power
series of $z$ and $y$ are identical.

In the previous sections, we studied the monochromator performance
using a beam propagation method. We performed simulations in the
framework of a simple model. Toroidal VLS grating was represented by
a combination of a planar grating with a fixed line spacing and a
lens after the grating. A BPM code was used to describe the
propagation of a beam with an arbitrary initial field distribution
through a paraxial system which was a combination of free-space,
lens and plane grating with fixed line spacing. The problem to be
solved now, is how to account for third order aberrations in the
frame of the BPM code. In the case of a point source this problem
has simple solution. Propagation from point source for a distance
$r_1$ from the grating, reflection from the grating, and subsequent
propagation to the image plane along a distance $r_2$ from the
grating becomes similar to propagation through a plane grating with
fixed line spacing, an ideal thin lens and an additional
transparency at the lens position, which changes the phase of the
reflected beam according to

\begin{eqnarray}
\delta \phi =k_0 C_{30} z^3 + k_0 C_{12} z y ^2 = k_0 C_{30}
\left(\frac{x_2}{\theta_d}\right)^3 + k_0 C_{12}
\left(\frac{x_2}{\theta_d}\right) y_2^2~ , \label{delphitre}
\end{eqnarray}

where $x_2$ and $y_2$ are the coordinates of the wavefront
immediately behind the grating.  Generally, in order to obtain the
output field distribution at a distance $z$ away from the input in
the paraxial approximation, we need to convolve the input field
distribution with the spatial impulse response

\begin{eqnarray}
h(x- x', y-y', z) \sim \exp\left[- ik_0\frac{(x-x')^2 +
(y-y')^2}{2z}\right] ~, \label{conh2}
\end{eqnarray}
and the above-described method to account for third order
aberrations is not applicable. The case with finite source size is
more complicated, and should be studied separately. However, in the
far-zone approximation, which is our case of practical interest, the
ratio between  source size and  beam size in the plane immediately
in front of the grating is relatively small. Also, the image size is
much smaller than the beam size in the plane immediately behind the
grating. As a result, the algorithm described above is applicable.
In this way, third order aberrations can be included into our study.

Fig. \ref{abberation_resolution_slit} shows a plot of the resolving
power derived from BPM code accounting for aberrations. Clearly, the
performance is limited by aberrations only at low photon energies
close to $300$ eV. Fig. \ref{abberation_on_slit} illustrates the
influence of aberrations on the lineshape at the photon energy of
$300$ eV. The profile is highly asymmetric, owing to aberrations
dominated by the primary coma.

\subsection{Beamline efficiency}

\begin{figure}
\begin{center}
\includegraphics[width=0.75\textwidth]{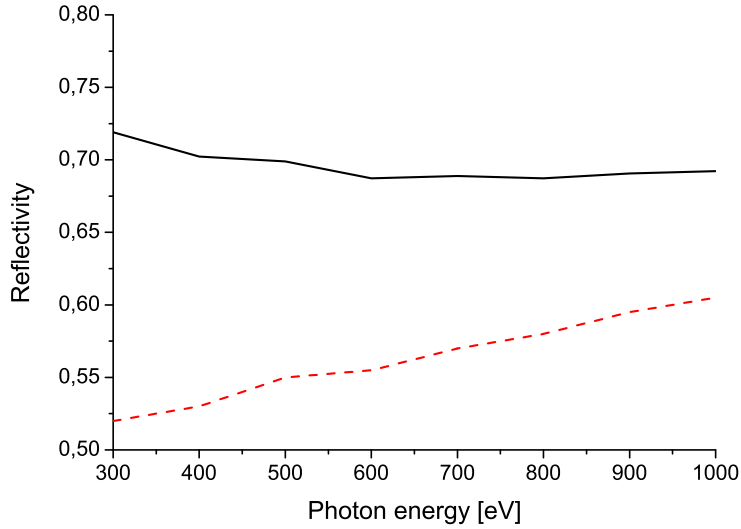}
\end{center}
\caption{Reflectivity of the post-grating optical components of the
beamline. The solid line shows the combined effect of the last two
fixed-angle mirrors. The dashed line represents the reflectivity of
the rotation mirror $M1$.} \label{gr_m_refl}
\end{figure}

\begin{figure}
\begin{center}
\includegraphics[width=0.75\textwidth]{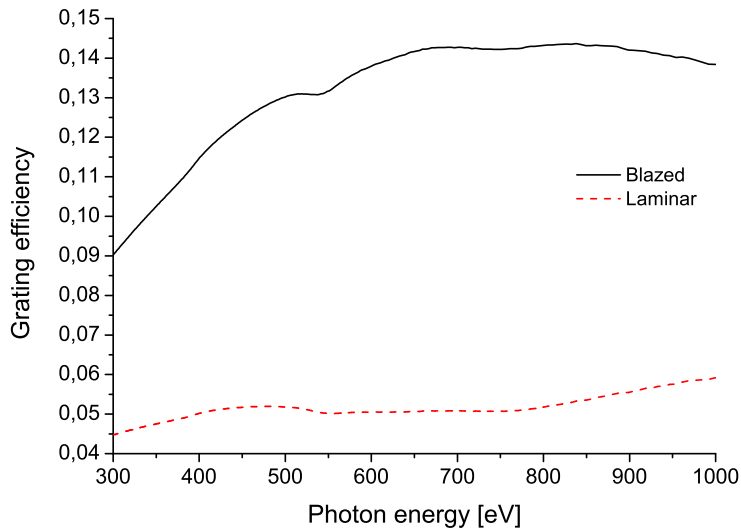}
\end{center}
\caption{First order efficiency for two different groove profiles.
In both cases, the groove density is $1100$ lines/mm, Pt coating is
assumed. The incidence angle is $1^\circ$. Lamellar grating
(rectangular): $11$ nm groove depth, 50\% duty cycle. Blazed
grating: $1,2^\circ$ blaze angle, $90^\circ$ anti-blaze angle.}
\label{bl_vs_lam}
\end{figure}
\begin{figure}
\begin{center}
\includegraphics[width=0.75\textwidth]{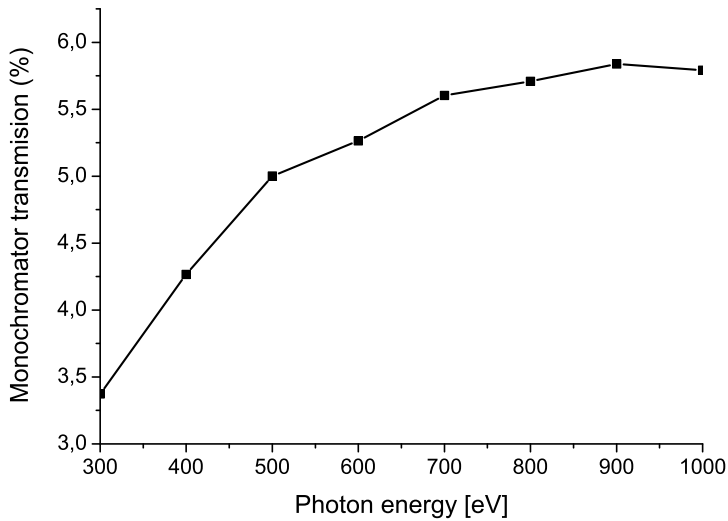}
\end{center}
\caption{Overall efficiency of the monochromator beamline without
exit slit as a function of the photon energy. The grating and the
three mirrors are platinum-coated.} \label{transmition}
\end{figure}

It is important to calculate the expected total reflectivity of the
monochromator beamline. The reflectivity of the mirrors was
calculated using the code CXRO \cite{CXRO}.  Mirrors are assumed to
be platinum-coated. The reflectivity of post grating optical
components are shown in Fig. \ref{gr_m_refl} as a function of the
photon energy. The combining effect of two fixed angle mirrors
cannot be neglected. In the soft X-ray range, platinum has
reflectivity of about $92 \%$ at $0.86$ degree grazing angle. The
compound loss over two last reflections is thus appreciable. The
most significant single factor in the post grating efficiency is, as
expected, the low reflectivity of the rotating plane mirror $M1$.
This is because the first post-grating mirror operates at a
relatively large incident angle of about $2$ degrees. The grating
efficiency was calculated using the code GSolver 5.2 \cite{GSOL}.
For comparison, Fig. \ref{bl_vs_lam} shows the first order
efficiency for two typically-used grating profiles, blazed and
laminar. Both gratings have a groove density of $1120$ lines/mm, are
considered to operate at an incidence angle of $1^\circ$, and have
had their geometry optimized for maximum efficiency in the first
order. Gratings are also assumed to be platinum-coated. The blazed
grating was optimized by adjusting the blaze angle and the laminar
grating by adjusting the groove depth. A laminar profile is widely
used due to its good suppression of the second and higher
diffraction orders. A blazed profile is preferable from the point of
view of absolute efficiency. Since the necessity of high seed power
at the entrance of the second undulator and harmonic contributions
are not an issue, the blazed profile has been chosen. The total
reflectivity of the beamline with blazed grating is shown in Fig.
\ref{transmition}.  This reflectivity refers to the
$\pi$-polarization component. The beamline reflectivity for the
$\sigma$-polarization component is not significantly different.

\subsection{Energy tuning and optical delay}

\begin{figure}[!h]
\begin{center}
\includegraphics[width=0.75\textwidth]{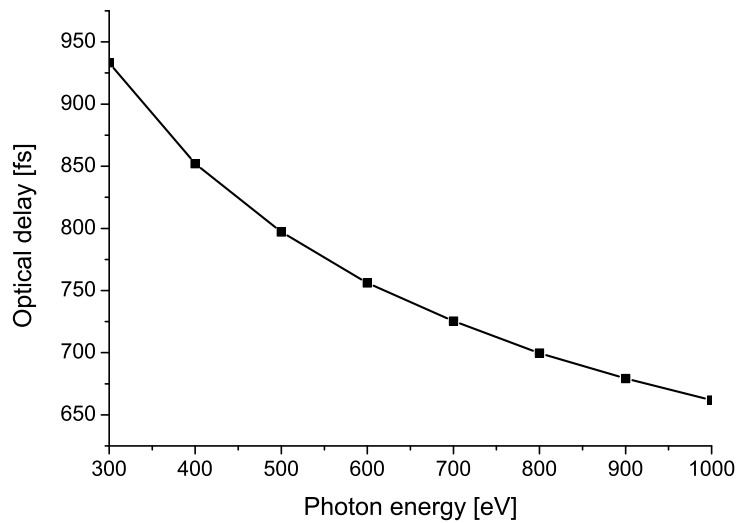}
\end{center}
\caption{Optical delay caused by the use of monochromator as a
function of photon energy.} \label{opt_delay}
\end{figure}
$
\\
\\
$
\begin{figure}[!h]
\begin{center}
\includegraphics[clip,width=0.75\textwidth]{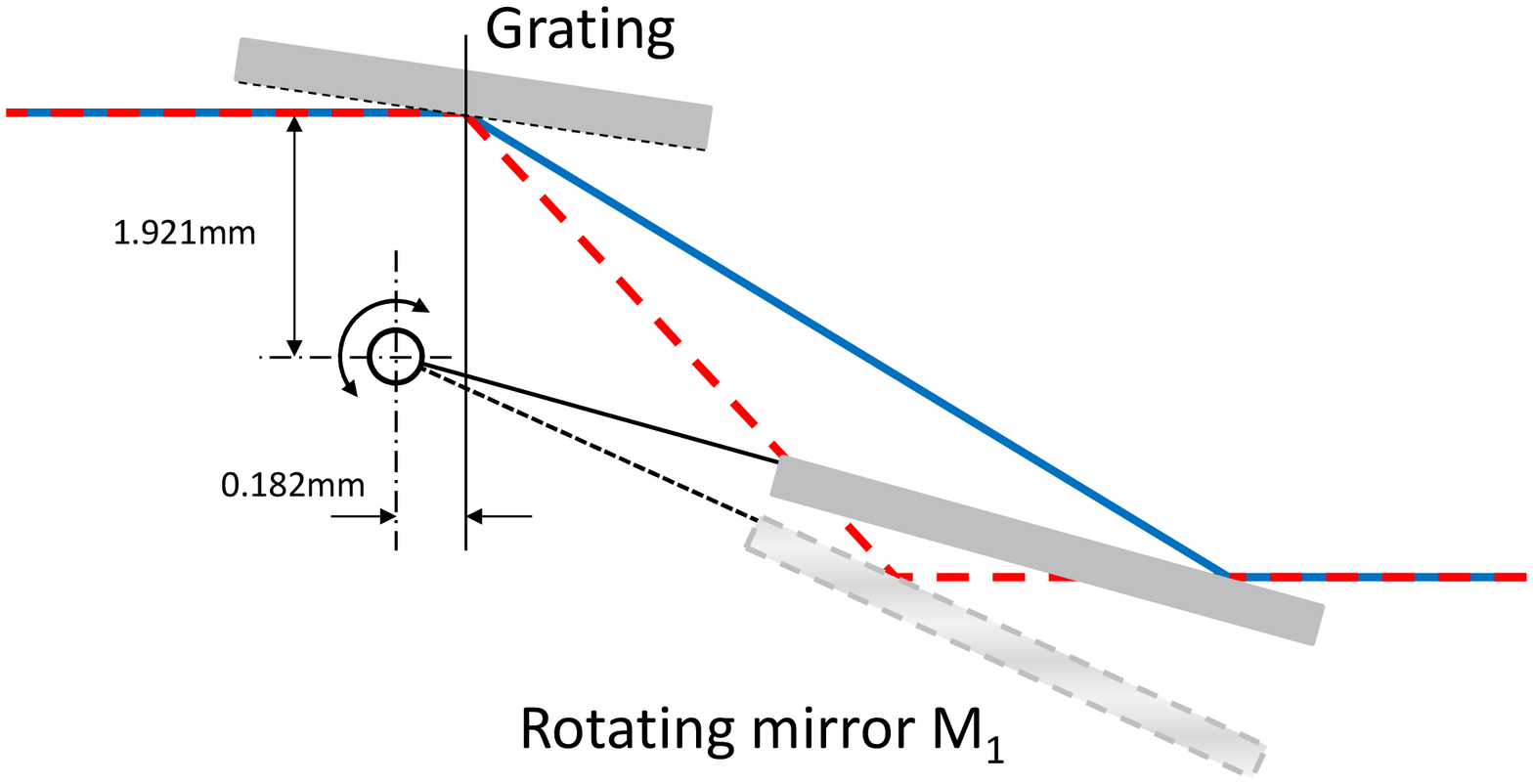}  
\end{center}
\caption{Principle of photon energy tuning. A plane scanning mirror,
$M_1$, is rotated to maintain a fixed exit beam direction and focal
spot at the exit slit. The mirror $M_1$ and grating are
schematically shown for two photon energies: 1 keV (solid) and 0.5
keV (dashed line).} \label{M1_rot}
\end{figure}

In order to maintain a constant direction of the exit beam, a
scanning post grating mirror is placed in the diffracted beam, and
rotated to direct the beam towards the exit slit. Thus, with a fixed
grating and exit slit one can maintain a good focus over a wide
photon energy range by simply translating and rotating a plane
mirror to aim the diffracted light at the exit slit. Translating the
mirror during rotation scanning can be achieved by pivoting the
mirror at a point above the center of the mirror. During the energy
tuning the beam walks along the surface of $M1$, as shown in Fig.
\ref{M1_rot}. The optical delay caused by the use of the grating
monochromator is about $0.7$ ps and its energy-dependence is shown
in Fig. \ref{opt_delay}. The delay is not constant, but varies with
the energy due to the fact that X-rays reflect off the post-grating
mirror $M1$ at different points, and take different optical paths as
the energy is tuned. The image on the slit plane is also found to
vary by $1~\mu$m in the dispersion direction, amounting to a change
of $1~\mu$rad in the angle of incidence, which is small compared to
the divergence of the beam at the slit plane. As such, the impact of
this effect on the monochromator performance is negligible.

\subsection{Effects of mirror surface errors}

\begin{figure}
\begin{center}
\includegraphics[clip,width=0.75\textwidth]{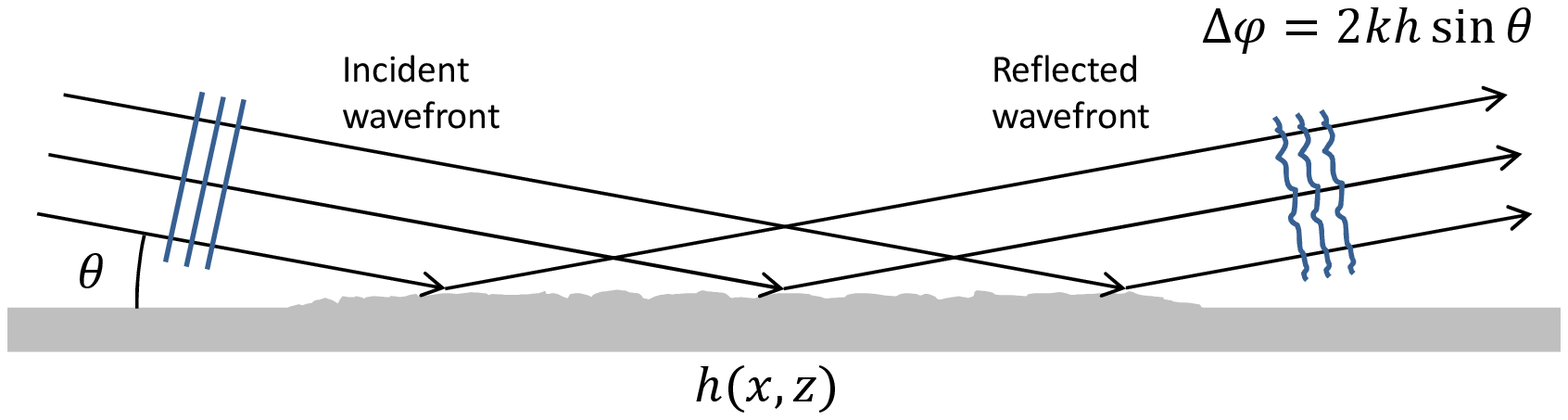} 
\end{center}
\caption{Thin-shifter-like behavior of surface roughness for small
mean square of surface displacement, adapted from \cite{BART}.}
\label{roughness_scheme}
\end{figure}
\begin{figure}
\begin{center}
\includegraphics[clip,width=0.75\textwidth]{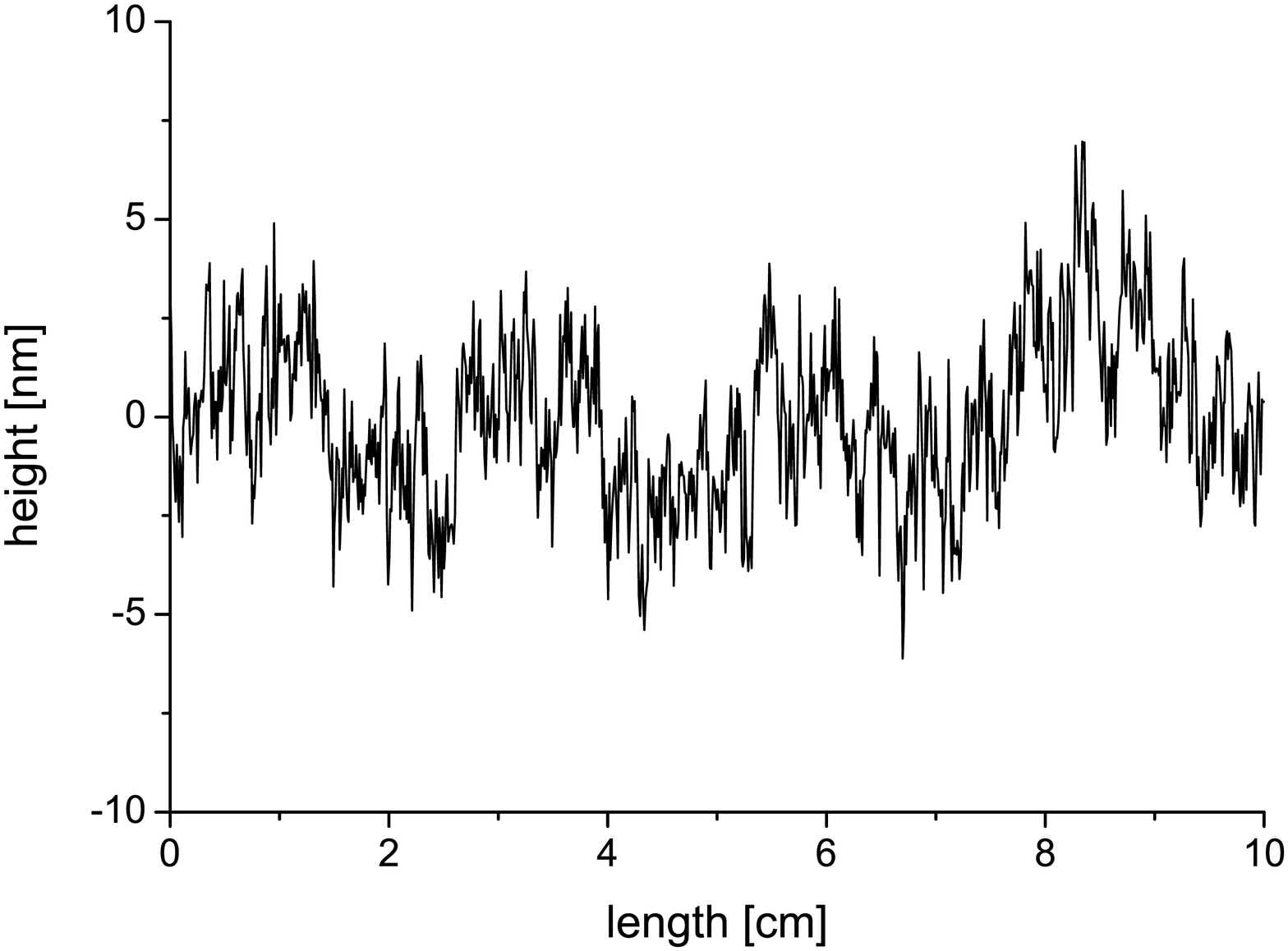} 
\includegraphics[clip,width=0.75\textwidth]{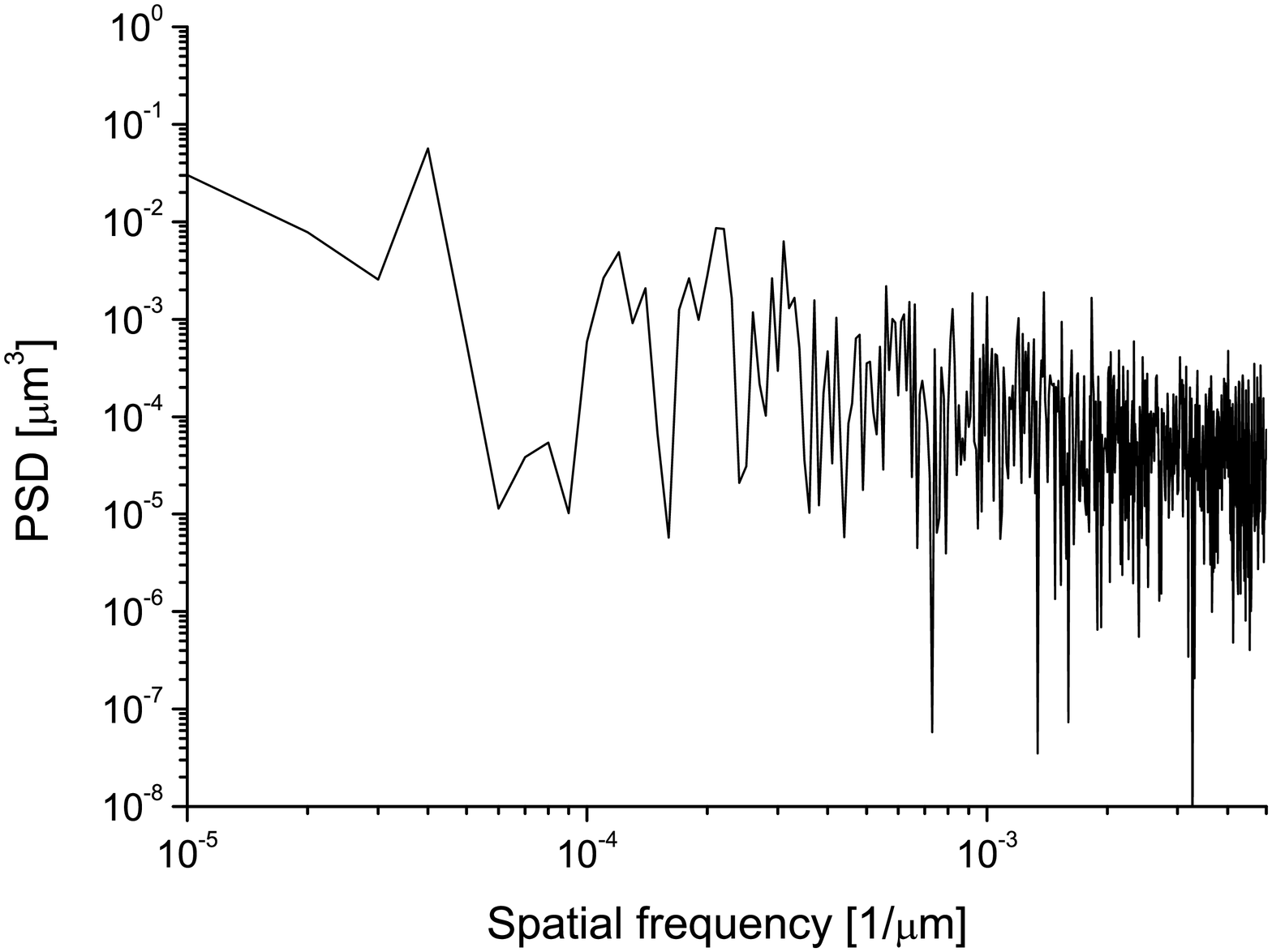} 
\end{center}
\caption{Distribution of residual height error and one-dimensional
power spectral density (1D PSD)  for the mirrors. The upper graph
shows the height error profile for a $10$ cm-long plane mirror M1.
The lower graph is the 1D PSD corresponding to the profile.}
\label{roughness_profile}
\end{figure}
\begin{figure}
\begin{center}
\includegraphics[clip,width=0.75\textwidth]{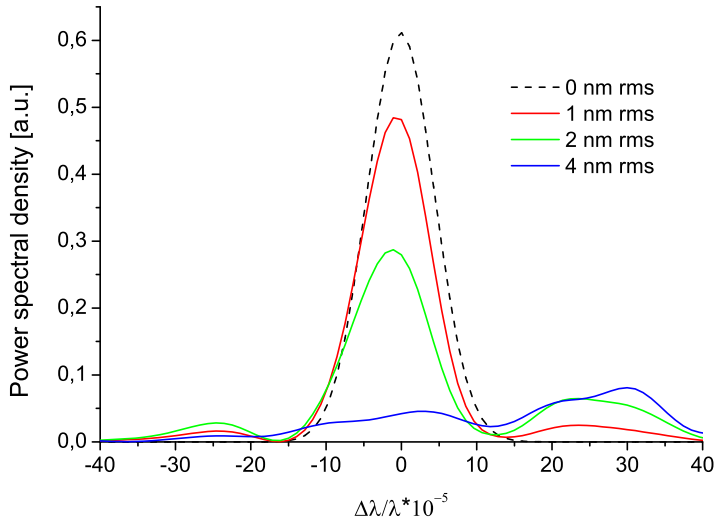} 
\end{center}
\caption{Effect of surface roughness on the monochromator
performance. Simulations of the instrumental function at closed slit
for different root-mean-square of surface displacements are shown in
the figure. Here $\hbar\omega=1$ keV.} \label{roughness_slit}
\end{figure}
\begin{figure}
\begin{center}
\includegraphics[clip,width=0.75\textwidth]{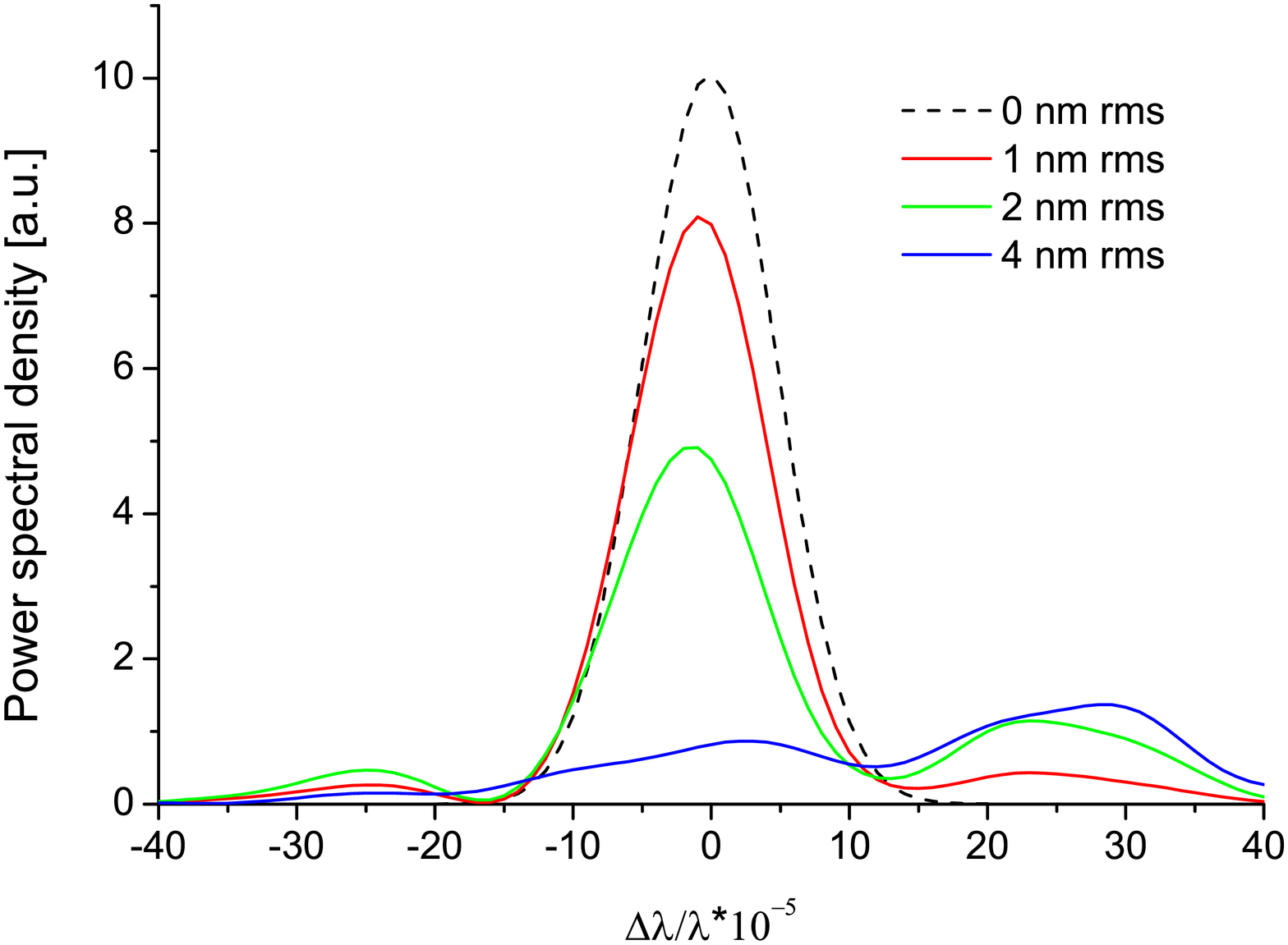} 
\end{center}
\caption{Effect of surface roughness on the monochromator
performance. Simulations of the instrumental function without exit
slit for different root-mean-square of surface displacements. Here
$\hbar\omega=1$ keV.} \label{roughness_noslit}
\end{figure}

A very important issue is the preservation of the radiation
wavefront from the source to the entrance of the second undulator.
Estimates of the requirements on the mirror for grating
monochromators are usually based on ray-tracing codes for incoherent
light sources. Since the XFEL beam will be almost transform-limited,
one needs to perform simulations of the effect of the mirror
imperfections by wavefront propagation codes. It is easy to
demonstrate that an error $\delta h$ on the optical surface will
perturb the wavefront of a phase $\phi$, according to

\begin{eqnarray}
\phi = \frac{4 \pi \delta h}{\lambda} \sin \theta_i~, \label{phidh}
\end{eqnarray}
where $\theta _i$ is the angle of incidence with respect to the
surface. In the case of a grating, the phase shift can be expressed
in terms of incidence and diffracted angles:

\begin{eqnarray}
\phi = 2\pi (\sin \theta_i +  \sin \theta_d) \frac{\delta
h}{\lambda}~.\label{gratphidh}
\end{eqnarray}
In practice, $\phi$ represents the deformation of the wavefront in
the propagation direction divided by the wavelength.

A reflection from the mirror becomes similar to the propagation
through a transparency at the mirror position, which just changes
the phase of the reflected beam without changing its amplitude,
\cite{BART} (see Fig. \ref{roughness_scheme}). For the shifter model
to be applicable, the phase change must be small, i.e. $|\phi| <<
1$. Optical elements were modeled as a phase shifters, and the
problem of simulating a monochromator was reduced to the proper
description of the phase shifters and of the propagation of the
wavefront in vacuum between the phase shifters. The main wavefront
distortion at the slit position and at the entrance of the second
undulator originates from the grating and the plane mirror $M1$.
Applying the Marechal criterion, i.e. requiring a Strehl ratio
larger than $0.8$, and treating the errors from the different optics
independently, we obtain the following condition for the rms height
error $h_\mathrm{rms}$ \cite{HEIM}:

\begin{eqnarray}
2 h_\mathrm{rms} \theta_i \sqrt{N} < \lambda/14 ~, \label{Mare}
\end{eqnarray}
where $\theta_i$ is the grazing angle of incidence and $N$ is the
number of optical elements. The most tight requirements  corresponds
to shortest wavelength. The grating operates at a fixed incidence
angle $\theta_i = 1$ degree, and at $1$ keV photon energy, the
diffraction angle is about $3.2$ degrees. This corresponds to an
incidence angle $(\theta_i+ \theta_d)/2 = 2.2$ degree for the mirror
$M1$. From Marechal criterion we conclude that a height error
$h_\mathrm{rms} = 1$ nm should be sufficient for diffraction-limited
monochromatization at the photon energy of $1$ keV. This is a very
tight requirement. State-of-the art manufacturing achieves routinely
rms values of $2$ nm for $10$ cm-long mirrors and one needs to
perform detailed simulations of the surface error effect for
understanding the requirements on the roughness.

The surface errors were generated from power spectral density (PSD)
functions described in mirrors specifications. The part of the PSD,
which makes the most significant contribution to the overall rms
height error, is the low spatial-frequency part. An example of
profile and PSD of mirror surface errors is shown in Fig.
\ref{roughness_profile}. Due to the very small incident angle, the
beam footprint is much larger in the tangential direction than in
the sagittal direction. The lowest spatial frequency that
contributes is in order of $\theta_i/w \sim 1/$cm in the tangential
direction. Here $w$ is the beam size at the optical element. It
follows that the grating and the mirrors will disturb the wavefront
mainly in the tangential direction. Simulations were performed using
a BPM code. The surface figure can be directly mapped onto the
optical field coordinate system using the geometrical transformation
described above. Examples of the simulated focus at the exit slit
for $1$ nm, $2$ nm and $4$ nm rms quality optics are presented in
Fig. \ref{roughness_slit}. Non Gaussian tails are seen on the sides
of the instrumental function. The influence of surface errors on the
resolution in the case when the slit is absent is shown in Fig.
\ref{roughness_noslit}. An rms roughness of order of $2$ nm seems at
present to be acceptable for the self-seeding setup.

\subsection{Ray-tracing results}

\begin{figure}
\begin{center}
\includegraphics[width=0.75\textwidth]{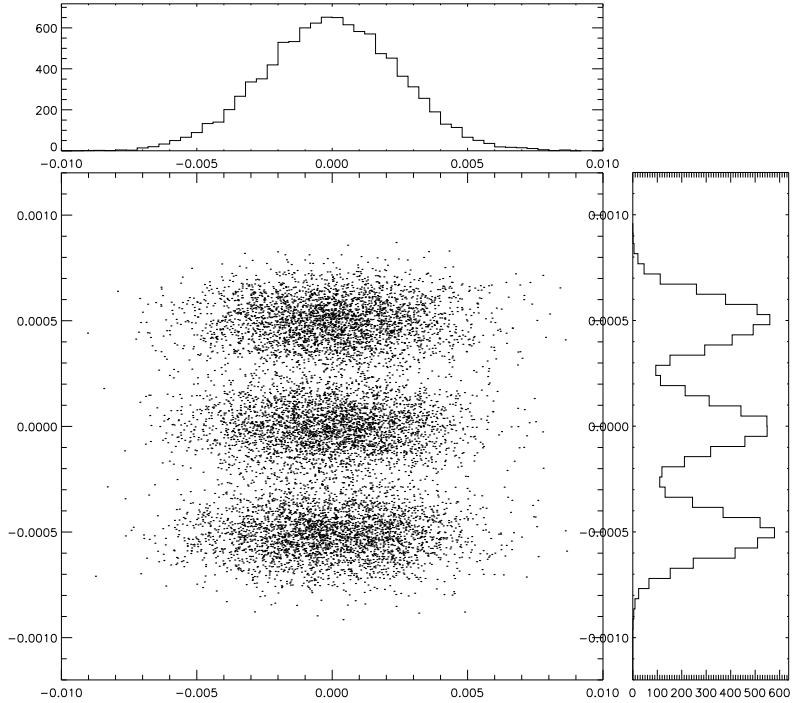}
\end{center}
\caption{Results from ray tracing simulations at the plane of the
exit slit for three photon energies $999.8$ eV, $1000.0$ eV and
$1000.2$ eV, obtained from the ray-tracing program SHADOW. The
histograms show the number of rays as a function of $x$ and $y$
coordinates. From the separation of the photon energies, a resolving
power of $5000$ would be expected.} \label{ray_1000_slitl}
\end{figure}
\begin{figure}
\begin{center}
\includegraphics[width=0.75\textwidth]{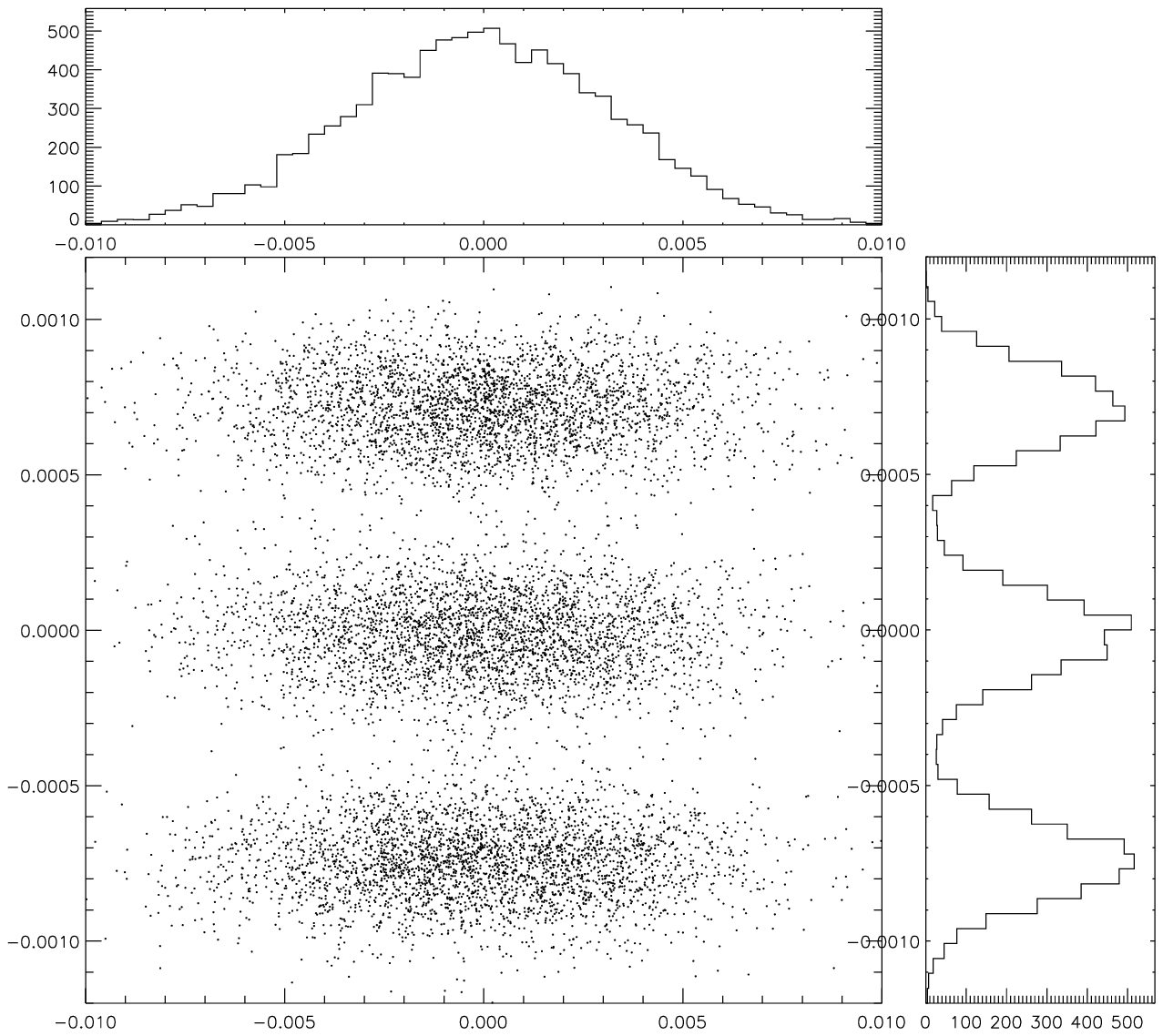}
\end{center}
\caption{Results from ray tracing simulations at the plane of the
exit slit for three photon energies $499.9$ eV, $500.0$ eV and
$500.1$ eV, obtained from the ray-tracing program SHADOW. The
histograms show the number of rays as a function of $x$ and $y$
coordinates. From the separation of the photon energies, a resolving
power of $5000$ would be expected.} \label{ray_500_slitl}
\end{figure}
\begin{figure}
\begin{center}
\includegraphics[width=0.75\textwidth]{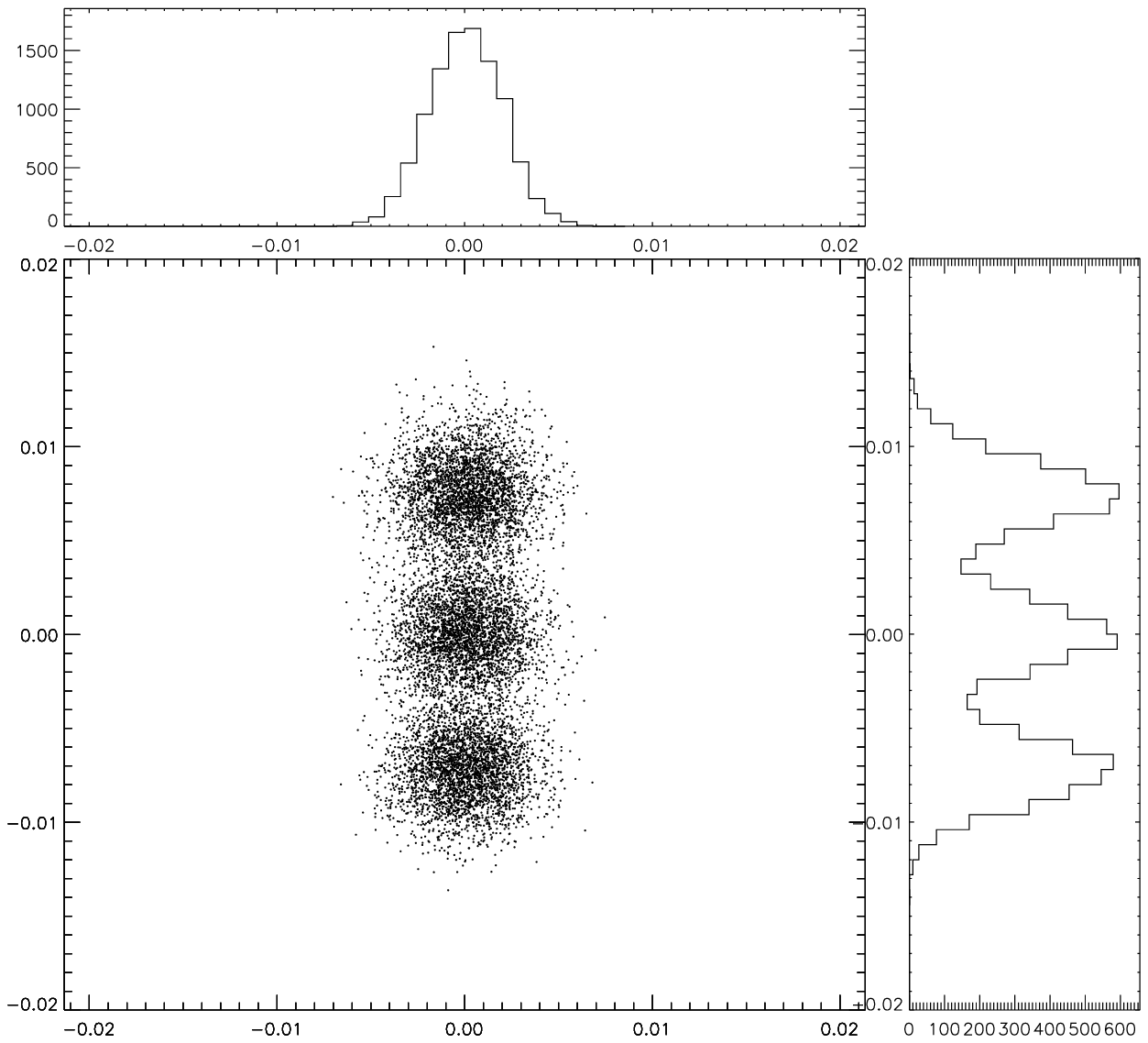}
\end{center}
\caption{Results from ray tracing simulations at the entrance of the
second undulator for three photon energies $999.8$ eV, $1000.0$ eV
and $1000.2$ eV, obtained from the ray-tracing program SHADOW. The
histograms show the number of rays as a function of $x$ and $y$
coordinates. From the separation of the photon energies, a resolving
power of $5000$ would be expected.} \label{ray_1000_undl}
\end{figure}
\begin{figure}
\begin{center}
\includegraphics[width=0.75\textwidth]{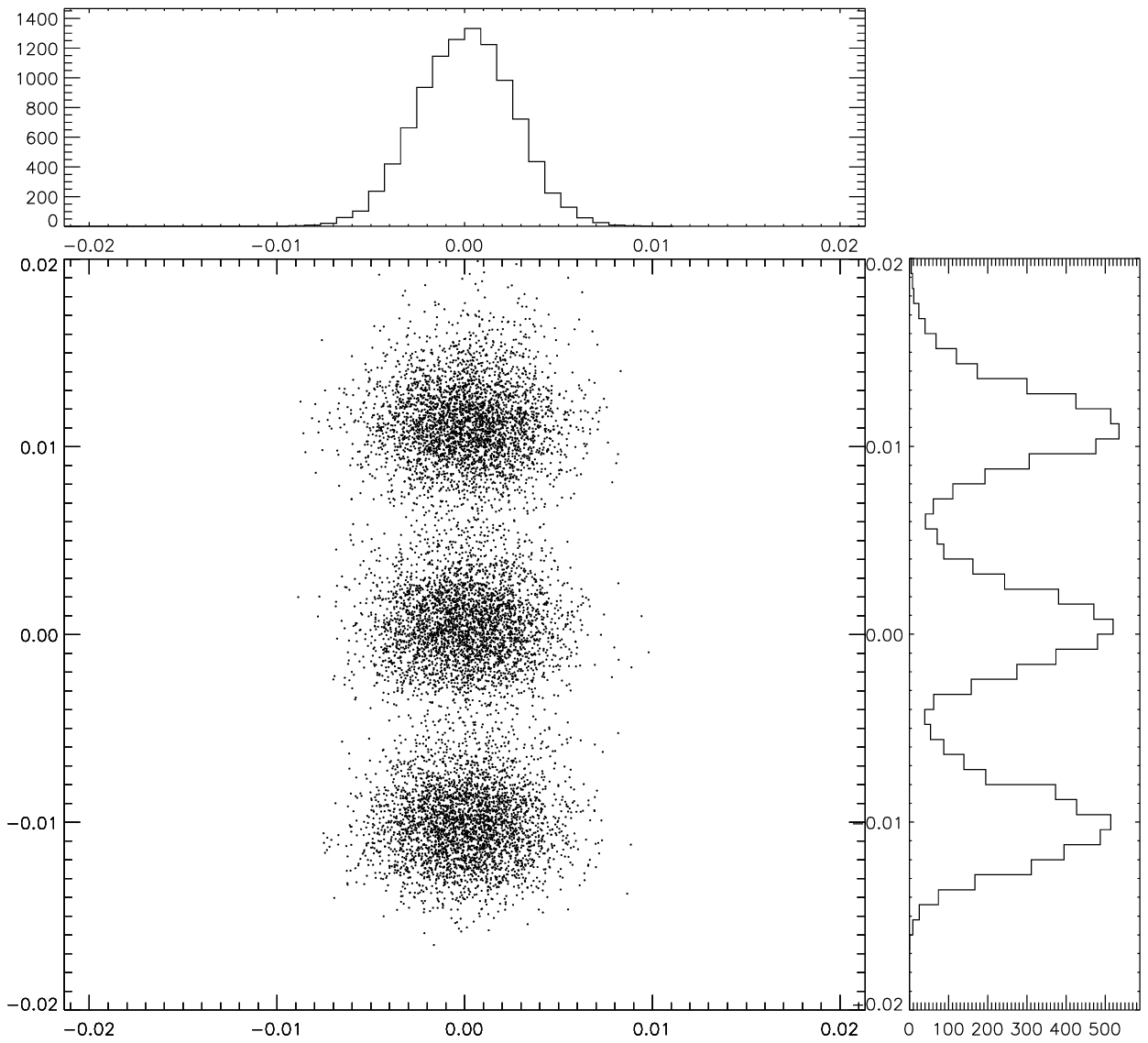}
\end{center}
\caption{Results from ray tracing simulations at the entrance of the
second undulator for three photon energies $499.9$ eV, $500.0$ eV
and $500.1$ eV obtained from the ray-tracing program SHADOW. The
histograms show the number of rays as a function of $x$ and $y$
coordinates. From the separation of the photon energies, a resolving
power of $5000$ would be expected.} \label{ray_500_undl}
\end{figure}
\begin{figure}
\begin{center}
\includegraphics[width=0.75\textwidth]{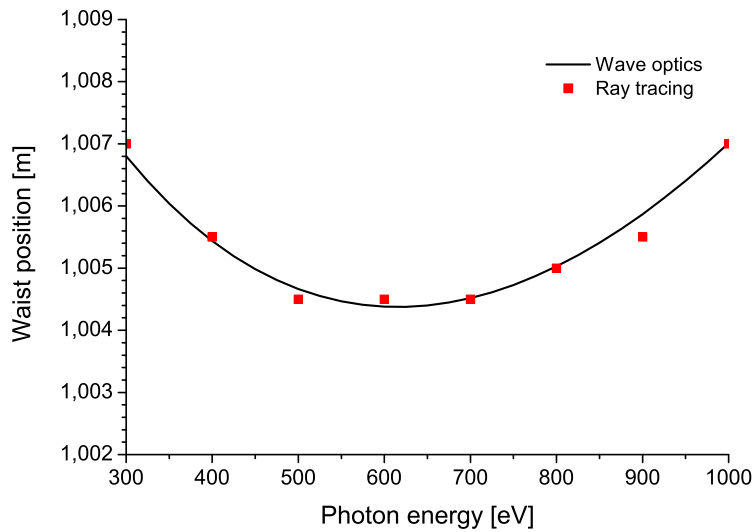}
\end{center}
\caption{Focusing at the slit position. Variation of the focus
location as a function photon energy.} \label{ray_waistpos}
\end{figure}
\begin{figure}
\begin{center}
\includegraphics[width=0.75\textwidth]{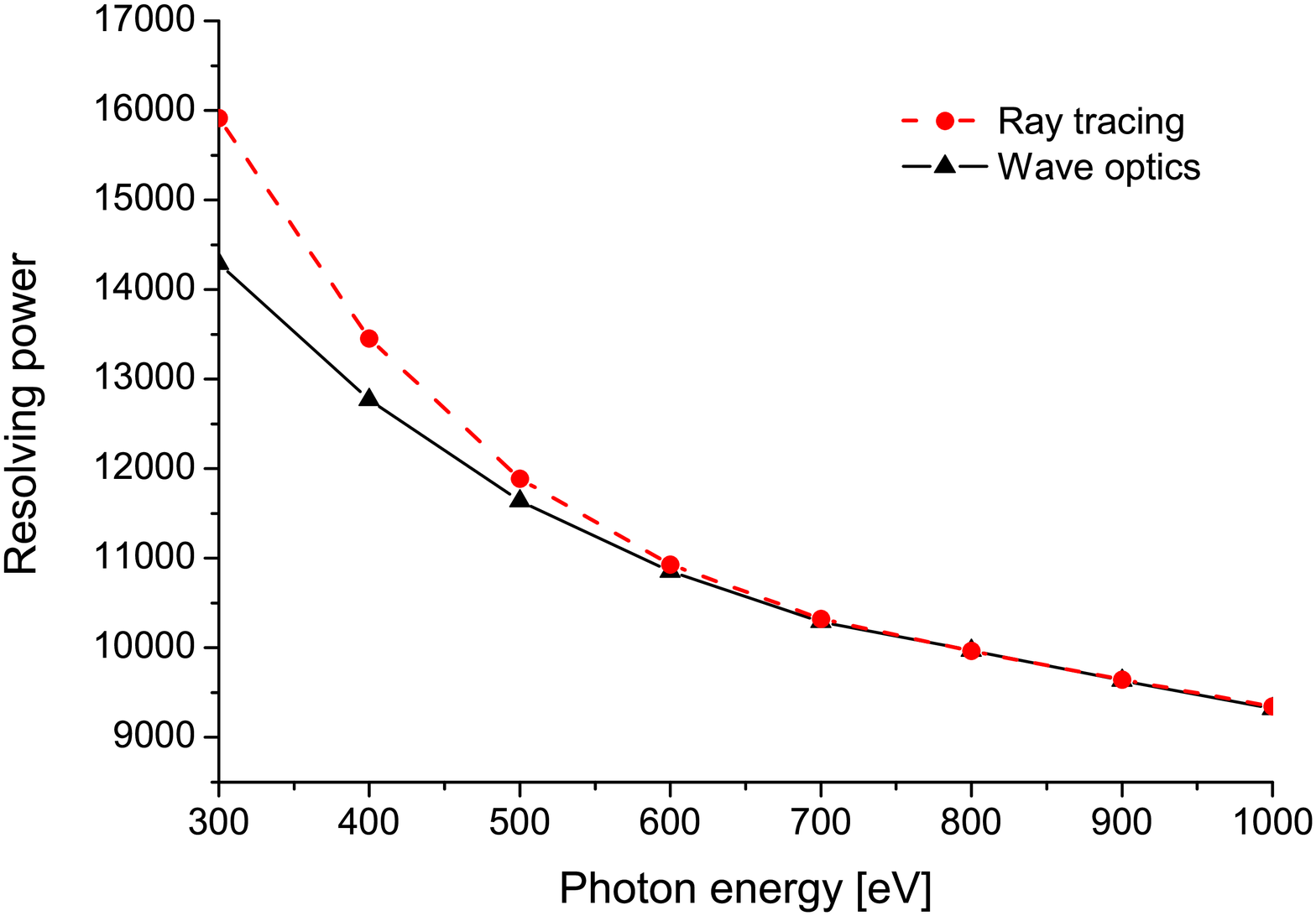}
\end{center}
\caption{Resolving power of the grating monochromator at closed
slit, as a function of the photon energy. Squares are calculated
using wave optics and including aberrations, and circles are
calculated with ray-tracing code.} \label{ray_resoll}
\end{figure}
The optical system was simulated using the ray-tracing code SHADOW
\cite{SHAD} in order to evaluate the performance of the
monochromator. The reason for modeling the monochromator using
ray-tracing is the need to check the results found by wave optics
calculations, especially minimization of aberrations. The source has
been modeled as a Gaussian-shaped beam and with Gaussian divergence
distribution as a function of wavelength, since the XFEL source will
be nearly transform-limited.  We performed ray-tracing simulations
at three different photon energies, $1000.2$ eV, $1000.0$ eV and
$999.8$ eV. We assumed the rms value of $17.1~ \mu$m for the source
size, and $5.57~\mu$rad for the beam divergence\footnote{We call the
product of the rms divergence angle by the rms source size the
emittance $\epsilon$ of the photon beam, $\epsilon = \sigma' \sigma
= \lambda/4\pi$.  For a two-dimensional distribution the definition
of the Gaussian beam emittance applies to each direction.}. No
figure errors (i.e. no slope errors) were included. In this way, the
FEL source has been propagated over grating and exit slit, and then
over the refocusing mirror. Ray-tracing results at the plane of the
exit slit for photons of $1000.2$ eV, $1000.0$ eV, and $999.8$ eV
(when the monochromator is tuned to $1000$ eV) are shown in
Fig.\ref{ray_1000_slitl}.  The histograms show that in the
dispersive (tangential) and non-dispersive (sagittal) direction the
distribution is almost Gaussian. As it can be seen from the figure,
the focusing properties of the monochromator at this photon energy
are excellent, and resolution is larger than $5000$. The
corresponding ray-tracing results for $0.5$ keV photons are shown in
Fig. \ref{ray_500_slitl}. The figure shows that the same focusing
quality is obtained also in this case. Actually, this is the case
for all photon energies in the range between $0.3$ keV and $1$ keV
due to the VLS toroidal grating parameters, minimizing defocusing
and coma aberrations. Fig. \ref{ray_1000_undl} and Fig.
\ref{ray_500_undl} display results of ray-tracing simulations at the
entrance of the second undulator.

The location of the beam focus, shown in Fig. \ref{ray_waistpos},
was found to vary with the energy around the slit. Fig.
\ref{ray_waistpos} also shows a comparison between results found
with ray tracing and wave optics calculations. Fig. \ref{ray_resoll}
summarizes the energy resolution obtained from the linear dispersion
and the FWHM of the spot size at the exit slit plane as a function
of the photon energy. Comparing to the optimal slit distance found
by wave optics calculations, it is seen that there is a very good
agreement between coherent and incoherent models. However a
complete, straightforward analysis of the full self-seeding
monochromator setup can be only be performed in terms of wave
optics.

\subsection{Heat load}

\begin{table}
\caption{Heat load on the grating}

\begin{small}\begin{tabular}{ l c c}
\hline & ~ Units &  ~ \\ \hline
Photon energy       & keV                 & 0.8     \\
Incident power      & W                   & 4.5   \\
Absorbed power      & W                   & 1.4    \\
Footprint           & cm$^2$              & 0.02   \\
Power density       & W/mm$^2$            & 0.7 \\
\hline
\end{tabular}\end{small}
\label{tt3}
\end{table}
\begin{figure}
\begin{center}
\includegraphics[width=0.75\textwidth]{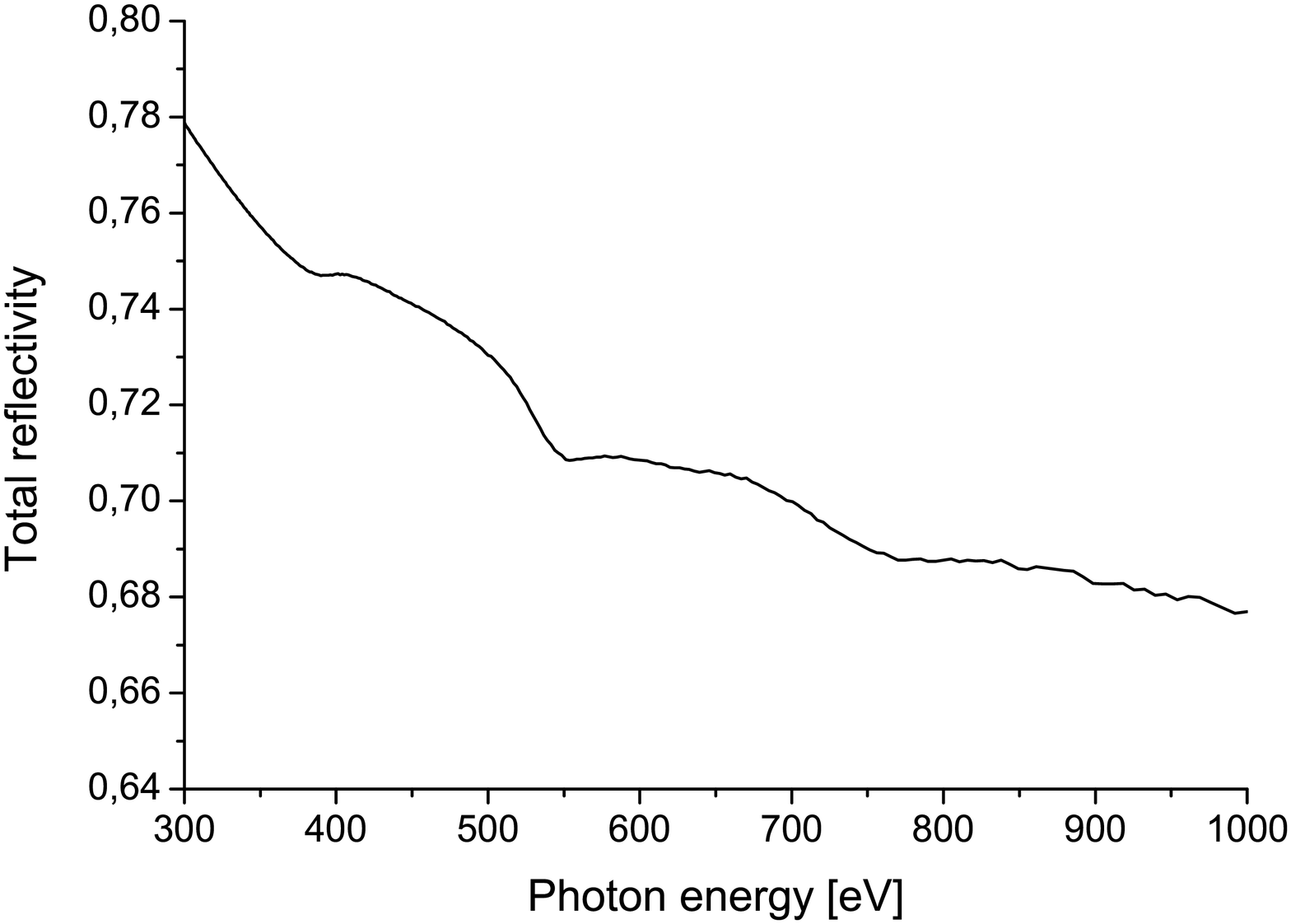} 
\end{center}
\caption{Total reflectivity of the grating as a function of photon
energy.} \label{gr_reflect}
\end{figure}
X-ray optics can survive the high average power during one
macropulse from the European XFEL. The heat load can be evaluated by
considering the absorbed power on the grating. In table \ref{tt3} we
consider the case for an electron beam with 0.1 nC charge. The power
absorbed by the grating can be evaluated by taking into account the
parameters of the first undulator source (see section
\ref{sec:FELsim} and Fig. \ref{IN1}) and the total reflectivity of
the grating, Fig. \ref{gr_reflect}. The results presented here refer
to the case of an impinging pulse train composed by $2700$ FEL
pulses of $0.001$ mJ each, at the photon energy of $800$ eV. The
power is averaged over the $0.6$ ms of the pulse train, which is the
most extreme approximation. A similar power density, $0.6$ W/mm$^2$-
$1.3$ W/mm$^2$ absorbed in the grating has been reported in previous
studies for the European XFEL \cite{TRAN}.

\subsection{Single shot damage}

While the average  absorbed power on the grating power is still
moderate, the peak power within the single pulse from the first SASE
undulator will be in the range of a fraction of GW. At these power
levels, the main issue may be no longer related with thermal
distortion, but rather with the possibility of ablation of the
grating surface, which would result in permanent damage. Ablation
depends on the radiation dose per pulse, which can be quantified as
the energy absorbed in the volume defined by the projected beam area
on the optical element and by one attenuation length, which is the
depth into the material, measured along the surface normal, where
the radiation intensity falls to $2,72$ times of its value at the
surface. Normalized to one atom this energy corresponds to the
atomic dose near the surface \cite{XTDR}

\begin{eqnarray}
D = \frac{E_\mathrm{pulse} (1 - R) \theta_i}{2\pi \sigma^2
l_\mathrm{att} n_A}~, \label{dose}
\end{eqnarray}
where $E_\mathrm{pulse}$ is the energy in one radiation pulse, $R$
is the reflectivity, $l_\mathrm{att}$ is the attenuation length,
$\sigma$ is the rms of the Gaussian beam intensity distribution
immediately in front of the grating, and $n_A$ denotes the
element-specific density of atoms.  For Pt coating we find $n_A \sim
6.4 \cdot 10^{22}$ cm$^{-3}$, $R \sim 0.7$, $l_\mathrm{att} \sim 2$
nm\footnote{at $2$ degrees grazing incidence. We also assume that
the incident angle is $1$ degree, and that the blaze angle is about
$1$ degree too.}, $\sigma \sim 0.05$ mm, and $E_\mathrm{pulse} \sim
0.001$ mJ. The calculated dose reaches up to $15$ meV/atom. This is
about $30$ times below the melting threshold for Pt \cite{FENG2}.
The grating is therefore safe from damage.

\section{\label{sec:FELsim} FEL simulations}

With reference to Fig. \ref{mon_lay_ov}, we performed a feasibility
study with the help of the FEL code Genesis 1.3 \cite{GENE} running
on a parallel machine. We will present a feasibility study for the
SASE3 FEL line of the European XFEL, based on a statistical analysis
consisting of $100$ runs. The overall beam parameters used in the
simulations are presented in Table \ref{tt1}.

\begin{table}
\caption{Parameters for the mode of operation at the European XFEL
used in this paper.}

\begin{small}\begin{tabular}{ l c c}
\hline & ~ Units &  ~ \\ \hline
Undulator period      & mm                  & 68     \\
Periods per cell      & -                   & 73   \\
Total number of cells & -                   & 21    \\
Intersection length   & m                   & 1.1   \\
Energy                & GeV                 & 10.5 \\
Charge                & nC                  & 0.1\\
\hline
\end{tabular}\end{small}
\label{tt1}
\end{table}

\begin{figure}[tb]
\includegraphics[width=0.5\textwidth]{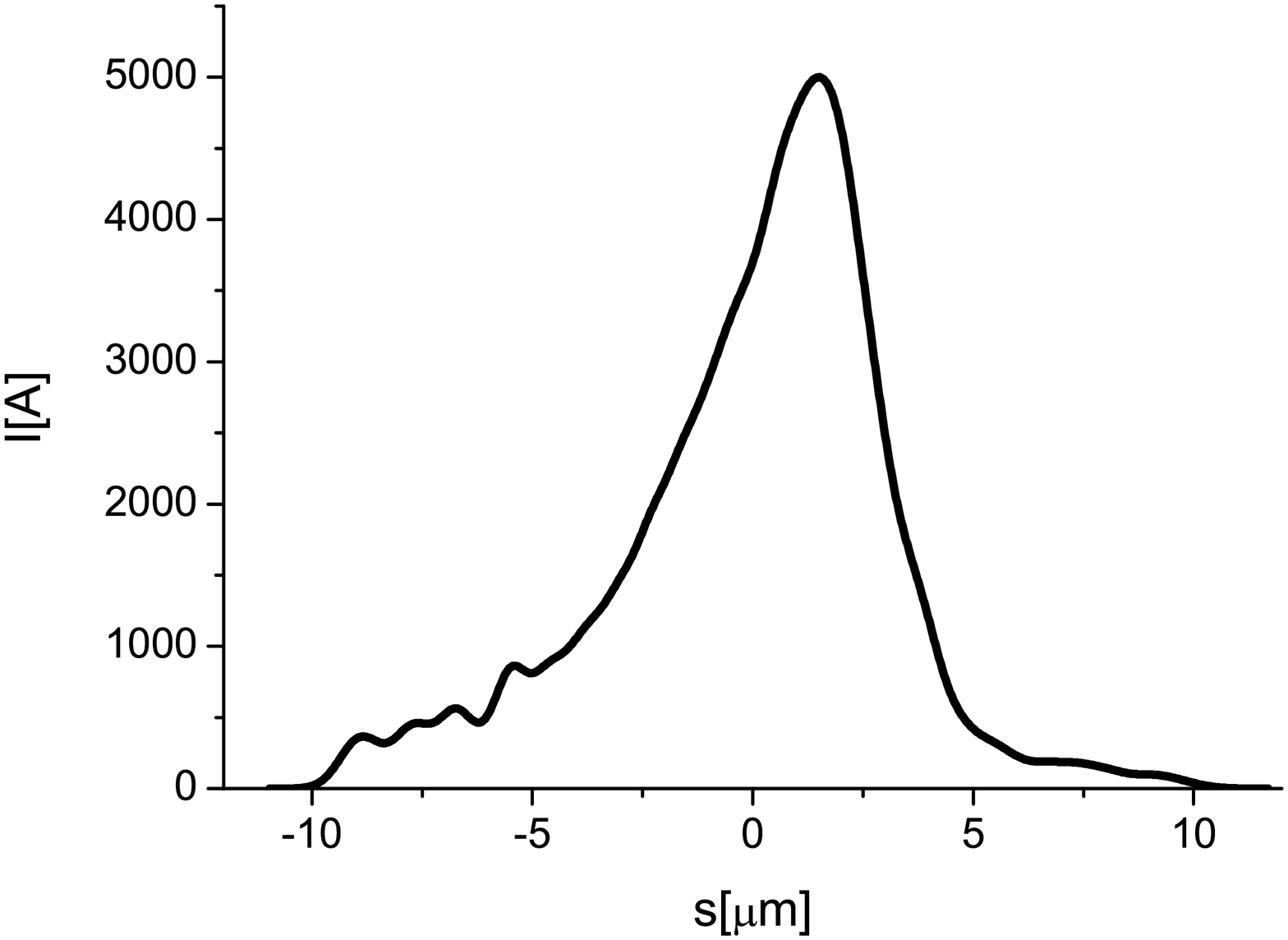}
\includegraphics[width=0.5\textwidth]{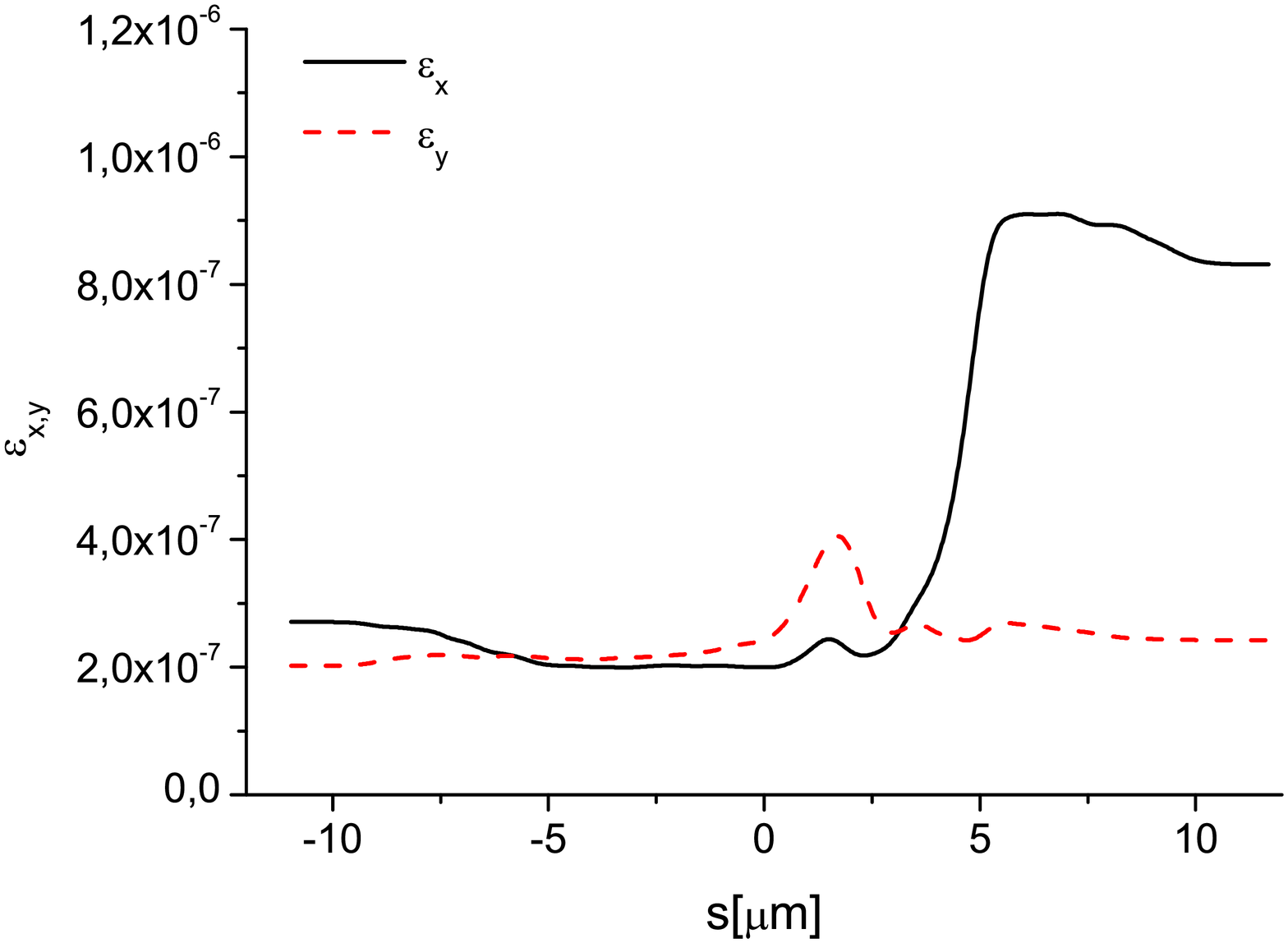}
\includegraphics[width=0.5\textwidth]{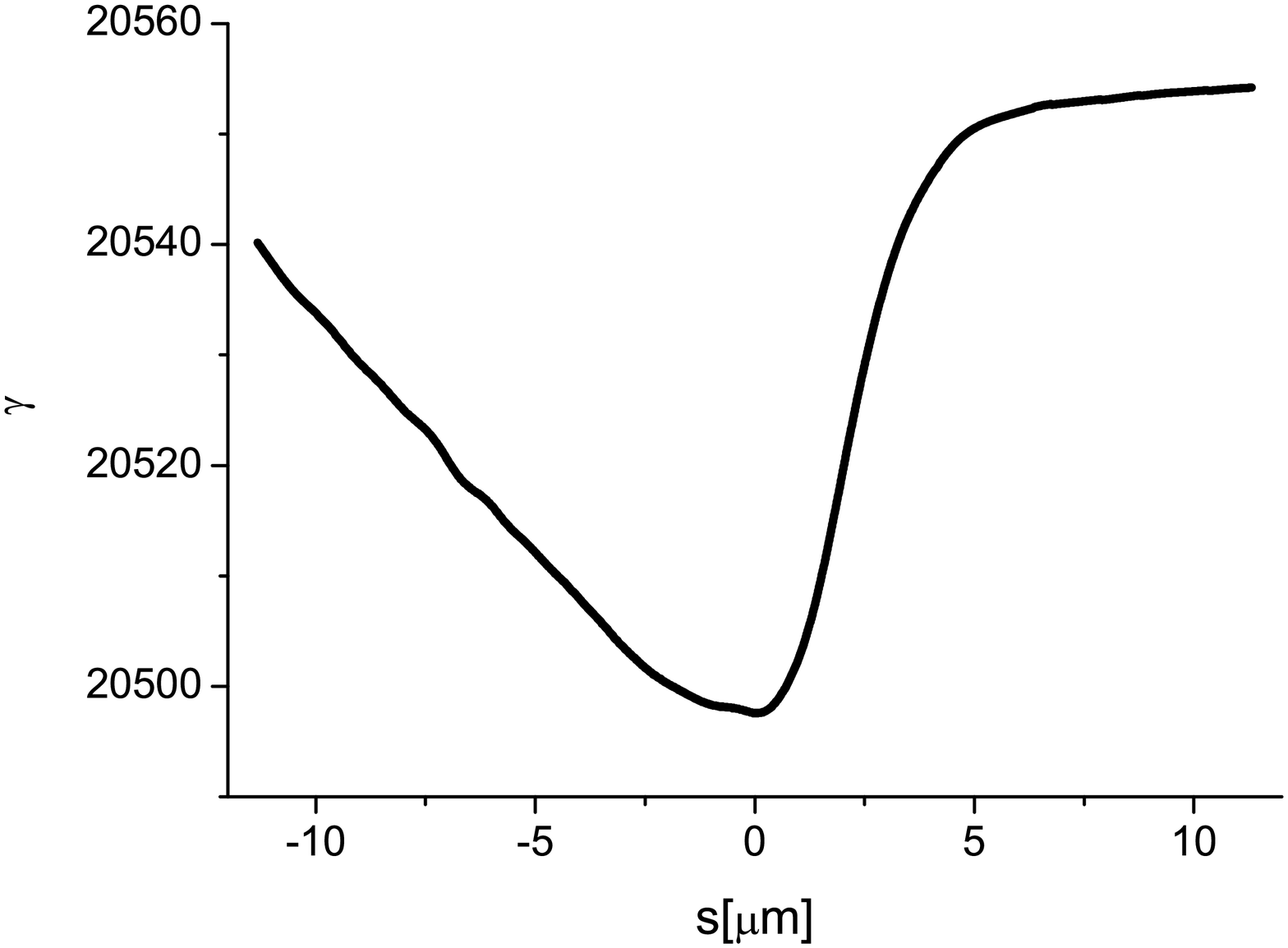}
\includegraphics[width=0.5\textwidth]{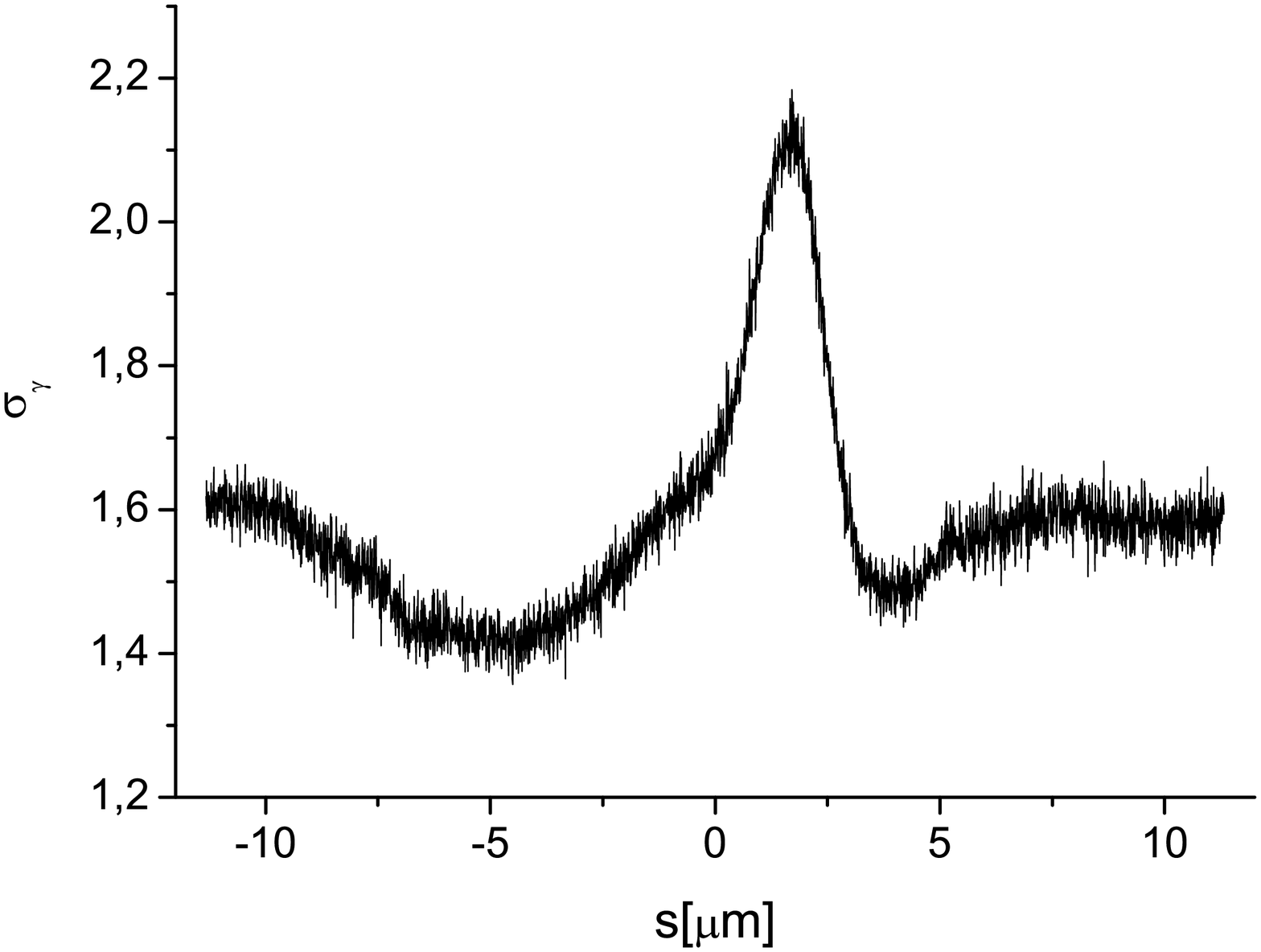}
\begin{center}
\includegraphics[width=0.5\textwidth]{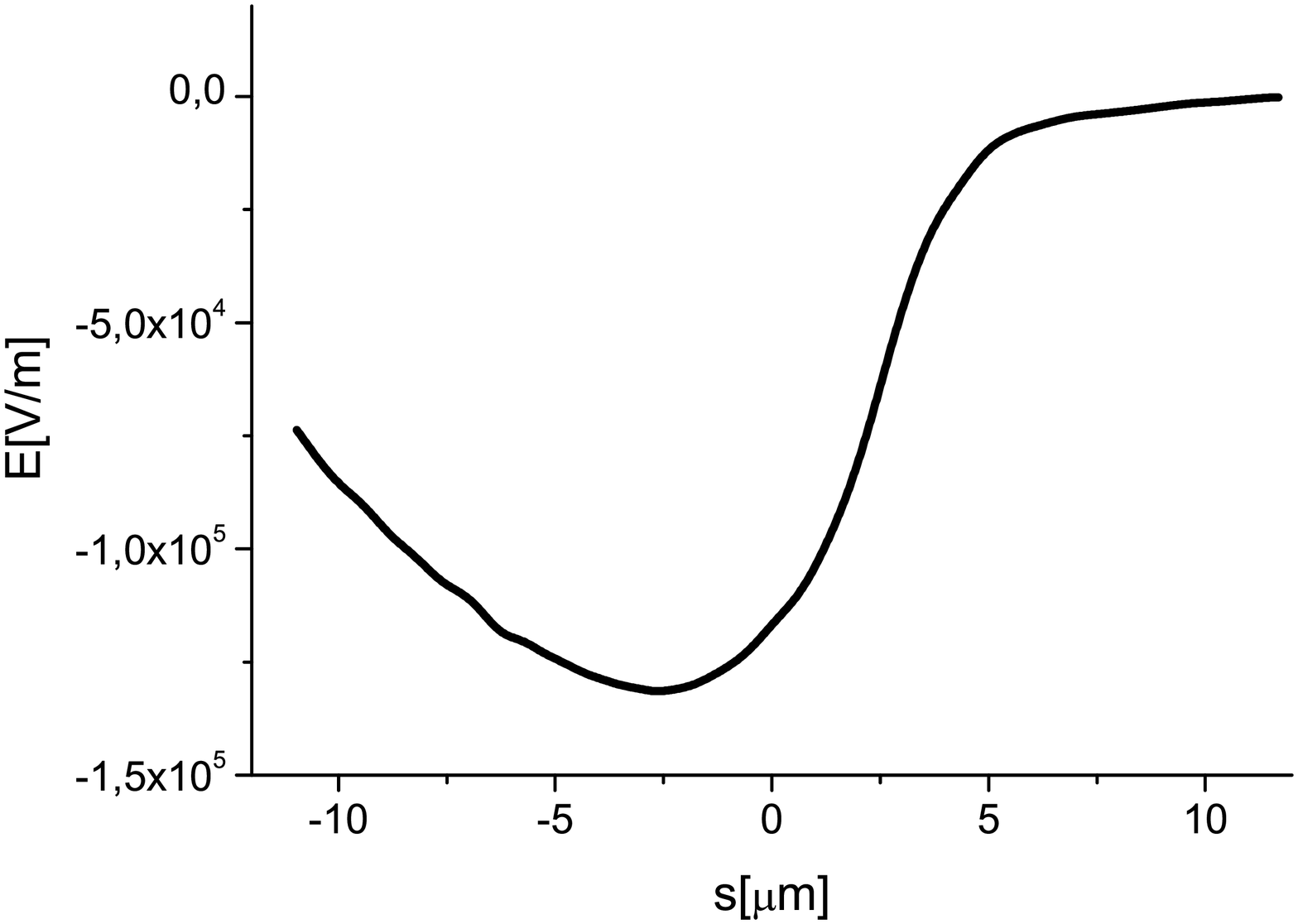}
\end{center}
\caption{Results from electron beam start-to-end simulations at the
entrance of SASE3. (First Row, Left) Current profile. (First Row,
Right) Normalized emittance as a function of the position inside the
electron beam. (Second Row, Left) Energy profile along the beam.
(Second Row, Right) Electron beam energy spread profile. (Bottom
row) Resistive wakefields in the SASE3 undulator.} \label{s2E}
\end{figure}
\begin{figure}[tb]
\includegraphics[width=0.75\textwidth]{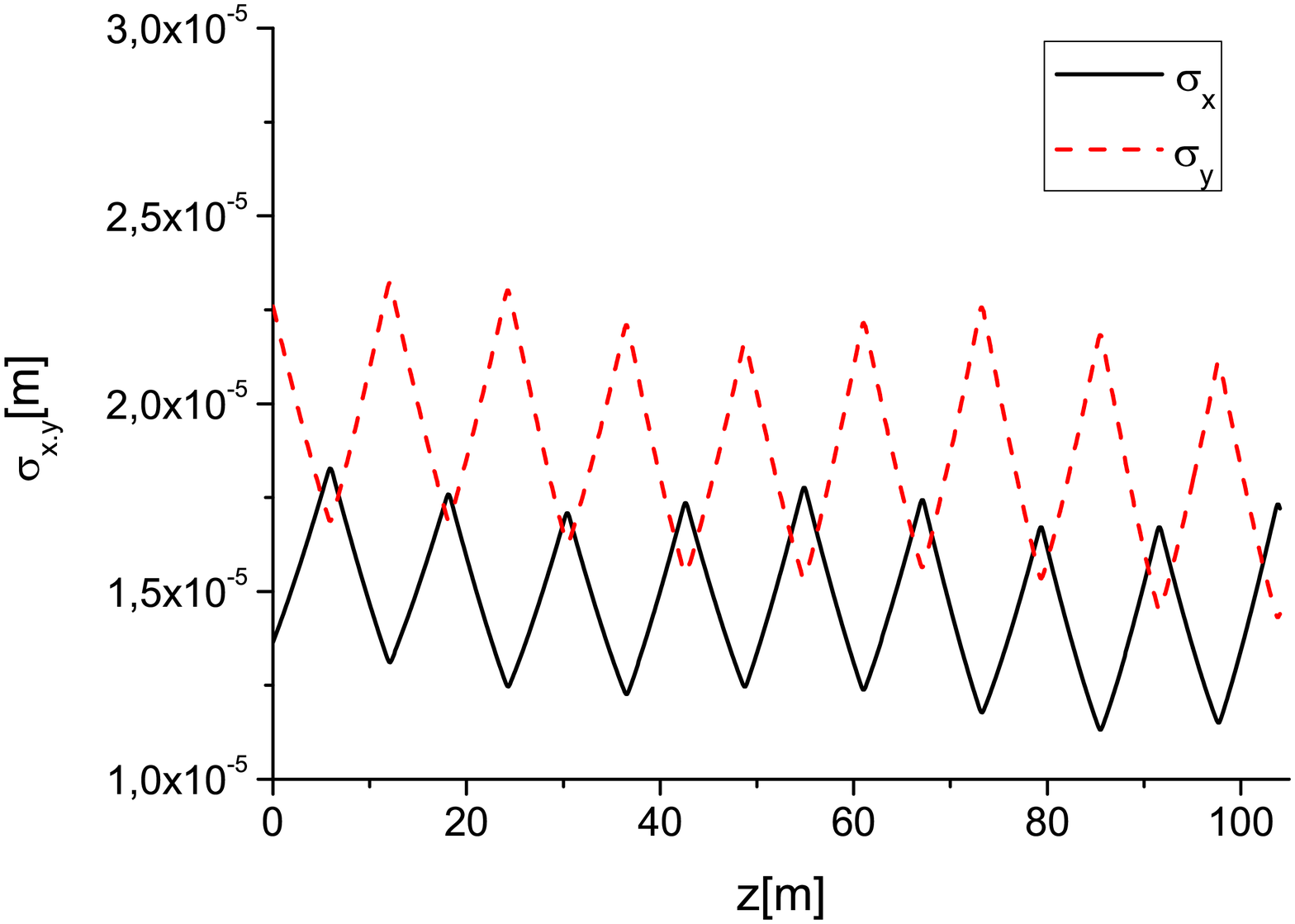}
\caption{Evolution of the horizontal and vertical dimensions of the
electron bunch as a function of the distance inside the SASE3
undulator. The plots refer to the longitudinal position inside the
bunch corresponding to the maximum current vale.} \label{sigma}
\end{figure}
The expected beam parameters at the entrance of the SASE3 undulator,
and the resistive wake inside the undulator are shown in Fig.
\ref{s2E}, \cite{S2ER}. The evolution of the transverse electron
bunch dimensions are plotted in Fig. \ref{sigma}.

\begin{figure}[tb]
\includegraphics[width=0.5\textwidth]{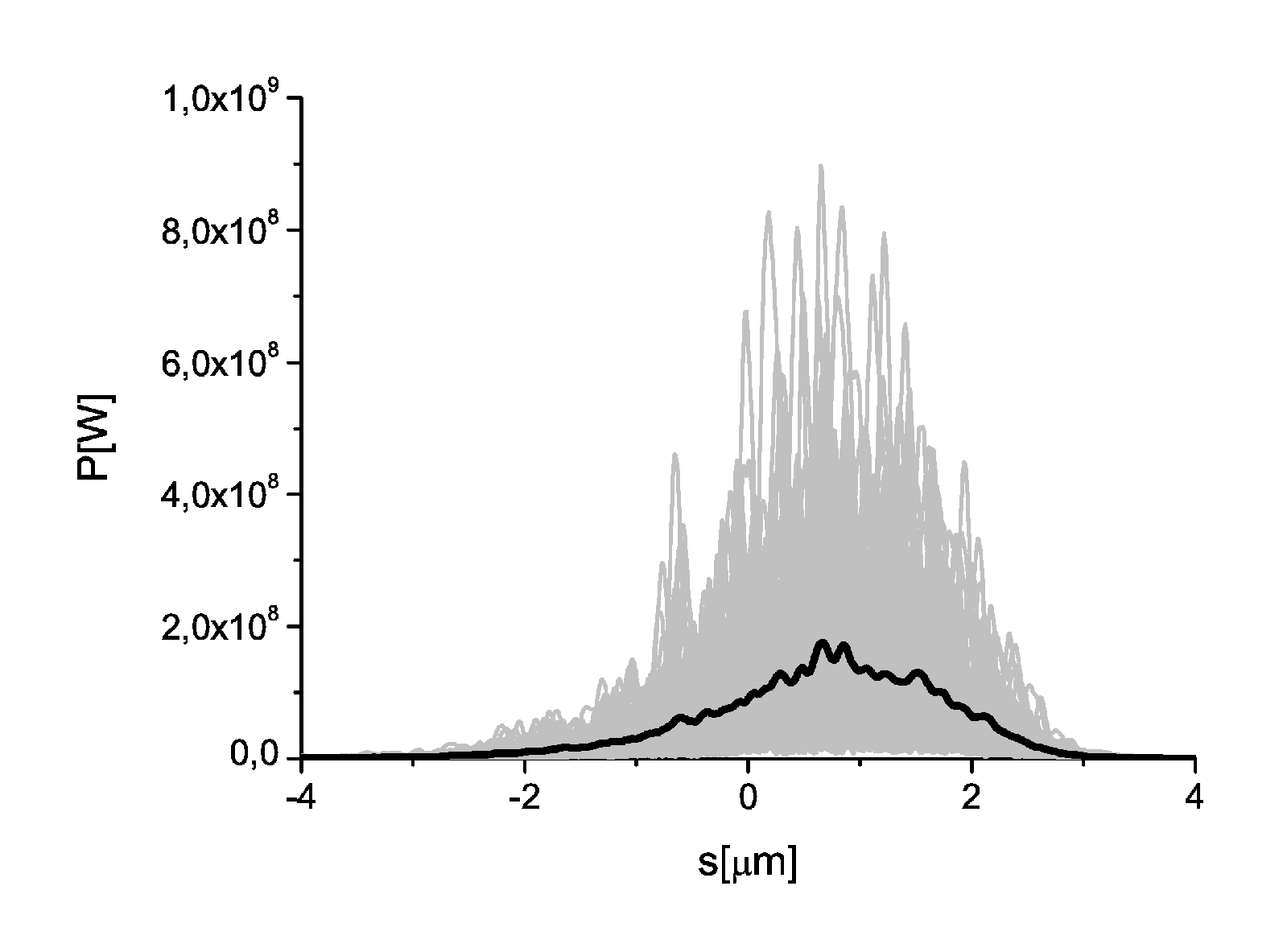}
\includegraphics[width=0.5\textwidth]{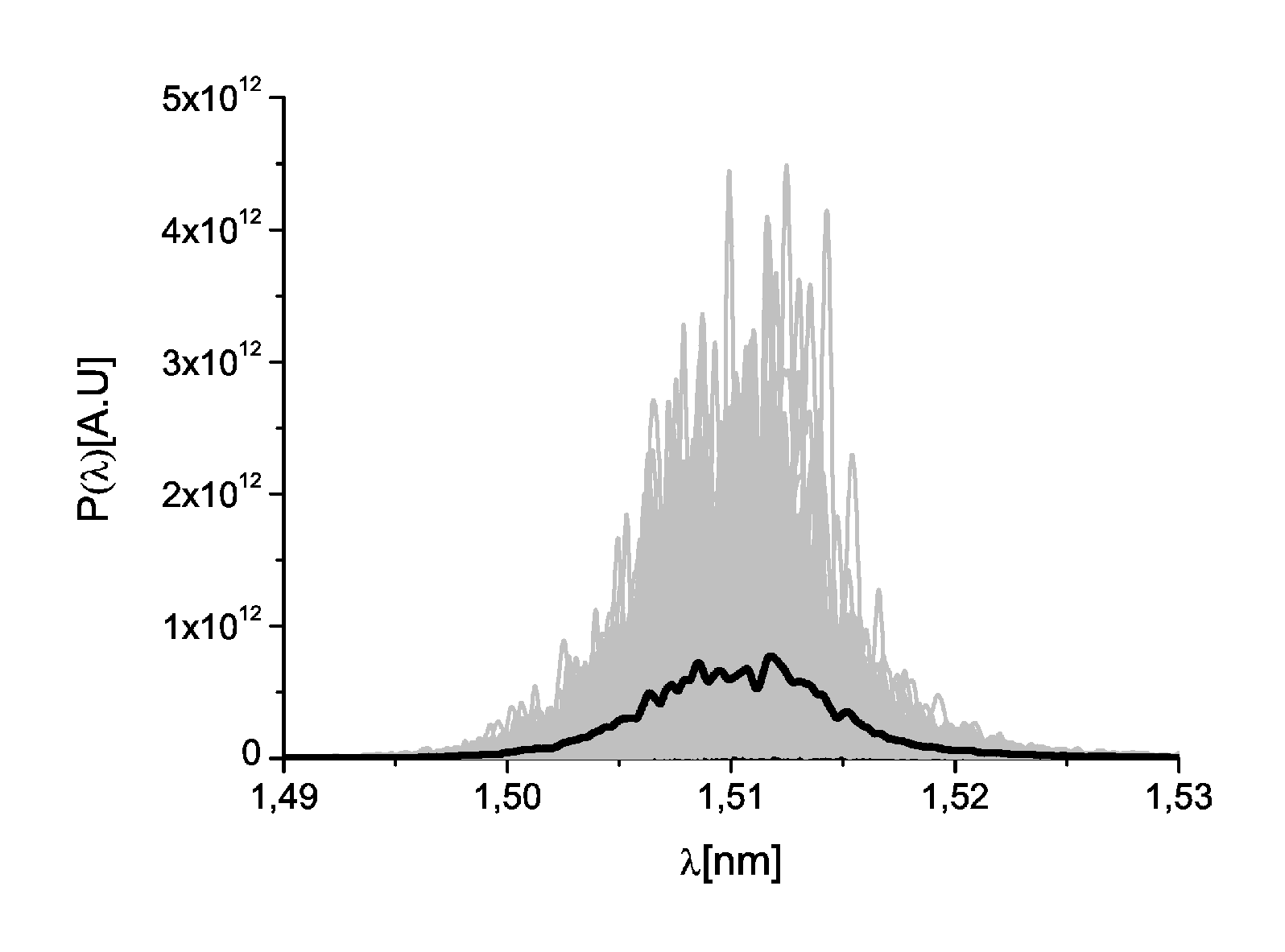}
\caption{Power distribution  and spectrum of the X-ray radiation
pulse after the first undulator. Grey lines refer to single shot
realizations, the black line refers to the average over a hundred
realizations.} \label{IN1}
\end{figure}
\begin{figure}[tb]
\includegraphics[width=0.5\textwidth]{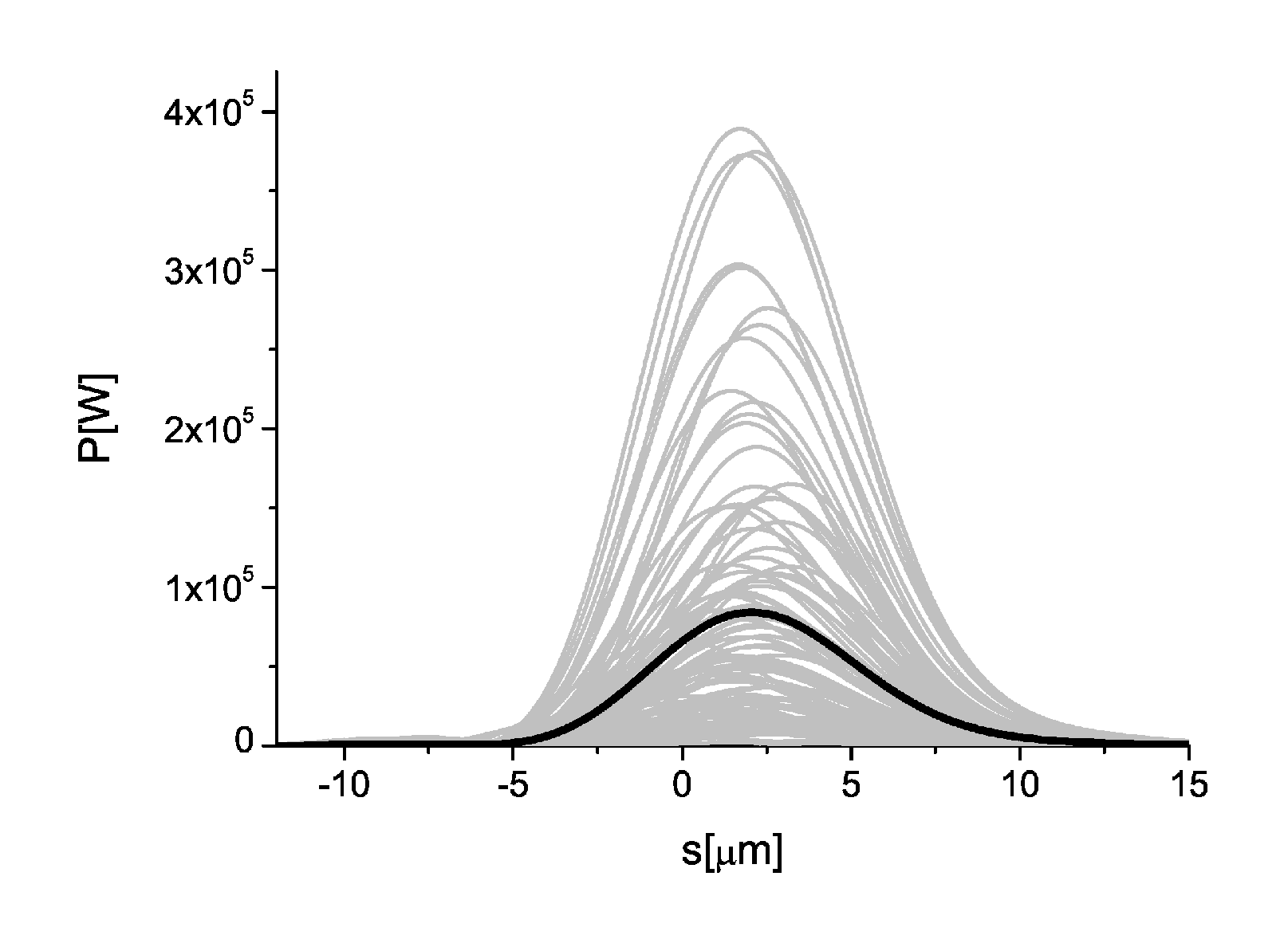}
\includegraphics[width=0.5\textwidth]{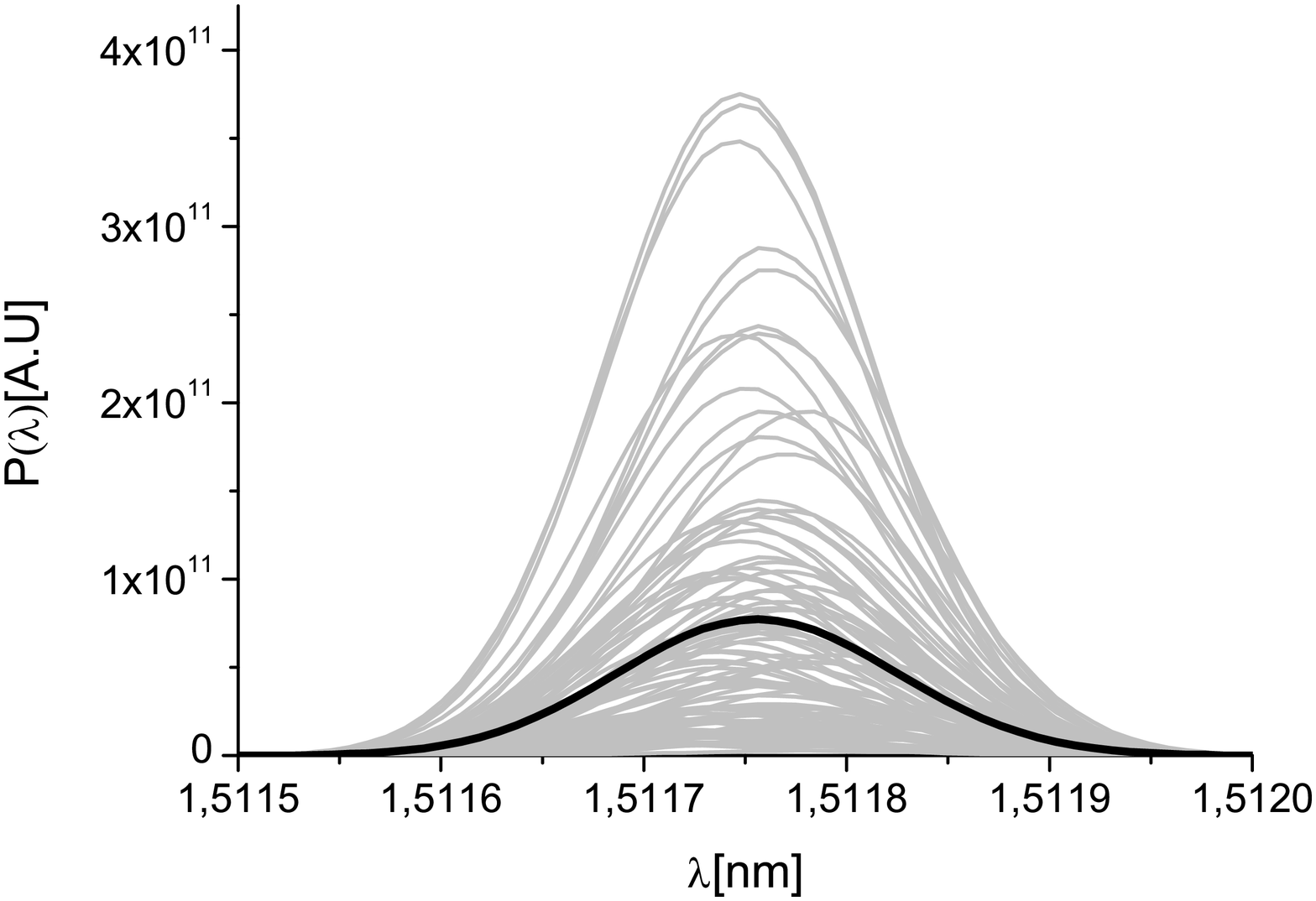}
\caption{Power distribution  and spectrum of the X-ray radiation
pulse after the monochromator. This pulse is used to seed the
electron bunch at the entrance of the output undulator. Grey lines
refer to single shot realizations, the black line refers to the
average over a hundred realizations.} \label{SEED}
\end{figure}
The SASE pulse power and spectrum after the first undulator is shown
in Fig. \ref{IN1}. This pulse goes through the grating
monochromator. The monochromator lineshape is presented in Fig.
\ref{lineprof}. At the exit of the monochromator, one obtains the
seed pulse, Fig. \ref{SEED}. As explained before, the monochromator
introduces only a short optical delay of about $0.7$ ps, which can
be easily compensated by the electron chicane. The chicane also
washes out the electron beam microbunching. As a result, at the
entrance of the second (output) undulator the electron beam and the
radiation pulse can be recombined.

\begin{figure}[tb]
\includegraphics[width=0.5\textwidth]{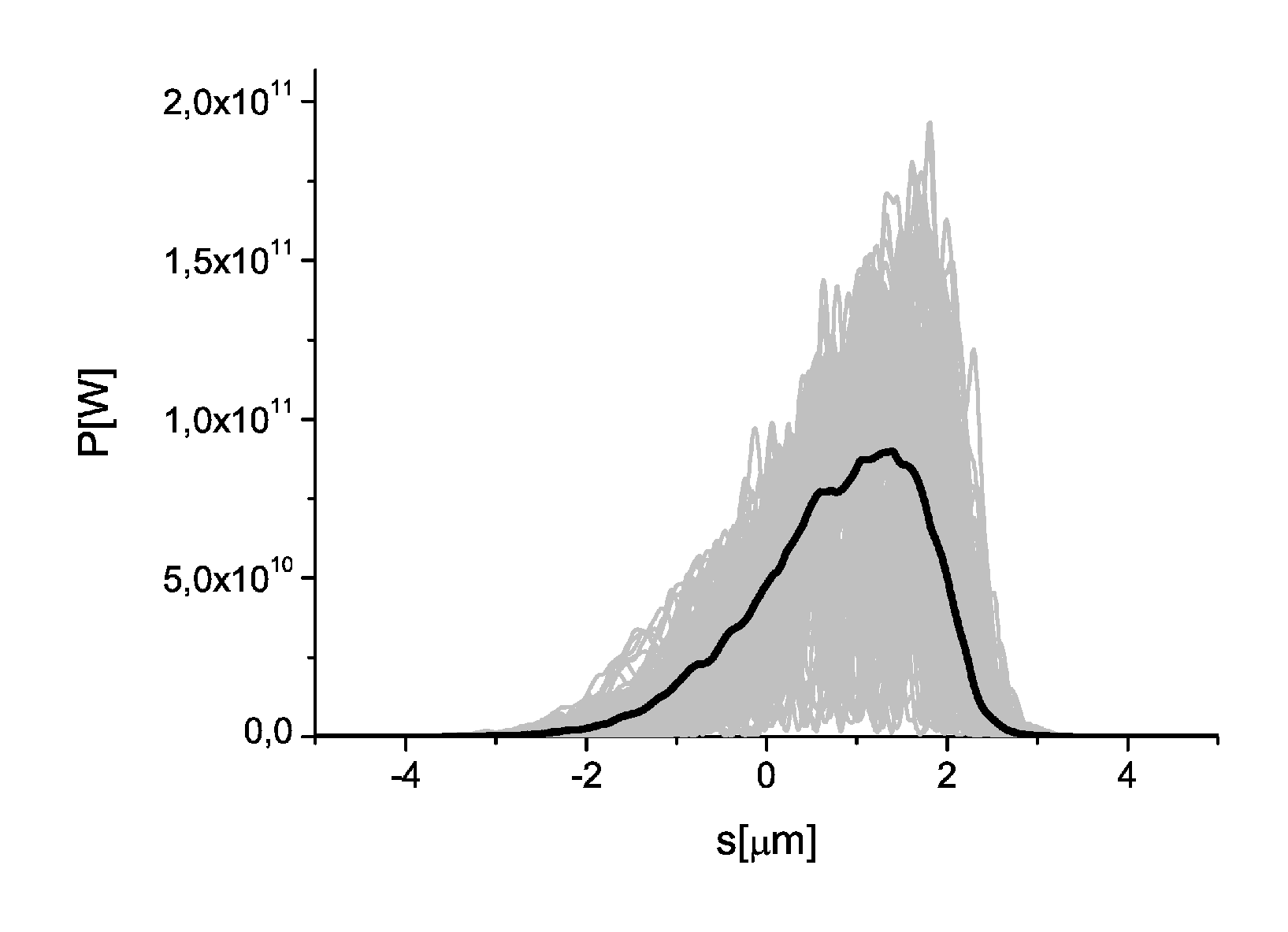}
\includegraphics[width=0.5\textwidth]{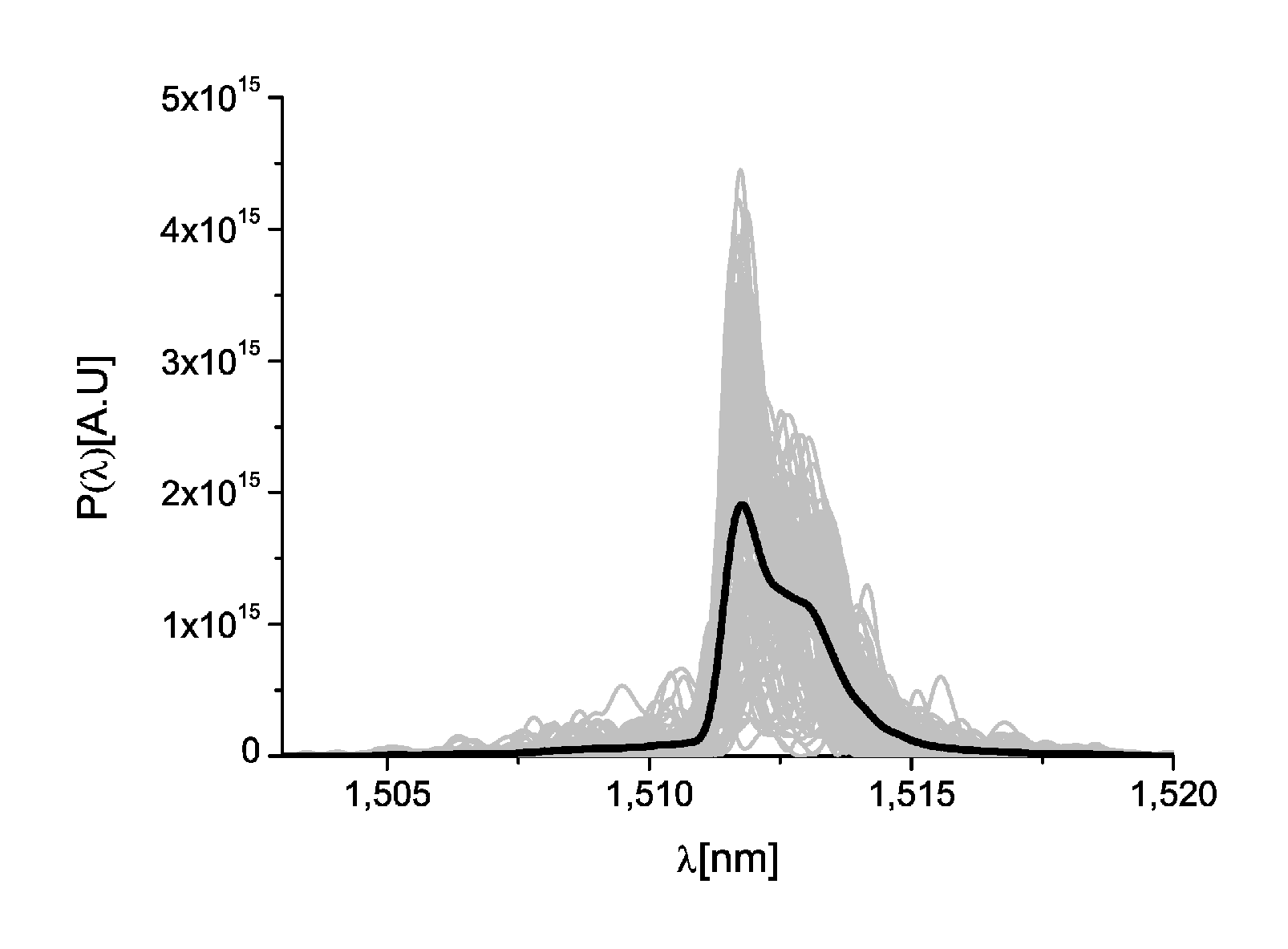}
\caption{Power distribution and spectrum of the X-ray radiation
pulse after the second undulator in the untapered case. Grey lines
refer to single shot realizations, the black line refers to the
average over a hundred realizations.} \label{OUT1}
\end{figure}
\begin{figure}[tb]
\includegraphics[width=0.5\textwidth]{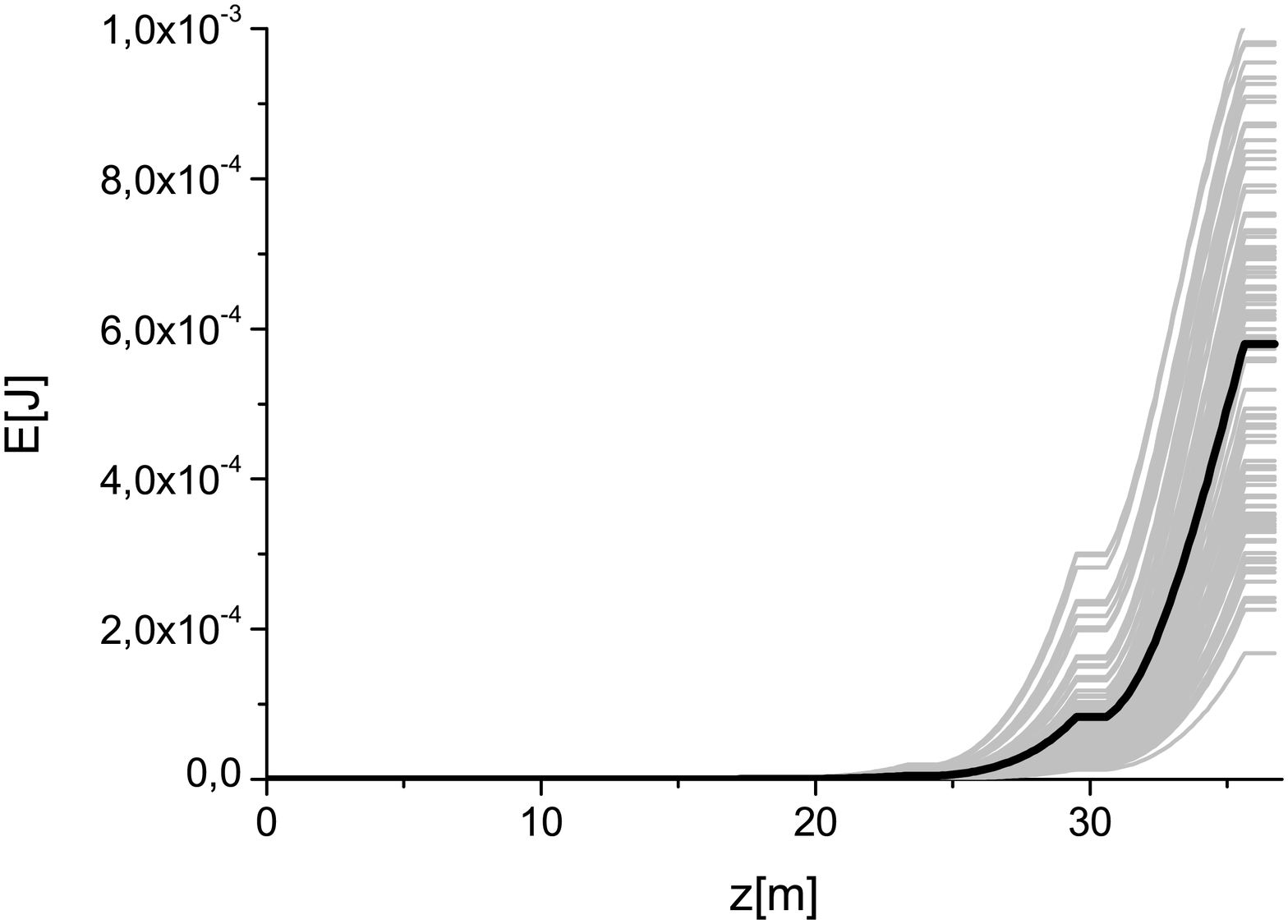}
\includegraphics[width=0.5\textwidth]{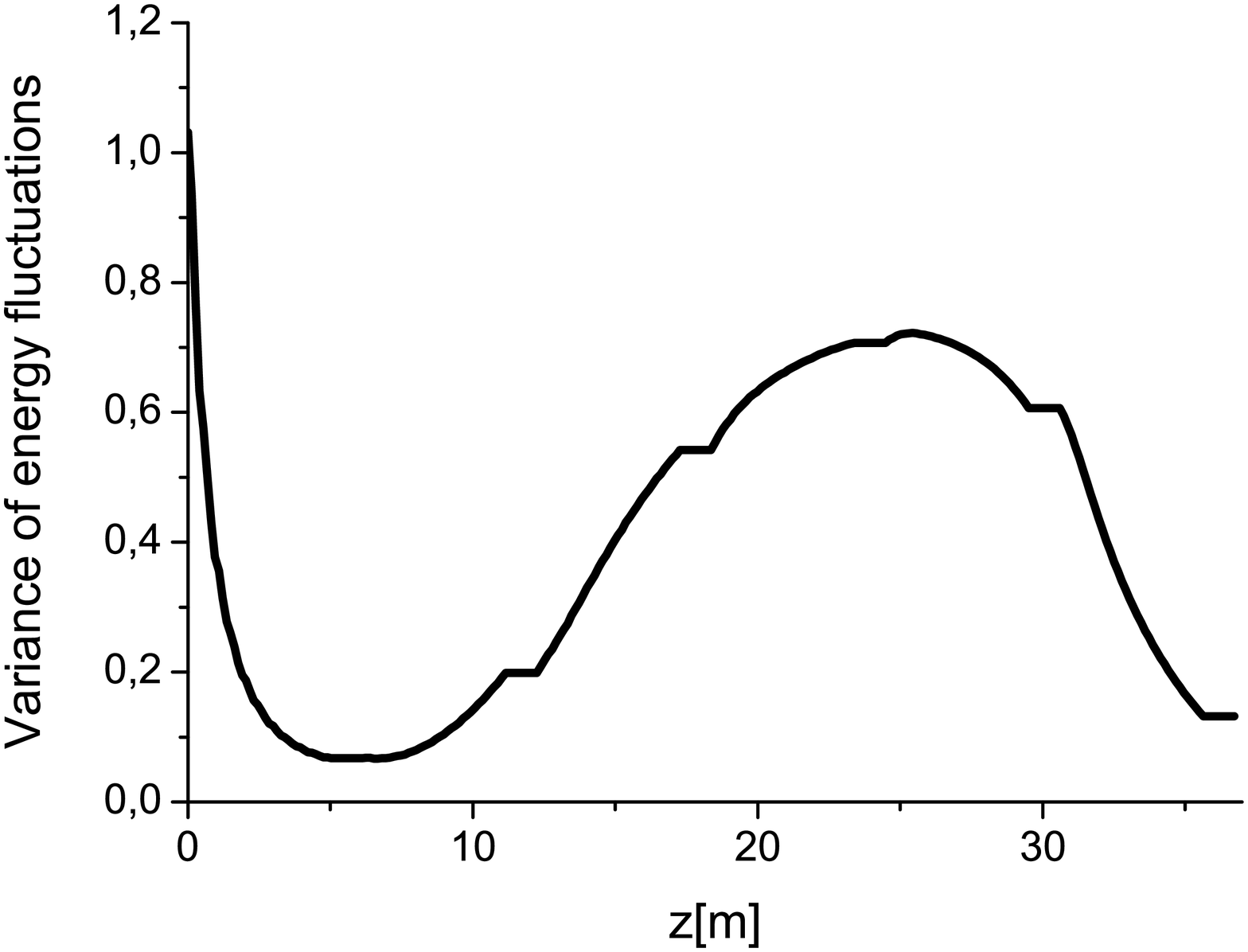}
\caption{Evolution of the energy per pulse and of the energy
fluctuations as a function of the undulator length in the untapered
case. Grey lines refer to single shot realizations, the black line
refers to the average over a hundred realizations.} \label{OUT2}
\end{figure}
\begin{figure}[tb]
\includegraphics[width=0.5\textwidth]{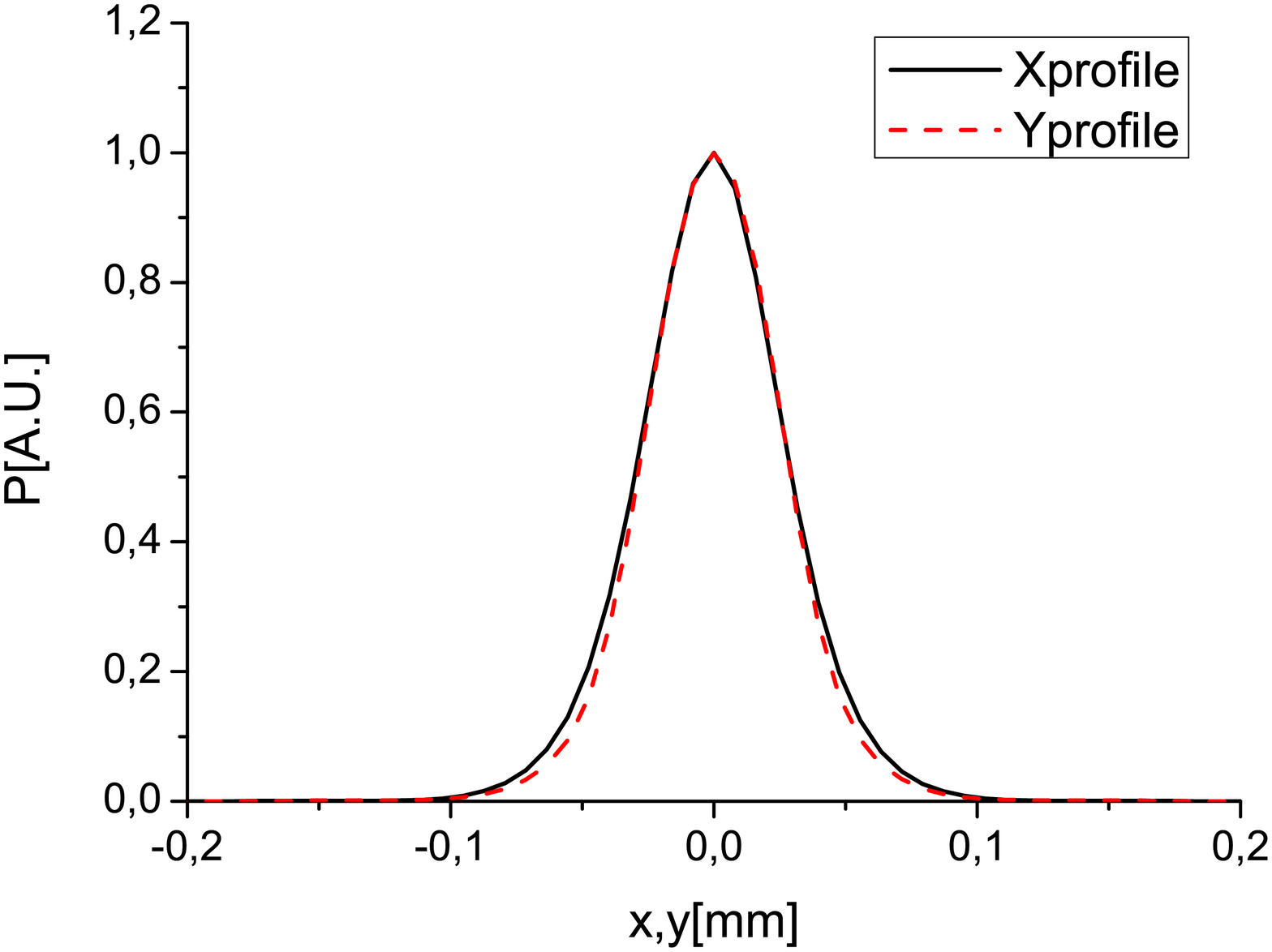}
\includegraphics[width=0.5\textwidth]{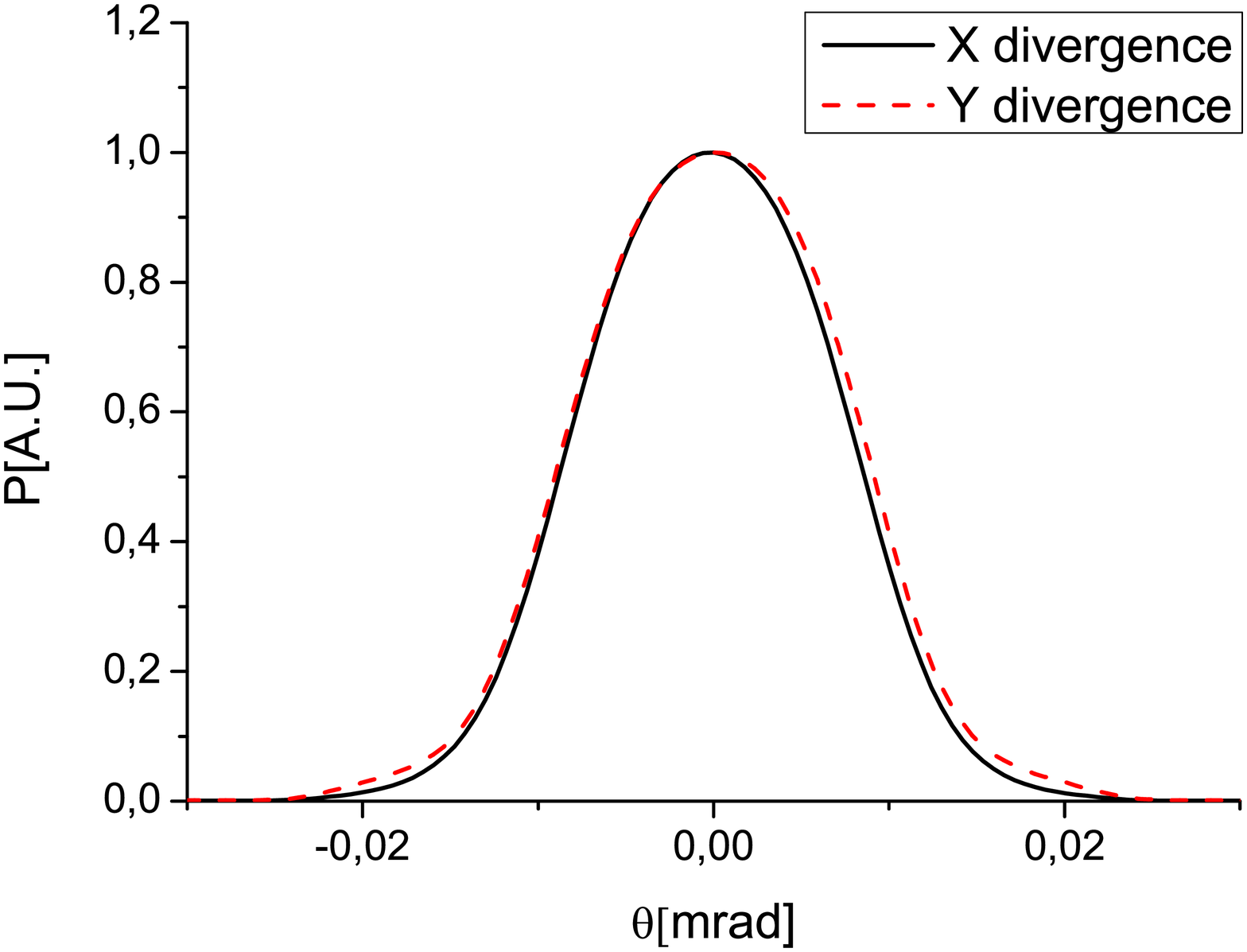}
\caption{(Left plot) Transverse radiation distribution in the
untapered case at the exit of the output undulator. (Right plot)
Directivity diagram of the radiation distribution in the case of
tapering at the exit of the output undulator.} \label{spot}
\end{figure}

If the output undulator is not tapered, one needs $7$ sections to
reach saturation. The best compromise between power and spectral
bandwidth are reached after $6$ sections, Fig. \ref{OUT1}. In this
case, the evolution of the energy per pulse and of the energy
fluctuations as a function of the undulator length are shown in Fig.
\ref{OUT2}. The pulse now reaches the $100$ GW power level, with an
average relative FWHM spectral width narrower than $10^{-3}$.
Finally, the transverse radiation distribution and divergence at the
exit of the output undulator are shown in Fig. \ref{spot}.

\begin{figure}[tb]
\begin{center}
\includegraphics[width=0.5\textwidth]{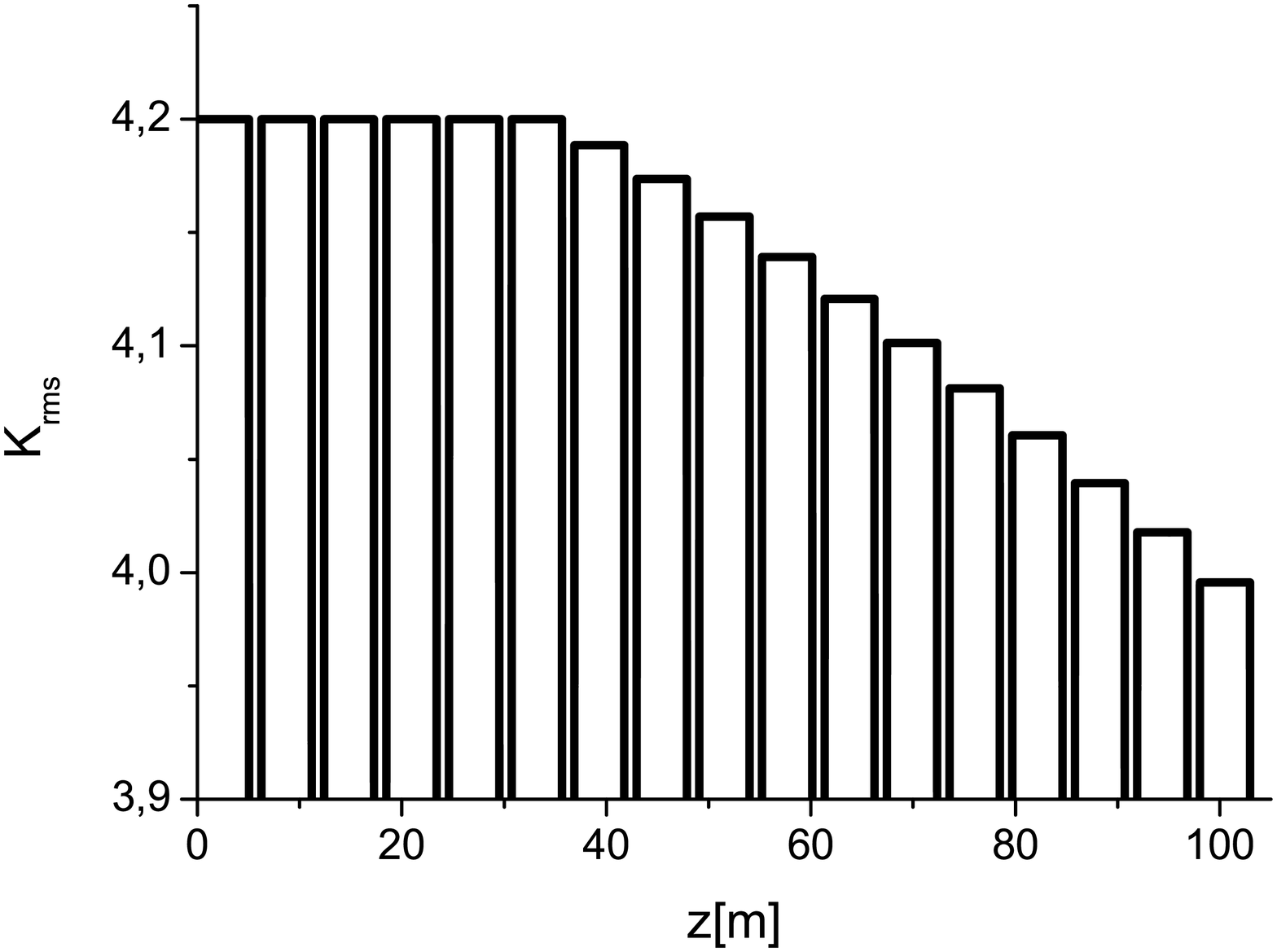}
\end{center}
\caption{Taper configuration for high-power mode of operation at
$1.5$ nm.} \label{krms}
\end{figure}
\begin{figure}[tb]
\includegraphics[width=0.5\textwidth]{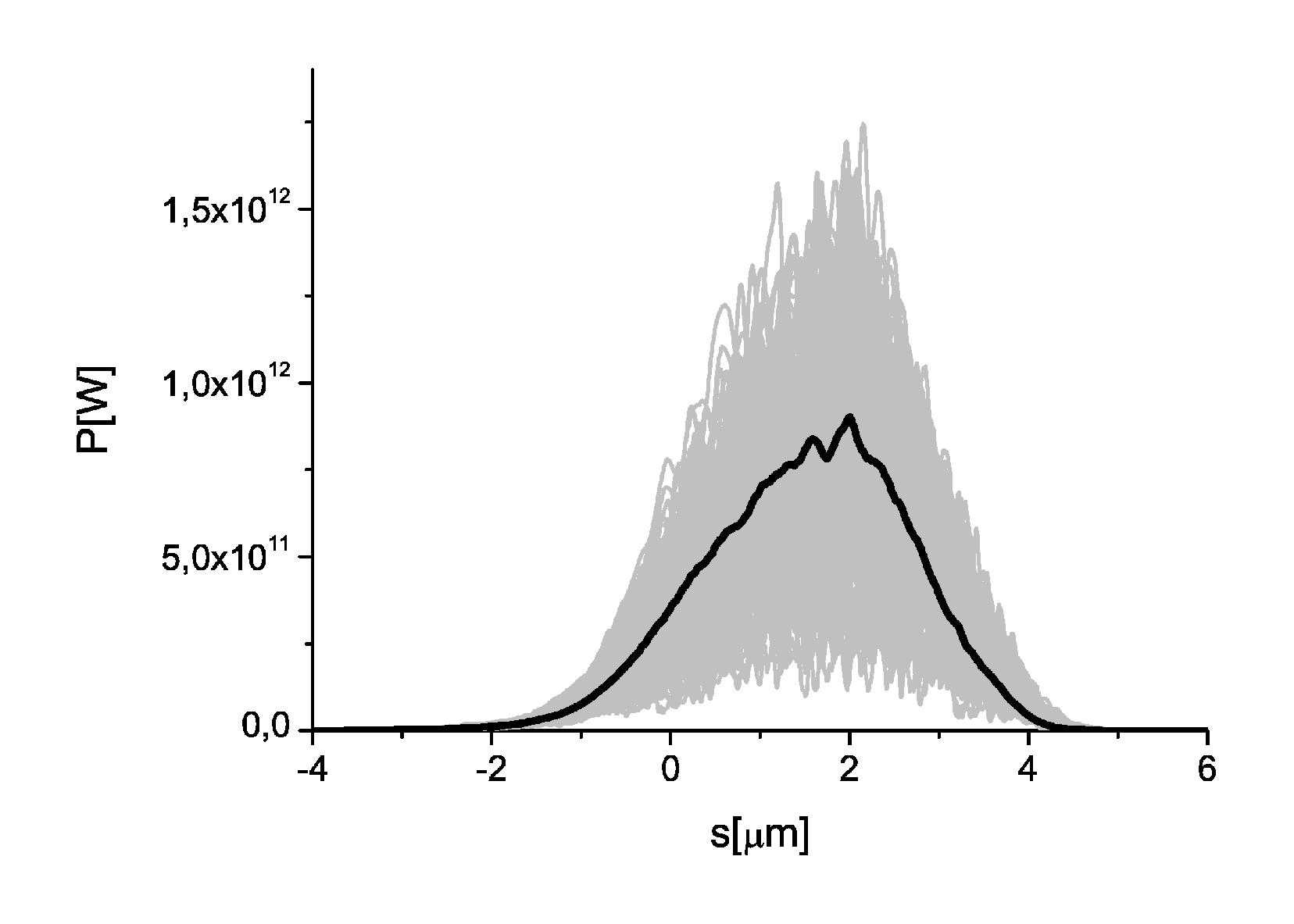}
\includegraphics[width=0.5\textwidth]{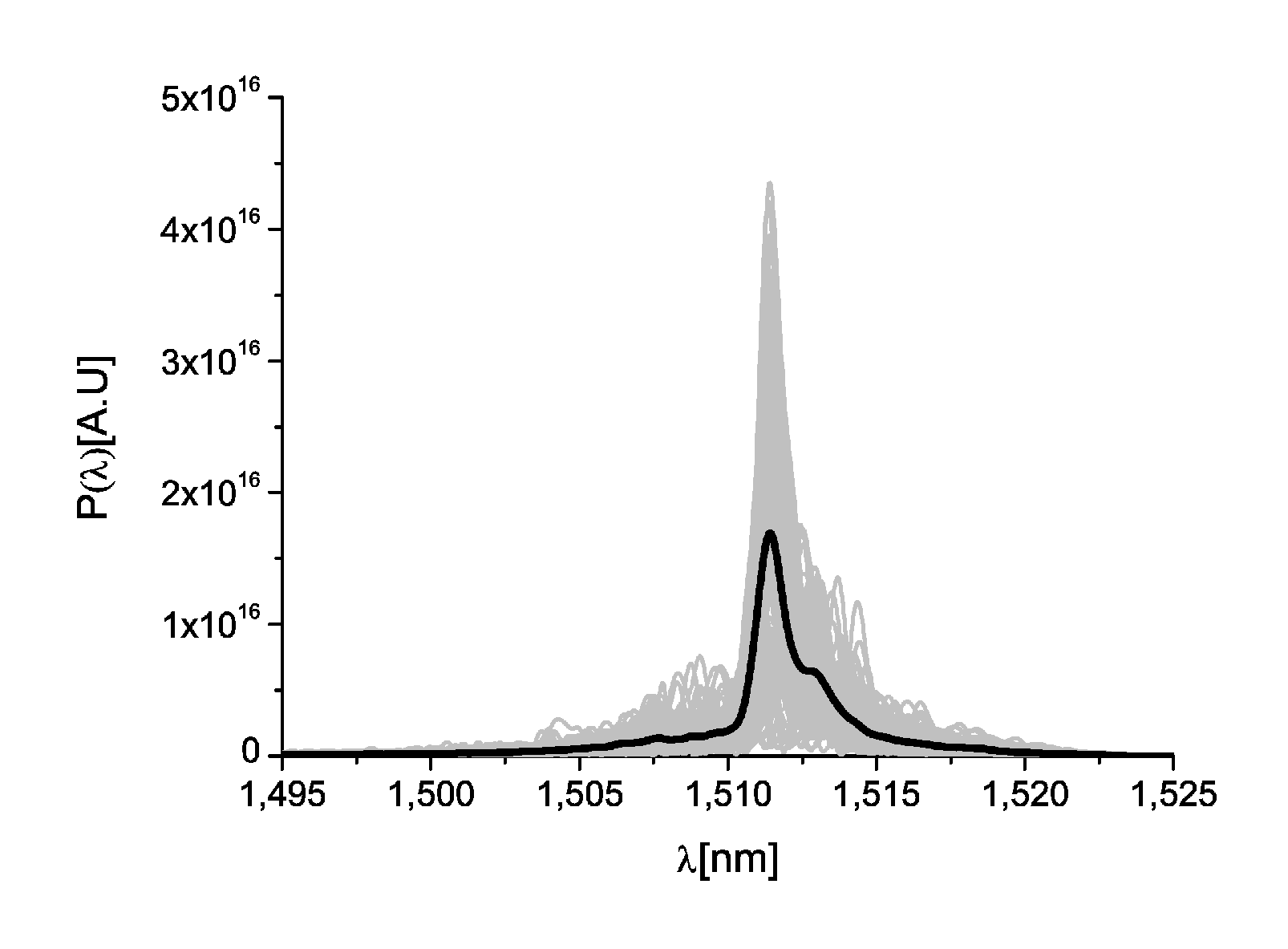}
\caption{Power distribution and spectrum of the X-ray radiation
pulse after the second undulator in the tapered case. Grey lines
refer to single shot realizations, the black line refers to the
average over a hundred realizations.} \label{OUTT1}
\end{figure}
\begin{figure}[tb]
\includegraphics[width=0.5\textwidth]{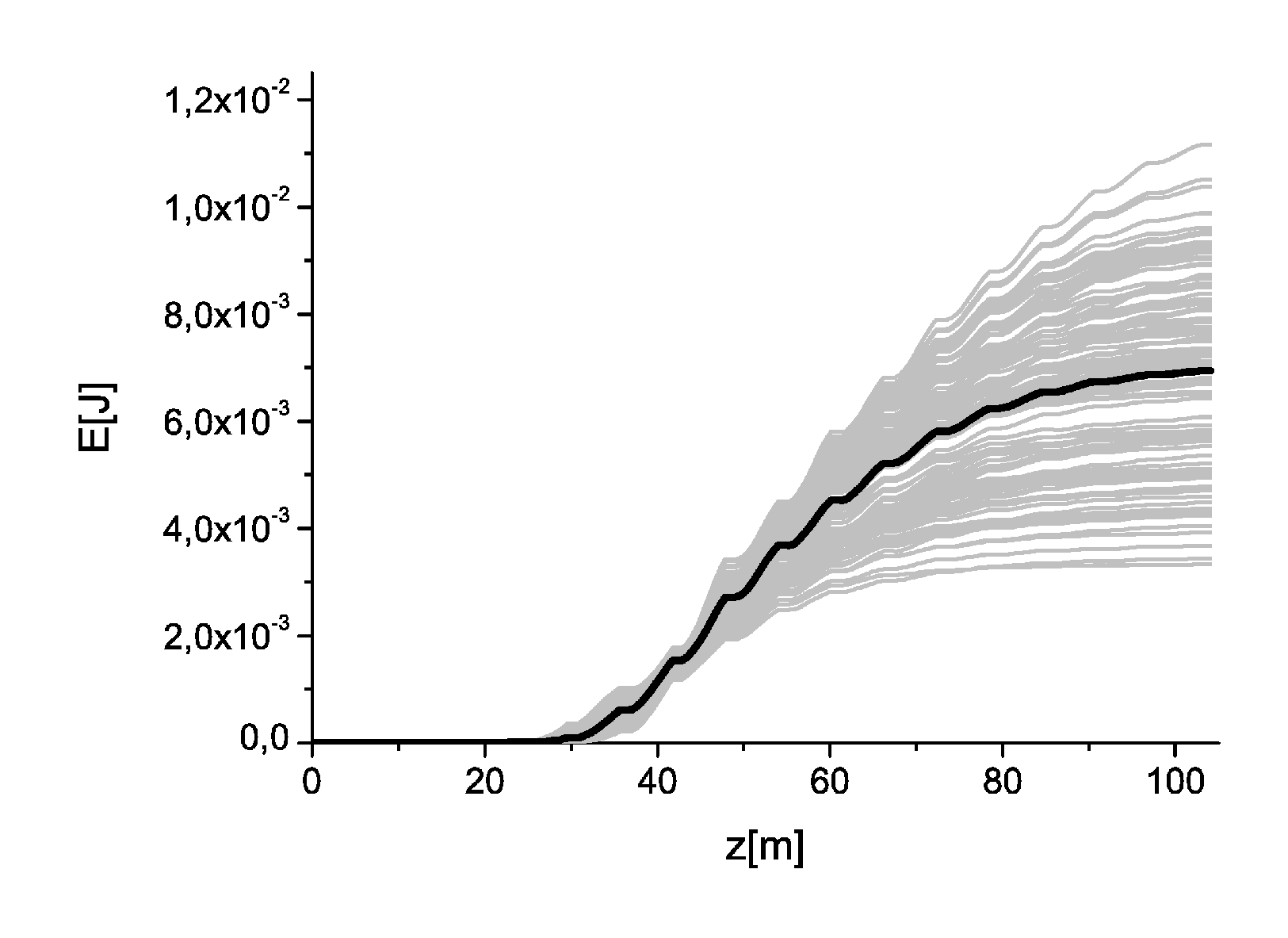}
\includegraphics[width=0.5\textwidth]{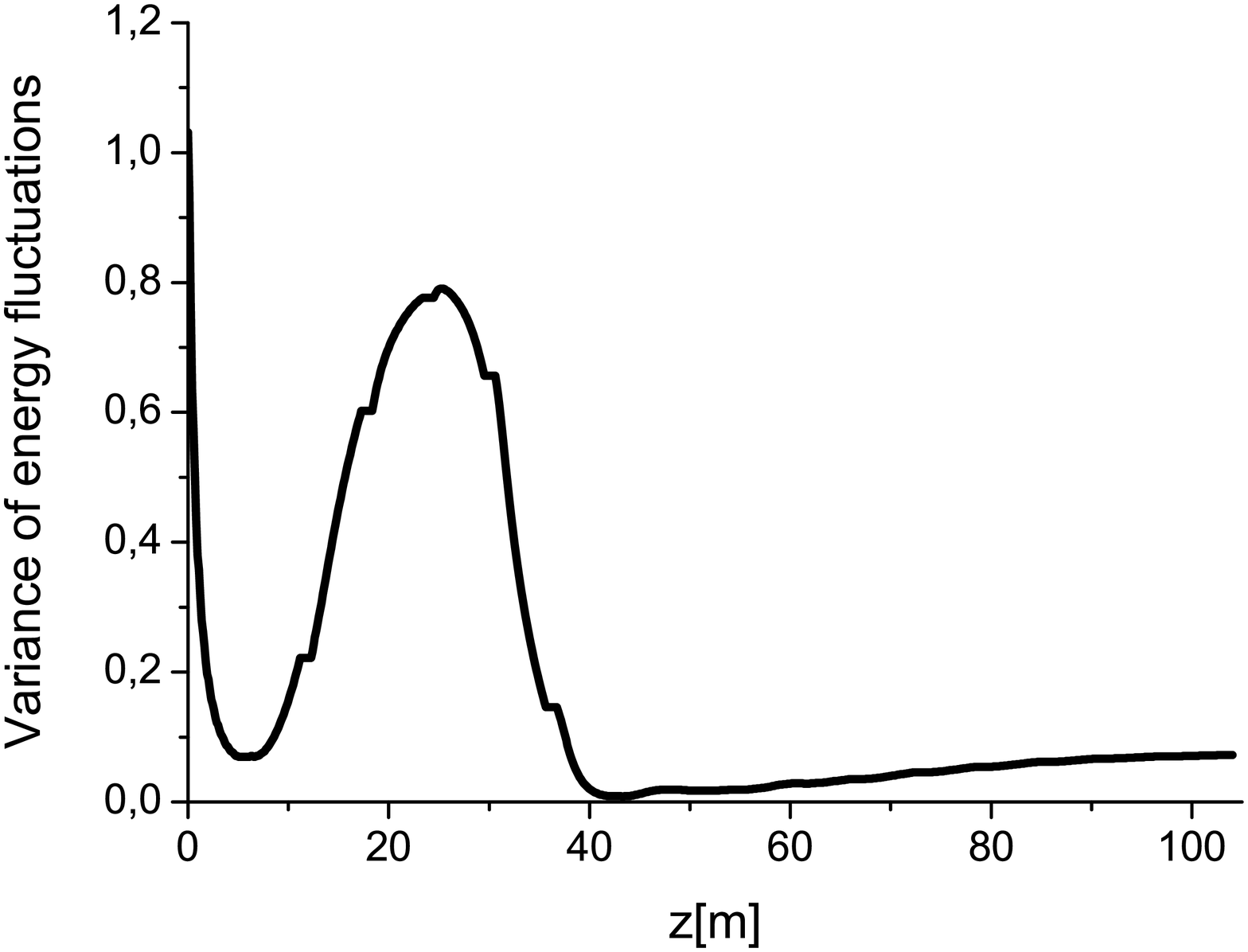}
\caption{Evolution of the energy per pulse and of the energy
fluctuations as a function of the undulator length in the tapered
case. Grey lines refer to single shot realizations, the black line
refers to the average over a hundred realizations.} \label{OUTT2}
\end{figure}
\begin{figure}[tb]
\includegraphics[width=0.5\textwidth]{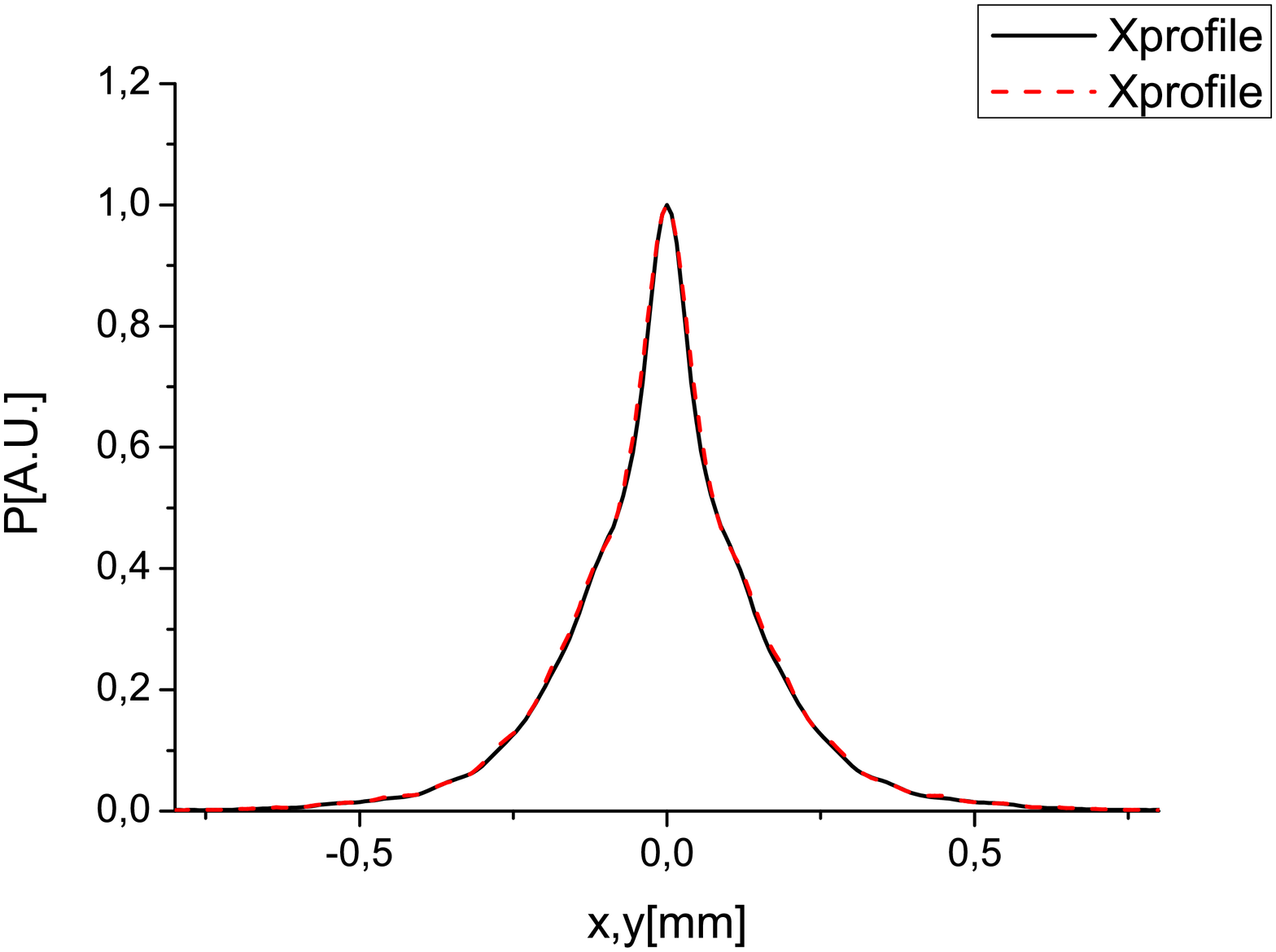}
\includegraphics[width=0.5\textwidth]{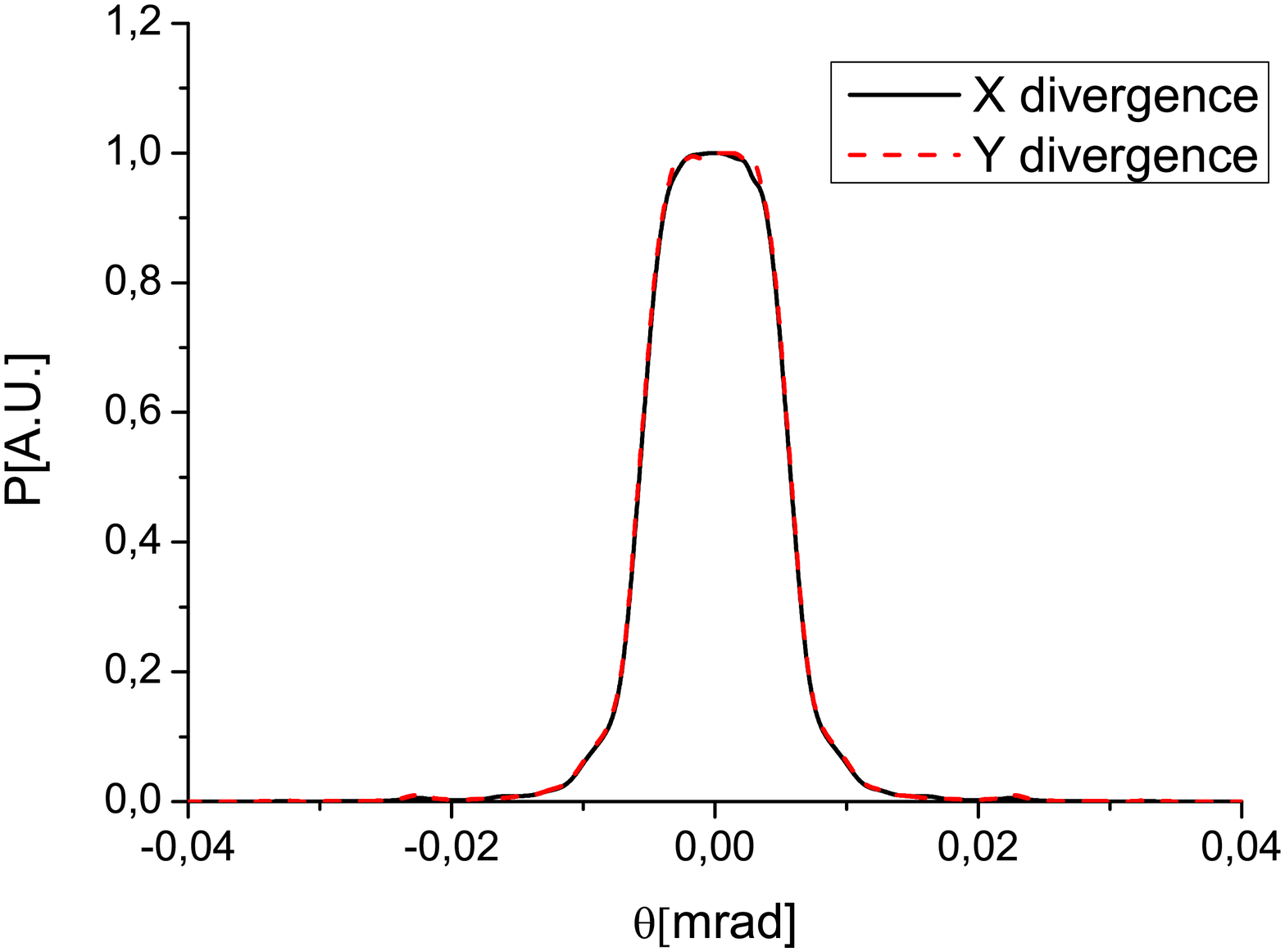}
\caption{(Left plot) Transverse radiation distribution in the case
of tapering at the exit of the output undulator. (Right plot)
Directivity diagram of the radiation distribution in the case of
tapering at the exit of the output undulator.} \label{spotT}
\end{figure}
The most promising way to increase the output power is via
post-saturation tapering. Tapering consists in a slow reduction of
the field strength of the undulator in order to preserve the
resonance wavelength, while the kinetic energy of the electrons
decreases due to the FEL process. The undulator taper could be
simply implemented as a step taper from one undulator segment to the
next, as shown in Fig. \ref{krms}. The magnetic field tapering is
provided by changing the undulator gap. A further increase in power
is achievable by starting the FEL process from the monochromatic
seed, rather than from noise. The reason is the higher degree of
coherence of the radiation in the seed case, thus involving, with
tapering, a larger portion of the bunch in the energy-wavelength
synchronism. Using the tapering configuration in Fig. \ref{krms},
one obtains the output characteristics, in terms of power and
spectrum, shown in Fig. \ref{OUTT1}. The output power is increased
of about a factor ten, allowing one to reach about one TW. The
spectral width remains almost unvaried, with an average relative
bandwidth (FWHM) narrower than $10^{-3}$. The evolution of the
energy per pulse and of the energy fluctuations as a function of the
undulator length are shown in Fig. \ref{OUT2}. The transverse
radiation distribution and divergence at the exit of the output
undulator are shown in Fig. \ref{spotT}. By comparison with Fig.
\ref{spot} one can see that the divergence decrease is accompanied
by an increase in the transverse size of the radiation spot at the
exit of the undulator.

\begin{figure}[tb]
\includegraphics[width=0.5\textwidth]{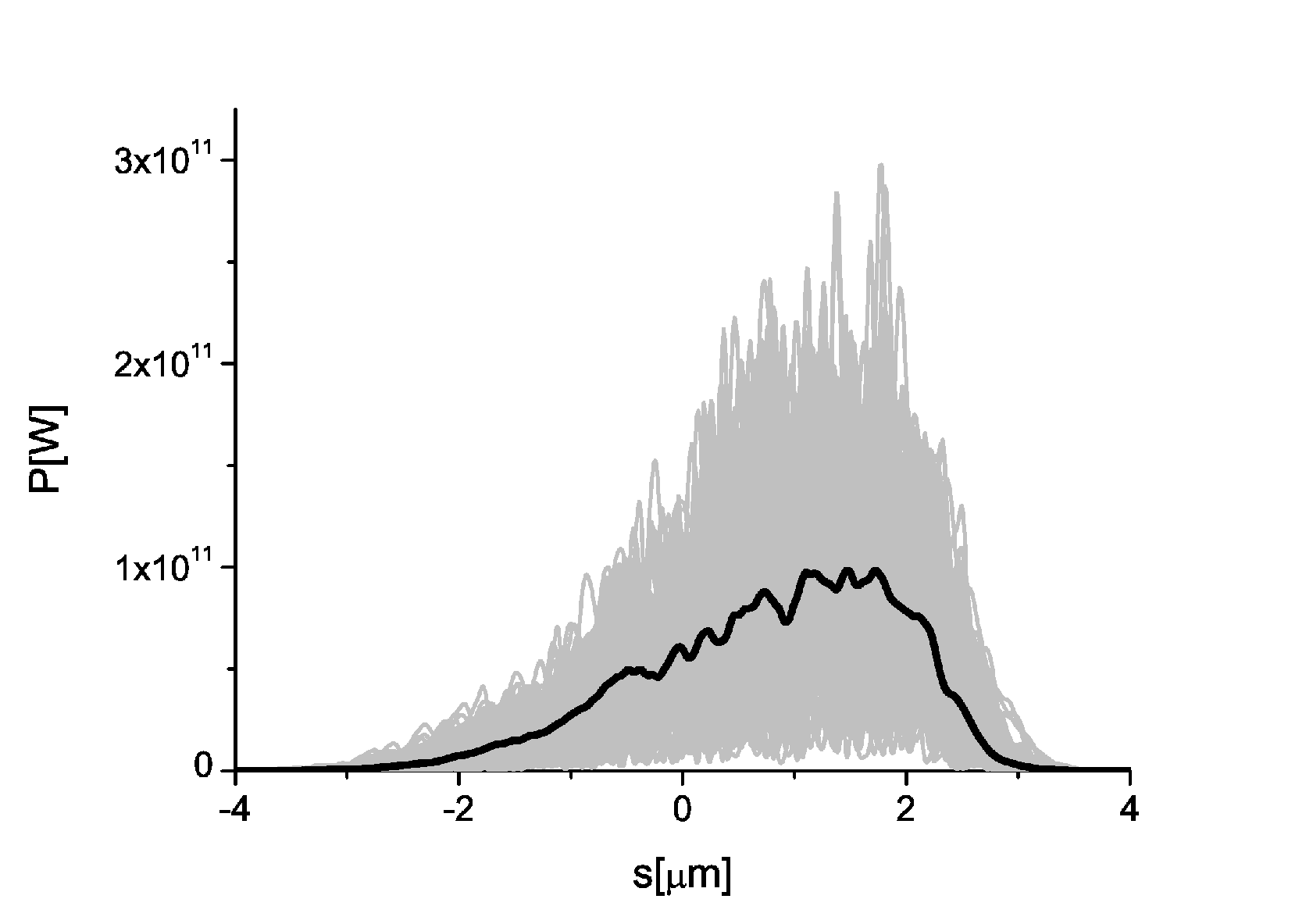}
\includegraphics[width=0.5\textwidth]{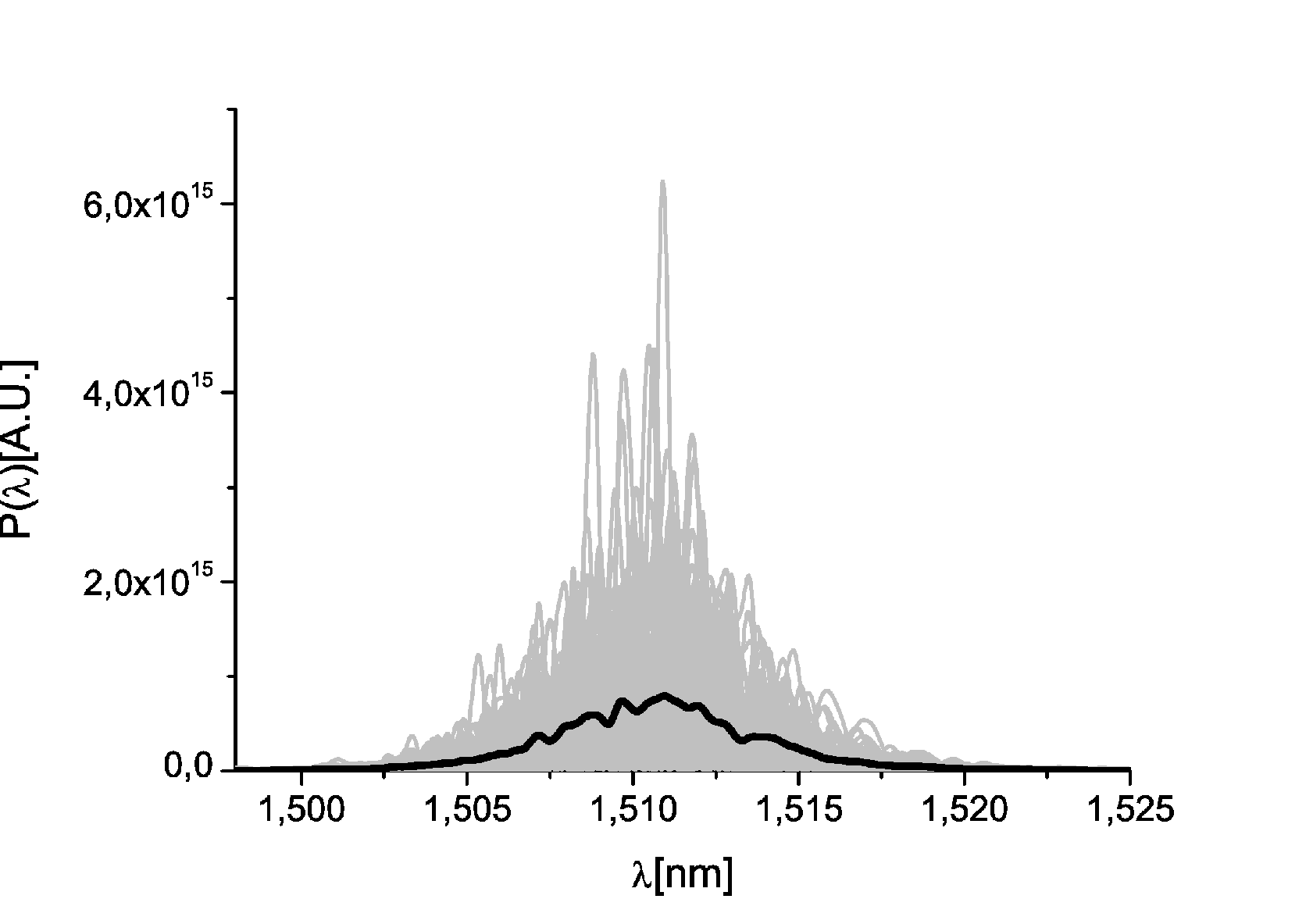}
\caption{Power distribution and spectrum of the baseline SASE X-ray
radiation pulse at saturation. Grey lines refer to single shot
realizations, the black line refers to the average over a hundred
realizations.} \label{SASEA}
\end{figure}
\begin{figure}[tb]
\includegraphics[width=0.5\textwidth]{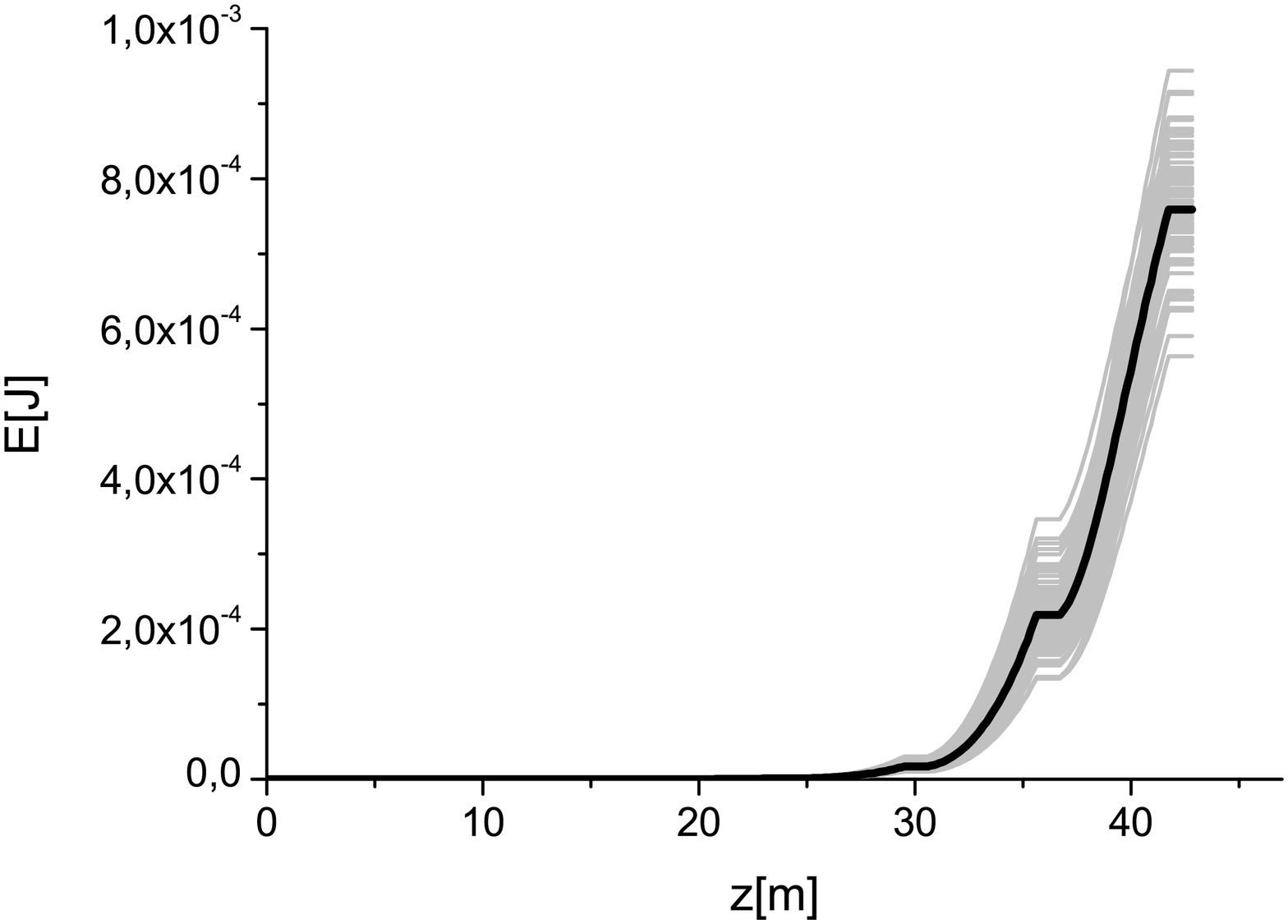}
\includegraphics[width=0.5\textwidth]{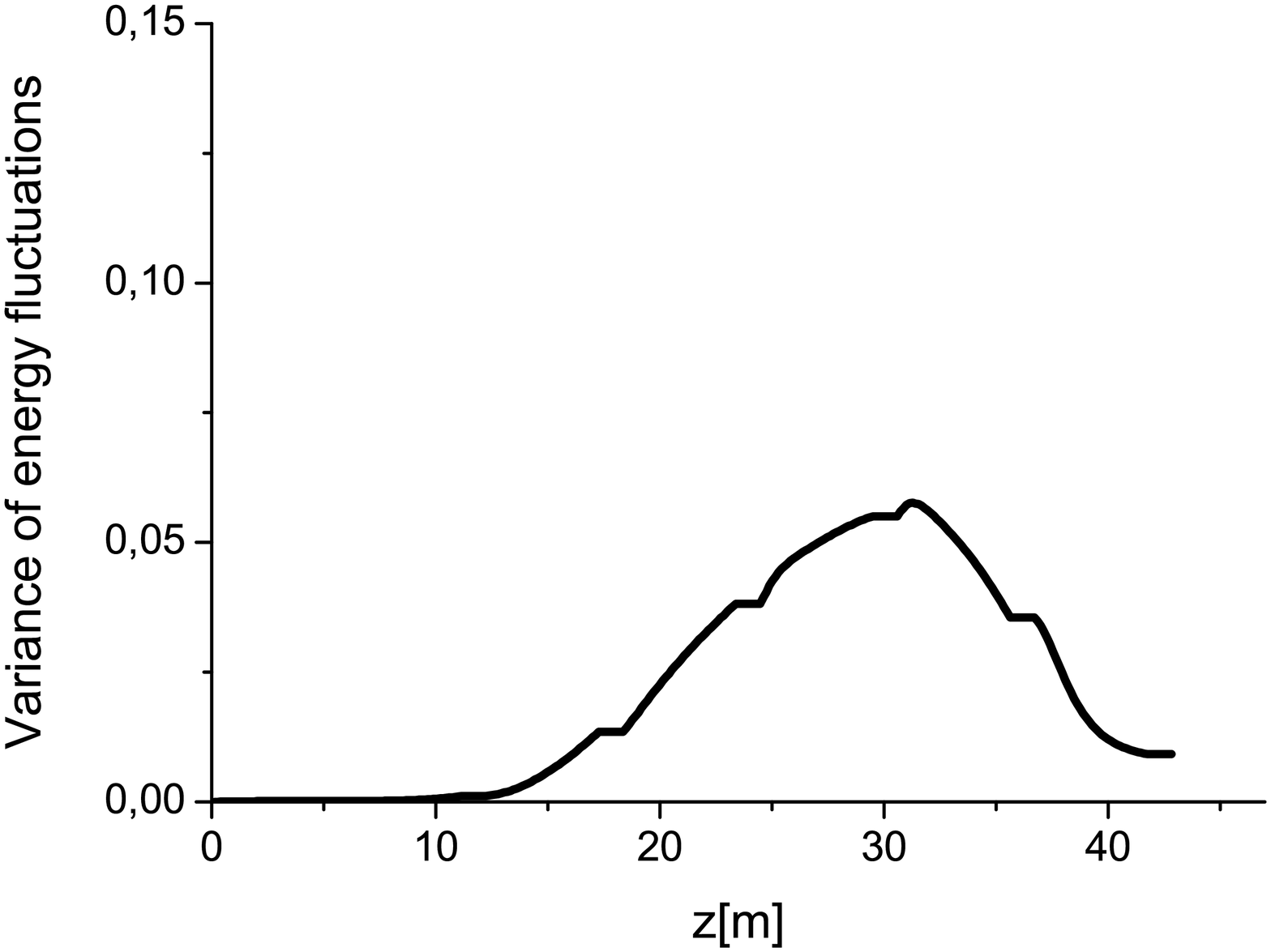}
\caption{Evolution of the energy per pulse and of the energy
fluctuations as a function of the undulator length in the case of
the baseline SASE pulse. Grey lines refer to single shot
realizations, the black line refers to the average over a hundred
realizations.} \label{SASEB}
\end{figure}
\begin{figure}[tb]
\includegraphics[width=0.5\textwidth]{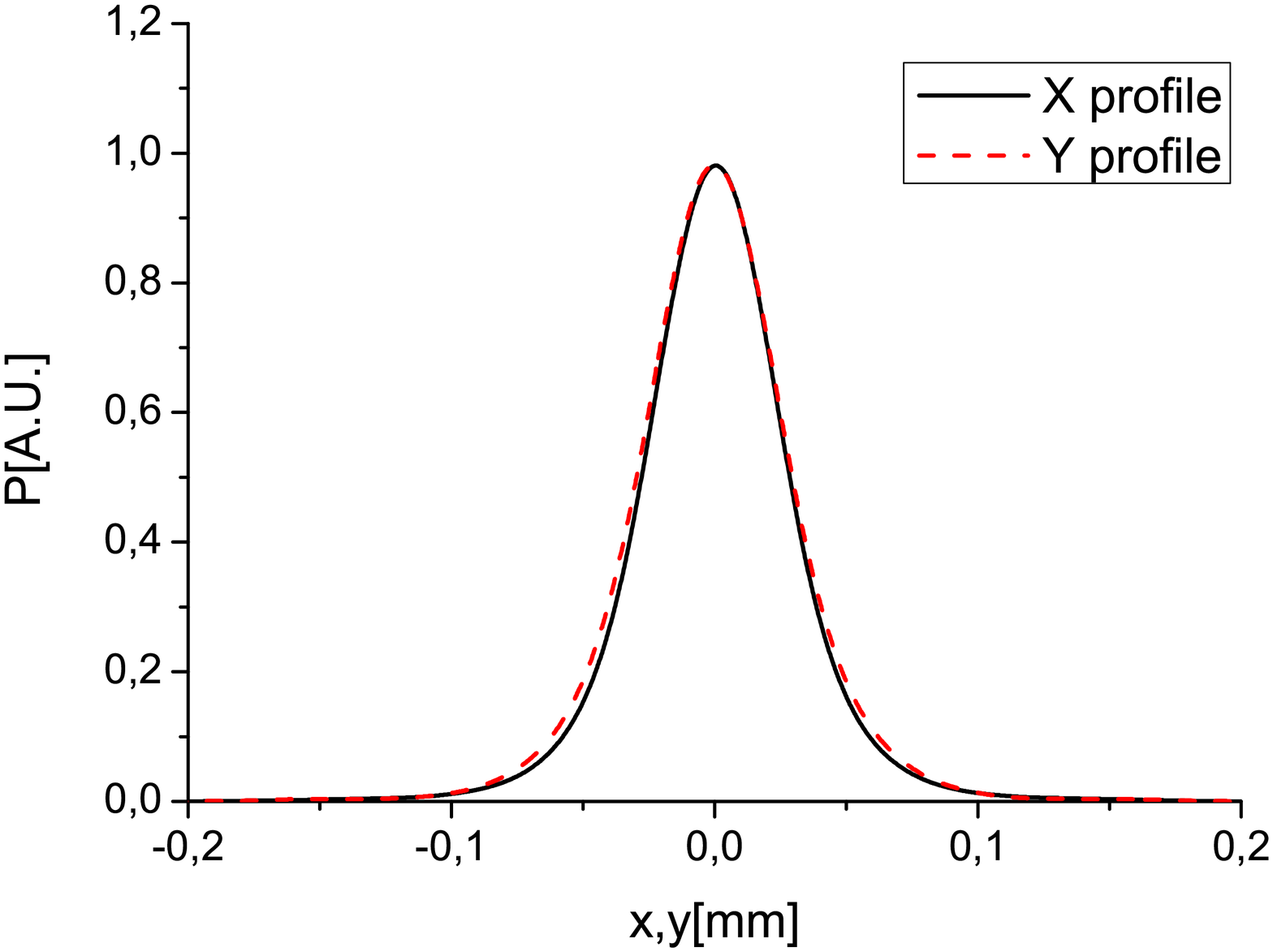}
\includegraphics[width=0.5\textwidth]{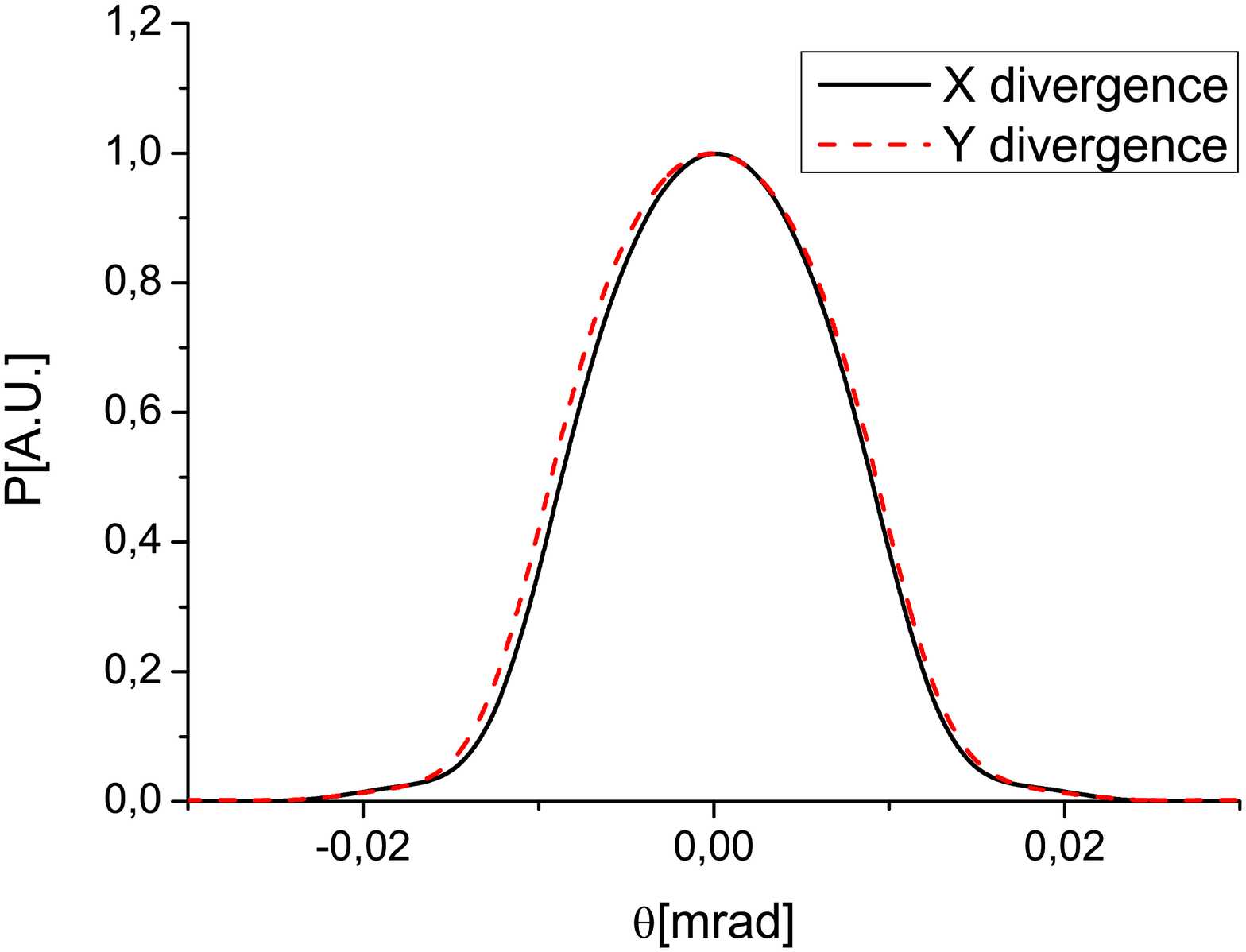}
\caption{(Left plot) Transverse radiation distribution in the case
of the baseline SASE pulse at saturation. (Right plot) Directivity
diagram of the radiation distribution in the case of the baseline
SASE pulse at saturation.} \label{SASEC}
\end{figure}
Finally, it is interesting to compare the results for the
self-seeded beam with the characteristics of the SASE pulse at SASE3
generated with the same electron beam. The output SASE
characteristics at saturation, in terms of power and spectrum, shown
in Fig. \ref{SASEA}. The evolution of the energy in the SASE pulse
and of the energy fluctuations as a function of the undulator length
are shown in Fig. \ref{SASEB}. The transverse radiation distribution
and divergence at saturation are shown in Fig. \ref{SASEC}.

\section{\label{concl} Conclusions}

Historically, self-seeding methods were first  proposed for the soft
X-ray region, and were based on the use of grating monochromators
\cite{SELF}, \cite{STTF}. However, self-seeding techniques were
first successfuly demonstrated in the hard X-ray region at the LCLS
\cite{EMNAT}, based on the use of a crystal monochromator. The
working principle for such monochromator was invented in
\cite{OURLB}, and resulted into a very compact self-seeding setup
design fitting within a single undulator module. During the last
three years, significant efforts were dedicated to both theoretical
investigation and $R\&D$ at the LCLS, leading to the design of a
compact self-seeding setup in the soft X-ray range, based on a
grating monochromator. The evolution of the design can be
reconstructed from \cite{FENG}-\cite{FENG3}, striving at the same
time for the needed resolution and compactness. The development of a
self-seeding grating monochromator with the same compactness of a
single-crystal monochromator is a challenging problem. However, the
final design of the LCLS  soft X-ray self-seeding setup has many
advantages. It is very compact and fits within one undulator module.
It is very simple and includes only four optical elements. It does
not include an entrance slit and during discussions, authors of
\cite{FENG3} pointed that this design might be operated  even
without exit slit by using the electron beam and the spatial
dispersion at the second undulator entrance  for spectral filtering
purpose.

In this article we present a technical study for a soft x-ray
self-seeding setup at the European XFEL. In particular we focus on
design and performance of a very compact self-seeding grating
monochromator, based on the LCLS design, which has been adapted to
the needs of the European XFEL. Usually, soft X-ray monochromators
operate with incoherent sources and their design is based on the use
of ray-tracing codes. However, XFEL beams are almost completely
transversely coherent, and in our case the optical system was
studied using a wave optics method in combination with FEL
simulations to evaluate the performance of the self-seeding scheme.
Our wave optics analysis takes into account the actual FEL beam
wavefront, third order aberrations and surface errors from each
optical elements. Wave optics together with FEL simulations are
naturally applicable to the study the influence of finite slit size
on the seeding efficiency. Most results presented in \cite{FENG3}
were obtained in the framework of a Gaussian beam model, in
combination with ray-tracing for Gaussian ray distribution. This is
a very fruitful approach, allowing one for studying many features of
the self-seeding monochromator by means of relatively simple tools.
Using our approach, we give a quantitative answer to the question of
the accuracy of the Gaussian beam model. It is also important to
quantitatively analyze the filtering process without exit slit. Wave
optics in combination with FEL simulations is the only method
available to this aim. We conclude that the mode of operation
without slit is superior to the conventional mode of operation, and
a finite slit size would only lead to a reduction of the
monochromator performance. We therefore propose an optimized design
based on a toroidal VLS grating and three mirrors, without exit
slit. The monochromator covers the range between $300$ eV and $1000$
eV, with a resolution never falling below $7000$, and introduces a
photon delay of only $0.7$ ps. This allows the entire self-seeding
setup to be fit into a single $5$ m-long undulator segment. The
overall performance of the setup is studied with the help of FEL
simulations, which show that, in combination with post-saturation
tapering, the SASE3 baseline at the European XFEL could deliver
TW-class, nearly Fourier-limited radiation pulses in the soft X-ray
range. Although we explicitly studied the a soft x-ray self-seeding
setup for the SASE3 undulator baseline at the European XFEL, the
same setup can be used without modifications also for the dedicated
bio-imaging beamline, a concept that was proposed in
\cite{BIO1}-\cite{BIO3} as a possible future upgrade of the European
XFEL. By exploiting third harmonic generation and fresh bunch
technique together with the self-seeding mode of operation
\cite{BZVI}-\cite{WU} one can extend the operation of the soft x-ray
self-seeding setup to the range between $1$ keV up to $3$ keV, thus
covering the sulfur K-edge without changes in the grating
monochromator design \cite{BIO3}. The X-ray beam will thus be
delivered in ultrashort pulses with 1 TW peak power within the
extended photon energy range between 0.3 keV up to 3 keV. For
operation at higher photon energies x-ray self-seeding setups based
on single crystal monochromators can be used.

\section{Acknowledgements}

We thank Daniele Cocco, Paul Emma, Yiping Feng, Jerome Hastings,
Philip Heimann and Jacek Krzywinski for useful discussions. We are
grateful to Massimo Altarelli, Reinhard Brinkmann, Henry Chapman,
Janos Hajdu, Viktor Lamzin, Serguei Molodtsov and Edgar Weckert for
their support and their interest during the compilation of this
work.

\end{document}